\documentclass[12pt]{article}

\usepackage{rotating}
\usepackage{longtable}
\usepackage{threeparttablex}
\usepackage{scicite}
\usepackage{times}
\usepackage[normalem]{ulem}
\usepackage{graphicx}
\usepackage{amssymb}
\usepackage{amsmath}
\usepackage{graphicx}
\usepackage{dcolumn}
\usepackage{bm}
\usepackage{comment}
\usepackage{siunitx}
\usepackage{graphicx}
\usepackage{xcolor}
\usepackage{xspace}
\usepackage{multirow}
\usepackage{CJKutf8}            
\usepackage{authblk}

\usepackage{hyperref}

\hypersetup{
    colorlinks=true,                
    breaklinks=true,                
    urlcolor= black,                
    linkcolor= blue,                
    bookmarksopen=false,
    filecolor=black,
    citecolor=blue,
    linkbordercolor=blue}

\date{}

\makeatletter
\renewcommand{\fnum@figure}{\textbf{Figure \thefigure}}
\renewcommand{\fnum@table}{\textbf{Table \thetable}}
\makeatother

\makeatletter

\makeatletter

\newenvironment{sciabstract}{%
\begin{quote} \bf}
{\end{quote}}

\title{ Star formation powers optical line emission from the CGM}

\author[1,2,3]{Huanian Zhang \begin{CJK*}{UTF8}{gkai}(张华年)$^*$\end{CJK*}} 
\author[4]{Lizhi Xie \begin{CJK*}{UTF8}{gkai}(谢利智)\end{CJK*}} 
\author[5]{Zhijie Qu \begin{CJK*}{UTF8}{gkai}(屈稚杰)\end{CJK*}} 
\author[6]{Dennis Zaritsky} 
\author[7,8]{Luis C. Ho} 
\author[9,10]{Min Bao \begin{CJK*}{UTF8}{gkai}(鲍敏)\end{CJK*}} 
\author[11,12]{Fabio Fontanot}
\author[13,11]{Michaela Hirschmann} 
\author[14]{Chenxu Liu \begin{CJK*}{UTF8}{gkai}(刘辰旭)\end{CJK*}} 
\author[11,12]{Gabriella De Lucia} 
\author[15,16]{Enci Wang \begin{CJK*}{UTF8}{gkai}(王恩赐)\end{CJK*}} 
\author[1]{Haoyan Zheng \begin{CJK*}{UTF8}{gkai}(郑皓焱)\end{CJK*}}

\affil[1]{Department of Astronomy,  Huazhong University of Science and Technology, Wuhan, Hubei 430074, People’s Republic of China, $^*$Corresponding author: huanian@hust.edu.cn}
\affil[2]{Hubei Fundamental Research Center for Physics, Wuhan, Hubei 430074, China}
\affil[3]{Hubei Key Laboratory of Gravitation and Quantum Physics, Huazhong University of Science and Technology, Wuhan, Hubei 430074, China}
\affil[4]{Astrophysics Center, Tianjin Normal University, Binshuixidao 393, Xiqing, Tianjin, People’s Republic of China}
\affil[5]{Department of Astronomy, Tsinghua University, Beijing 100084, People’s Republic of China}
\affil[6]{Steward Observatory, University of Arizona, Tucson, AZ, USA}
\affil[7]{Kavli Institute for Astronomy and Astrophysics, Peking University, Beijing 100871, People’s Republic of China}
\affil[8]{Department of Astronomy, School of Physics, Peking University, Beijing 100871, People’s Republic of China}
\affil[9]{School of Physics and Technology, Nanjing Normal University, Nanjing 210023, People's Republic of China}
\affil[10]{School of Astronomy and Space Science, Nanjing University, Nanjing 210023, People’s Republic of China}
\affil[11]{INAF – Astronomical Observatory of Trieste, via G.B. Tiepolo 11, I-34143 Trieste, Italy}
\affil[12]{IFPU - Institute for Fundamental Physics of the Universe, via Beirut 2, 34151, Trieste, Italy}
\affil[13]{Dipartimento di Fisica G. Occhialini, Universit\`a degli Studi di Milano Bicocca, Piazza della Scienza 3, 20126 Milano, Italy}
\affil[14]{South-Western Institute for Astronomy Research, Yunnan University, Kunming 650500, People's Republic of China}
\affil[15]{Department of Astronomy, University of Science and Technology of China, Hefei, Anhui 230026, People’s Republic of China}
\affil[16]{School of Astronomy and Space Science, University of Science and Technology of China, Hefei, Anhui 230026, People’s Republic of China}

\begin{document}

\maketitle

\textbf{Abstract}
\begin{sciabstract}

Using integral field spectroscopy, we explore the disk-halo interface, or the inner circumgalactic medium (CGM), of individual galaxies by constructing and analyzing emission-line maps for a large sample (72) of normal, low-redshift galaxies spanning three orders of magnitude in stellar mass and four orders in star formation rate (SFR). We find a steep turnover occurring at $(1-2) R_e$ in the H$\alpha$, [O {\small II}], and [O {\small III}]  line emission radial profiles. Beyond this radius, the slope of the line emission radial profiles becomes shallower as the SFR of the central galaxy decreases, which might reflect the strength of the feedback processes.  The line emission fluxes at large radius ($(5-10) R_e$ or $\sim (0.1-0.25)r_{\rm vir}$) correlate with the galaxy's SFR, but not with its stellar mass. These findings suggest that ionizing photons escaping from star-forming regions in the central galaxy 
account for the observed emission line fluxes from the inner CGM, with escape fractions inferred from the [O {\small III}] and [O {\small II}] ratio. Different state-of-the-art theoretical models do not agree on the predicted dependence of cool gas on the SFR of  the central galaxies, highlighting the importance of CGM emission line measurements to distinguish between different subgrid models for star formation and feedback processes. 

Short Title: The inner CGM of a large sample of individual galaxies.

Teaser: Star formation 
drives the observed emission line fluxes from the inner CGM.

\end{sciabstract}

\section{Introduction}

The circumgalactic medium (CGM) of galaxies is the fundamental baryon reservoir for galaxies\cite{Behroozi2010, McGaugh2010, CGM2017, Peroux2020, CGM2023} and has been studied primarily in absorption along a limited number of sightlines taking advantage of the features it imprints on the spectra of bright background sources\cite{Tumlinson2013,Werk2014,CGM2017}. 
The detection of the CGM in emission lines, while observationally more challenging, allows for a more complete characterization of its properties by constructing two-dimensional distribution maps. 
The first detection of  H$\alpha$ and [N{ \small II}] $\lambda$6585 emission  extending to projected radii of $\sim$ 100 kpc from low-redshift, normal galaxies\cite{Zhang2016}
used massive stacks of Sloan Digital Sky Survey (SDSS)\cite{sdss, SDSSDR16} spectra and reached a surface brightness (SB) sensitivity of $\sim 10^{-20} \ {\rm erg} \ {\rm cm}^{-2} \ {\rm s}^{-1} \ {\rm arcsec}^{-2}$. Subsequent studies characterized the line-emitting, cool CGM within 50 kpc or 0.25 $r_{\rm vir}$ in low redshift, normal galaxies\cite{Zhang2018a, Zhang2018b, Zhang2019, Zhang2020a, Zhang2020b, Zhang2021, Zhang2022, Zhang2024a}, as a function of radius, stellar mass, environment, interaction, and orientation angle. Using long exposures with integral field spectrographs (IFS) on Multi Unit Spectroscopic Explorer (MUSE)\cite{MUSE2010}, the average extended emission in multiple lines (Mg {\small II} $\lambda \lambda$2796,2803, [O {\small II}]$\lambda \lambda$3727,3729) can be detected out to $\sim 25-45$ kpc with a similar SB sensitivity of $\sim 10^{-20} \ {\rm erg} \ {\rm cm}^{-2} \ {\rm s}^{-1} \ {\rm arcsec}^{-2}$ by stacking a sample of $\sim$ 600 galaxies at $z\approx1$\cite{Dutta2023, Dutta2024}, and bipolar outflows out to $\sim 10$ kpc traced by Mg {\small II} $\lambda \lambda$2796,2803 emission are detected by stacking a few tens of galaxies with stellar mass $>10^{9.5}$ M$_\odot$ at $z\approx1$\cite{Guo2023}.  These studies primarily use the Mg {\small II} $\lambda \lambda$2796,2803 emission line as a tracer. However, as a resonant line, it traces mostly the neutral gas. Consequently, a direct probe of the cool, ionized gas is still lacking. 

Using long exposures with an IFS on an 8-m or 10-m  telescopes,
one can study the cool CGM 
in specific systems where one might have anticipated stronger than typical CGM emission. Examples of such work include  the studies of   a compact massive galaxy (M$_* \sim 10^{10.3}$ M$_\odot$) with a star formation rate (SFR) of $\sim10$ M$_\odot$ yr$^{-1}$ at $z= 0.737$\cite{Pessa2024},  a starburst sub-$L^*$  galaxy (M$_* \sim 10^{10.05}$ M$_\odot$) with a SFR of $10\sim20$ M$_\odot$ yr$^{-1}$ at $z= 0.702$\cite{Zabl2021} that is probed by a quasar sightline, a starburst sub-$L^*$ galaxy (M$_* \sim 10^{9.9}$ M$_\odot$) with a SFR of 50 M$_\odot$ yr$^{-1}$ at $z= 0.6942$\cite{Burchett2021}, a massive galaxy (M$_* \sim 10^{11.1}$ M$_\odot$) at $z=0.459$ with a SFR of 245 M$_\odot$ yr$^{-1}$\cite{Rupke2019},  a nearby starburst sub-$L^*$ galaxy (M$_* \sim 10^{10.1}$ M$_\odot$) with a SFR of 12.1 M$_\odot$ yr$^{-1}$ at $z=0.01911$\cite{Nielsen2024}, a very low-mass blue compact dwarf galaxy (M$_* \sim 6\times 10^{6}$ M$_\odot$, SFR $\sim$ 0.7 M$_\odot$ yr$^{-1}$)\cite{Herenz2023} in the Local Universe, the nearby starburst M 82\cite{Lokhorst2022}, and a source-blind survey for emission from the CGM across a wide range of redshifts in one deep MUSE field\cite{Zhang2024b}. However, the missing piece was the ability to do similar studies on a large sample of individual, more representative galaxies across a wide range of stellar masses and star formation rates.  

Using the Mapping Nearby Galaxies at Apache Point Observatory (MaNGA)\cite{manga} integral field spectroscopic data, we map the cool CGM using multiple optical emission lines 
in a large sample of individual nearby galaxies at sufficiently high spatial resolution.
The MaNGA survey is a large integral field spectroscopic survey targeting $\sim$10,000 nearby galaxies with the hexagonal fiber-bundle integral field units (IFU)\cite{Drory2015}. Among the full survey, there are 72 galaxies at $0.015 < z < 0.146$ targeted with an IFU bundle with $>$ 91 fibers for which the field of view (FOV) is $>$ 5 times the half-light, or effective, radius ($R_e$) of the galaxy. For these galaxies we can probe the line emission originating from the inner CGM  ($\gtrsim 5R_e$ and within the MaNGA FOV, details in Methods). We construct the narrow-band images of the optical emission lines using velocity windows of $150 - 375$ km s$^{-1}$,  depending on the stellar mass of the galaxy (details in Methods),  across each MaNGA fiber bundle field of view. The stellar mass, star formation rate and specific star formation rate (sSFR) of our MaNGA galaxy sample span $\sim 3-4$  orders of magnitude, with which we can probe how the line emission from the cool CGM depends on galaxy properties  and can constrain its origin. We do not correct the emission-line fluxes for dust extinction. Reliable extinction measurements due to the CGM are challenging, and the effect is expected to be negligible \cite{Menard2010, Peek2015}, which report circumgalactic dust extinction on the order of just 0.03 mag in the V-band. In the analysis of these data that we present below, we adopt a $\Lambda$CDM cosmology with parameters $\Omega_m$ = 0.3, $\Omega_\Lambda =$ 0.7, $\Omega_k$ = 0, and $h = $ 0.7\cite{riess,Planck2018}.

\section{Results and Discussion}

\subsection{Surface Brightness and Ionization Radial Profile}

\begin{figure*}[ht!]
\begin{center}
\includegraphics[width = 0.475 \textwidth]{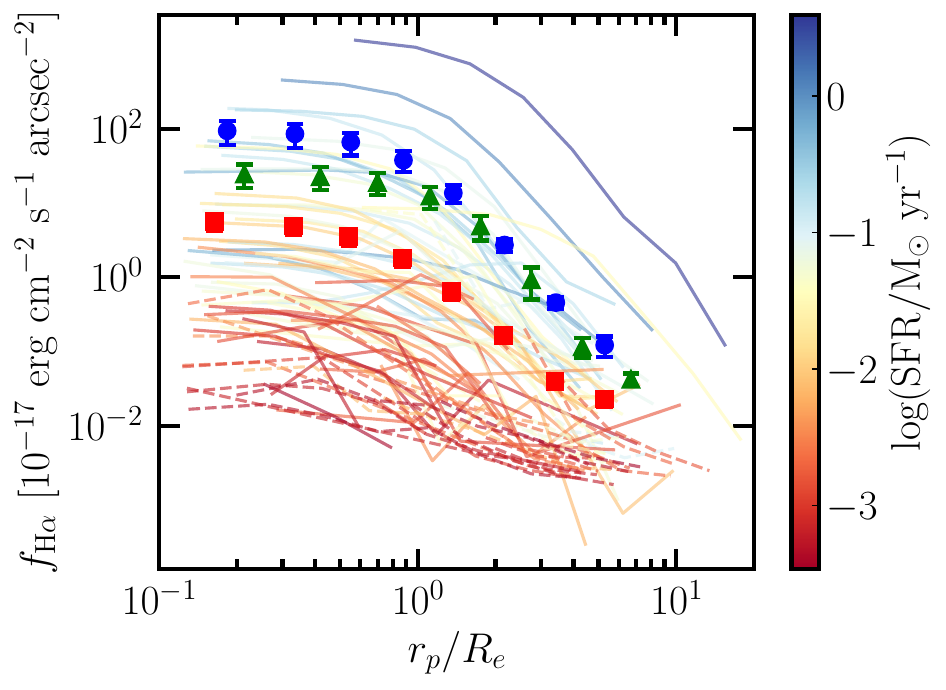}
\hspace{0.5 cm}
\includegraphics[width = 0.475 \textwidth]{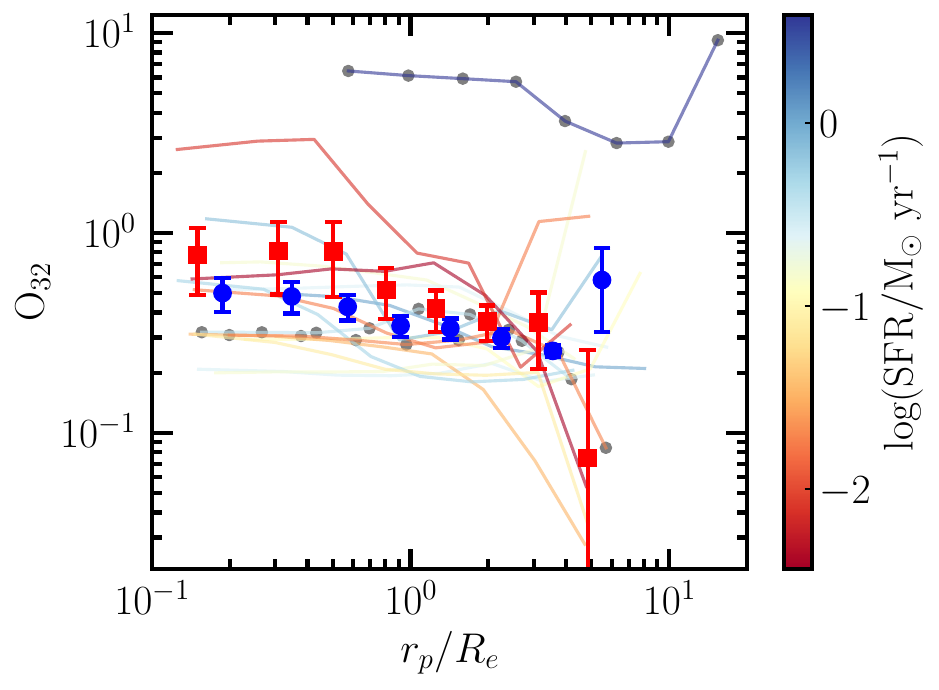}
\end{center}
\caption{{\bf The radial profile of H$\alpha$ emission SB and O$_{32}$.} The H$\alpha$ emission surface brightness radial profile  of all our MaNGA galaxies (left) and the ratio O$_{32}$ (right). In both panels, we color code the profiles by the galaxy's star formation rate. In the left panel, dashed lines denote non-detections where we adopt the 2$\sigma$ upper limit. The blue, green, and red points represent the mean stacked radial profiles for galaxies within SFR bins of $-1 < \log(\rm SFR/M_\odot \ yr^{-1})< 0 $, $-1.75 <\log(\rm SFR/M_\odot \ yr^{-1})<-1$ and $\log(\rm SFR/M_\odot \ yr^{-1})<-1.75$, respectively, using only galaxies with detections in all radial bins. In the right panel, we use  only two SFR bins of ($-1 < \log(\rm SFR/M_\odot \ yr^{-1})<0$ and $\log(\rm SFR/M_\odot \ yr^{-1})<-1$), again using only galaxies with both [O {\small III}] and [O {\small II}] detections at all radii. The little dots in the right panel indicate the three AGN galaxies } 
\label{fig:radialProf}
\end{figure*}

We present the H$\alpha$ emission surface brightness radial profiles for all of our MaNGA galaxies in  the left panel of Figure \ref{fig:radialProf}. In the three SFR bins of $-1 < \log(\rm SFR/M_\odot \ yr^{-1})< 0 $, $-1.75 <\log(\rm SFR/M_\odot \ yr^{-1})<-1$ and $\log(\rm SFR/M_\odot \ yr^{-1})<-1.75$, we also present the mean stacked profiles. Only profiles with 3$\sigma$ detection for all radial bins are included. There are 11, 8 and 7 galaxies in those three stacks, respectively. We find that the median stacks provide consistent results with the mean stacks. Our analysis clearly shows that the normalization of the H$\alpha$ emission surface brightness radial profile is positively correlated to the SFR of the galaxies, which will be further discussed in the following.
Moreover, we find a steep turnover occurring at $\sim(1-2) R_e$ in the detected emission profiles (solid lines). 
Although the trend that the slope beyond the turnover radius ($\gtrsim (1-2) R_e$) becomes shallower as the SFR decreases is not that obvious for individual galaxies (color lines), the stacked profiles (points with error-bars) show a much shallower slope  with $\sim 3 \sigma$ significance for galaxies in the lowest SFR bin compared to that in the highest SFR bin, which might reflect the strength of the accumulated feedback processes. The slopes of the three mean (median) stacked profiles beyond 1 $R_e$ are $-3.5\pm0.1$ ($-3.7\pm0.3$), $-3.5\pm0.3$ ($-3.0\pm0.3$), $-2.5\pm0.3$ ($-2.5\pm0.3$) for the three stacks, respectively. The other strong emission lines [O {\small II}]$\lambda \lambda$3727,3729 and [O {\small III}]$\lambda$5008 (referred to as [O {\small II}] and [O {\small III}] afterwards)  have similar radial profiles (see details in Methods). 

The ratio O$_{32} \equiv$ [O {\small III}]/[O {\small II}], which is commonly used to trace the gas ionization parameter\cite{Kewley2019}, is presented in the right panel of Figure \ref{fig:radialProf} for those galaxies with both $3\sigma$ [O {\small II}] and [O {\small III}] detections for all radial bins. The mean O$_{32}$ radial profiles for 10 (6) galaxies with SFR above (below) 0.1 M$_\odot$ yr$^{-1}$ are also presented. The galaxy with the highest SFR shows very distinct O$_{32}$ features, and is therefore not included in the stack. The mean stack is consistent with the median stack, except for the last radial bin for the high-SFR stack, with mean (median) O$_{32}$ value of $0.58\pm0.25$ ($0.24\pm0.25$). We observe  consistent radial trends in O$_{32}$ for galaxies with SFR above and below 0.1 M$_\odot$ yr$^{-1}$. 
The galaxy with the highest SFR (SFR $\sim$ 3.9 M$_\odot$ yr$^{-1}$, top dark blue line) is classified as having an active galactic nuclei (AGN) based on the mid-infrared color selection in the MaNGA-AGN catalog \cite{Comerford2020}. However, the line ratios presented in our BPT diagram \cite{bpt} are not consistent with the AGN being the ionization source (details see Methods). It exhibits two notable turnover points in the O$_{32}$ radial profile at $\sim 3 R_e$ and $\sim 10 R_e$, respectively, with an O$_{32}$ difference between neighboring radial measurements of $2.1\pm0.3$ and $6.3\pm1.2$, 
respectively. These inflection points might reflect transitions from the galaxy disk to the disk-halo interface, where ionization is influenced by escaping photons from the central star-forming regions, and then, at larger radius, to a region dominated by high escaping ionizing photons from the central galaxy, or possible by heat shocks or cooling gas inflows (lower metallicity). 
We caution, however, that the interpretation of O$_{32}$ is complicated by its additional dependence on metallicity\cite{Kewley2002}. We refer to the Methods section for more details on the metallicity gradient measurements for individual galaxies, where it has been possible to measure the gas metallicity. 

Our measurements of the O$_{32}$ ratio can also be used to study the distribution of the ionizing radiation that is able to escape from the star-forming galaxies and ionize the CGM and intergalactic medium (IGM) around them, referred to as the escape fraction. Studies using O$_{32}$ to  constrain the escape fraction suggest that the extreme O$_{32}$ values measured in Green Pea galaxies at $z = 0.1-0.3$ are due to a high escape fraction\cite{Leitet2013, Jaskot2013, Izotov2018, Flury2022}, although there exist additional complicating factors such as the galaxy stellar mass, the metallicity, the SFR and sSFR of the galaxy, the half light radius and the SFR surface density of the galaxy, etc. We measure similar values of O$_{32}$ for some of the high-SFR galaxies in our sample.
Adopting the best-fit scaling relation between the Lyman continuum (LyC) escape fraction and O$_{32}$ from the recent studies\cite{Izotov2018, Flury2022}, with the caution that the scatter in the scaling relation is large (as high as $1-2$ dex), and acknowledging the difficulties in measuring the escape fraction, we conclude that the inferred escape fractions are within the plausible limits of 1 to $2 $\% ($< 1$\%) out to a radius of $\sim 5 R_e$ in general for the high-SFR (low-SFR) galaxies in our sample, but as high as $\sim 35$\% for the galaxy with highest SFR out to $r_p \sim$ 13 kpc ($\sim 0.2 \ r_{\rm vir}$). 

\subsection{The inner CGM across the entire sample}

\begin{figure*}[ht!]
\begin{center}
\includegraphics[width = 0.825 \textwidth]{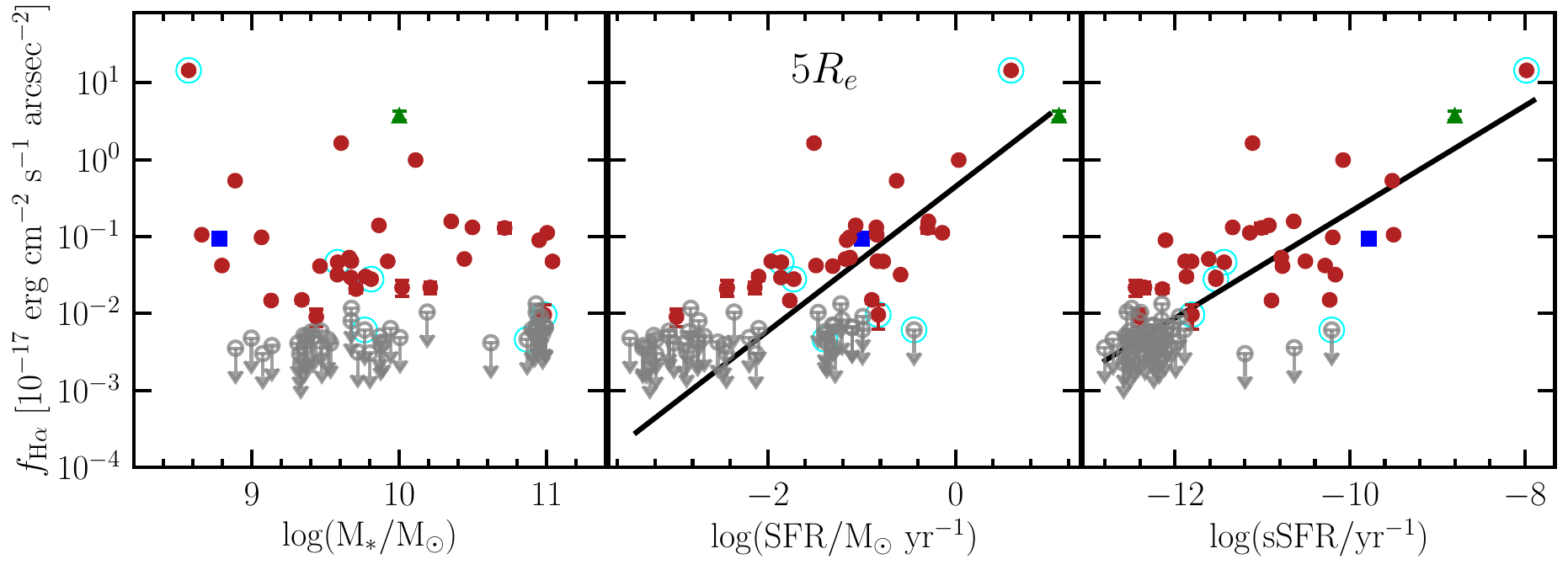}
\includegraphics[width = 0.825 \textwidth]{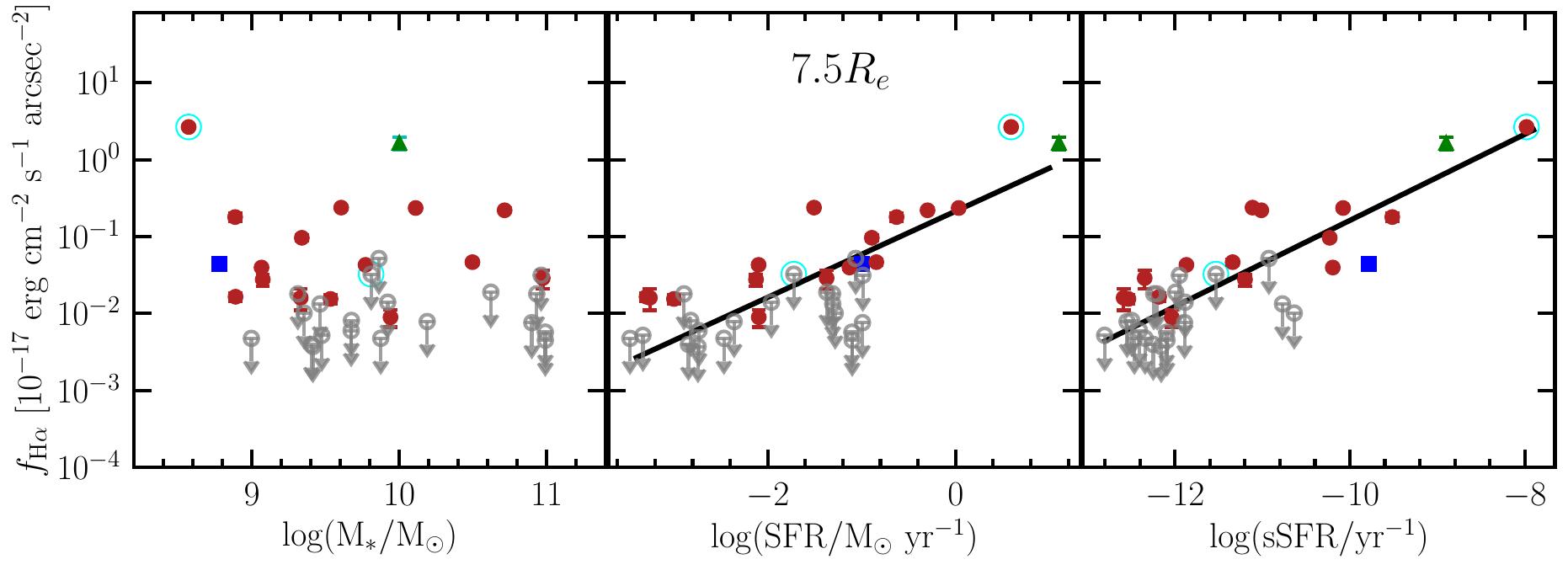}
\includegraphics[width = 0.825 \textwidth]{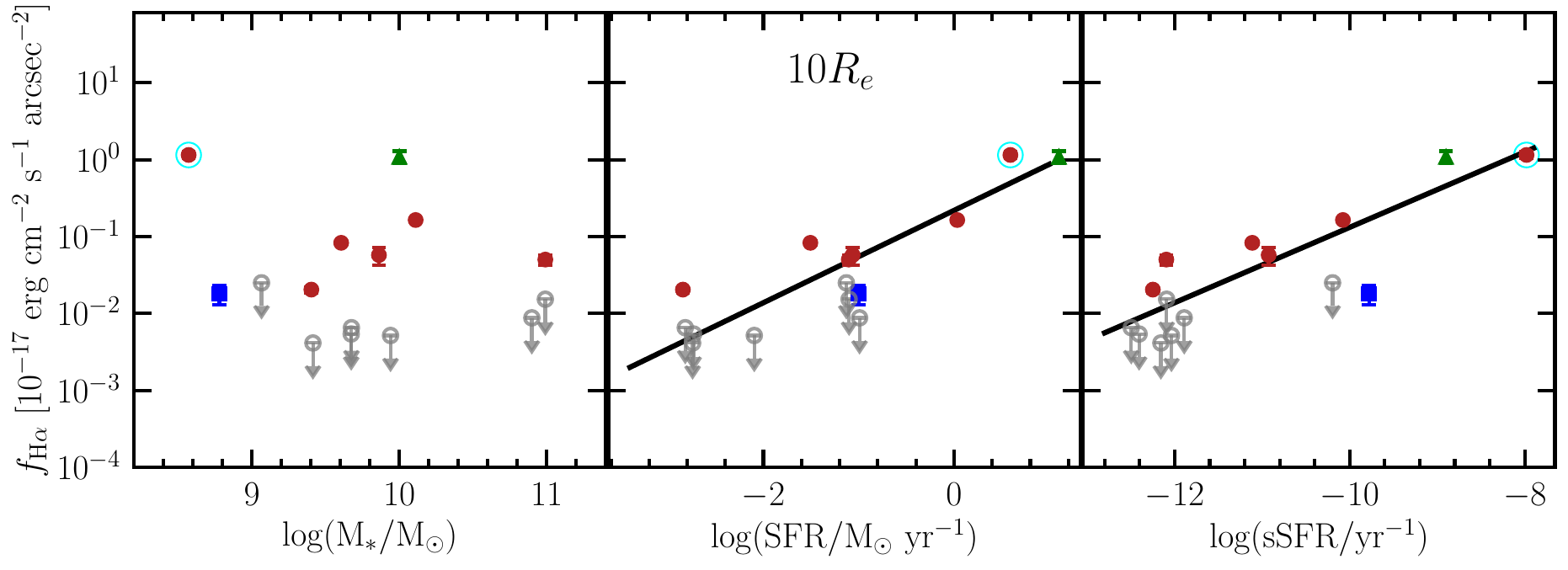}
\end{center}
\caption{{\bf The correlation between H$\alpha$ emission SB and galaxy's properties.} The H$\alpha$ emission surface brightness at $5 R_e$ ($4.5-5.5R_e$, top), $7.5 R_e$ ($7-8R_e$, middle) and $10 R_e$ ($9.5-10.5R_e$, bottom) as a function of M$_*$, SFR and sSFR. The solid red circle represents the H$\alpha$ emission detection, the open grey circle stands for H$\alpha$ emission non-detection and the $2\sigma$ upper limits are adopted. The blue square represents the H$\alpha$ emission measurement using MUSE data\cite{Zhang2024b} and the green triangle denotes the converted H$\alpha$ emission measurement from the H$\beta$ emission measurement for the starburst galaxy IRAS 08339+6517\cite{Nielsen2024}. The open cyan cycle represents the AGN hosts. The solid black line represents the best-fit model.}  
\label{fig:fluxStats}
\end{figure*}

Among the 72 target galaxies,  we find significant ($> 3 \sigma$) H$\alpha$ emission detection out to a radius of at least $5 R_e$ (mean measurements within  annuli of $(4.5-5.5) R_e$) for 31 and bare to moderate ($\sim 1-2 \sigma$) measurements for four additional ones. Among those 35, 15 of them have significant ($> 3 \sigma$) detections and four have bare to moderate ($\sim 1-2 \sigma$) measurements out to  $7.5 R_e$ (mean measurements within  annuli of $(7-8) R_e$). Among the total of 13 galaxies that can be covered out to a radius of $10 R_e$, six of them have significant ($\gtrsim 4\sigma$) H$\alpha$ emission out to a radius of $10  R_e$ (mean measurements within  annuli of $(9.5-10.5) R_e$) and one of them is poorly measured ($\sim 1\sigma$). The mean and median values of the H$\alpha$ emission measurements within those annulus are consistent with each other. All measurements with significance less than $2\sigma$ are treated as the upper limits. The mean (median) stellar mass of the galaxies with H$\alpha$ emission detection at $5 R_e$ is $10^{9.82}$ ($10^{9.71}$) M$_\odot$, which is slightly lower than those without H$\alpha$ emission detections (mean, median: $10^{9.97}$, $10^{9.77}$ M$_\odot$). The mean (median) SFR of the galaxies with H$\alpha$ emission detection at $5 R_e$ is $10^{-1.21}$ ($10^{-1.13}$) M$_\odot$ yr$^{-1}$, respectively, which is more than one order of magnitude higher than those without H$\alpha$ emission detections (mean, median: $10^{-2.21}$, $10^{-2.47}$ M$_\odot$ yr$^{-1}$), which is also the case for the sSFR of the galaxies (mean, median: $10^{-11.03}$, $10^{-11.11}$ yr$^{-1}$ for detections and $10^{-12.21}$, $10^{-12.27}$ yr$^{-1}$ for non-detections). The mean (median) stellar metallicity of the galaxies \cite{Sanchez2022} with an H$\alpha$  detection is $0.5\ Z_\odot$ ($0.7\ Z_\odot$), which is lower than that for those without an H$\alpha$ detection (mean, median: $0.8\ Z_\odot$, $0.9\ Z_\odot$). Six among these 72 target galaxies are classified as AGNs according to the MaNGA-AGN catalog \cite{Comerford2020}, denoted by cyan open cycles in Figure \ref{fig:fluxStats}. Four of them are selected purely by Wide-field Infrared Survey Explorer (WISE) \cite{WISE_Wright_2010,WISE_Mainzer_2011} mid-infrared color cuts, the remaining two are selected purely by radio emission detection. None of them has detected X-ray emission or the signature of broad emission lines. 

In Figure \ref{fig:fluxStats}, we present the H$\alpha$ emission surface brightness measurements at $5R_e$, $7.5 R_e$  and $10 R_e$ as a function of M$_*$, SFR and sSFR of the galaxies. For those galaxies without H$\alpha$ emission detection or the detection significance less than $2\sigma$, we adopt and plot the $2\sigma$ upper limit. The  H$\alpha$ emission surface brightness at all radii correlates well with either SFR or sSFR, with the Pearson correlation coefficient ($p$-value) of 0.74 ($3.54\times 10^{-7}$) and 0.64 ($3.17\times 10^{-5}$) for SFR and sSFR at $5 R_e$, respectively. Similar results hold for this correlation 
at $7.5 R_e$ and $10 R_e$. The galaxies without H$\alpha$ emission detections mainly occur at sSFR $\lesssim$ $10^{-12}$ yr$^{-1}$. In contrast, we do not find a significant trend between the H$\alpha$ emission surface brightness and the galaxy's stellar mass at all radii. The Pearson correlation coefficient and $p$-value for H$\alpha$ emission surface brightness versus stellar mass is 0.018 and 0.92, respectively, at $5 R_e$ (similar results for the galaxy subsample at $7.5 R_e$ and $10 R_e$), indicating that we can not rule out the null hypothesis that the values are randomly distributed with regard to stellar mass. We do not find any differences for the H$\alpha$ emission line fluxes or other properties including the mass-size and mass-SFR relation for those galaxies classified as AGN hosts except for the galaxy with the highest SFR, but the H$\alpha$ emission of the galaxy with the highest SFR still aligns with the correlation.  
We obtain similar results for the other strong emission lines ([O {\small II}] and [O {\small III}], see details in Methods). 
Expressing the relation between H$\alpha$ emission surface brightness and either SFR or sSFR as follows:
\begin{equation}
\begin{split}
    \log f_{\rm H\alpha} & = m_1 \log({\rm SFR/10^{-1}M_\odot yr^{-1}}) + b_1  \\ 
    & = m_2 \log({\rm sSFR/ 10^{-11}yr^{-1})} + b_2,
\end{split}
\label{eq:linear}
\end{equation}
where the H$\alpha$ emission surface brightness is in units of $10^{-17}$ erg cm$^{-2}$ s$^{-1}$ arcsec$^{-2}$, SFR is in units of M$_\odot$ yr$^{-1}$ and sSFR is in units of yr$^{-1}$, we present the best-fit parameters in Table \ref{tab:param}. The best-fit models, derived using all data points including the upper limits, are also included in Figure \ref{fig:fluxStats}.  Overall, both linear relations are well-fit, with the relation to SFR being slightly better. We notice that the slope as shown in Table \ref{tab:param} becomes shallower ($\gtrsim 3 \sigma$ significance) at larger radius (10$R_e$ versus 5$R_e$) for the correlation between  H$\alpha$ emission line surface brightness and SFR, which may suggest that the influence of the galaxy’s SFR on the cool gas decreases with increasing radius. Excluding these six AGNs has a negligible effect on the best-fit parameters presented in Table \ref{tab:param}. 

\begin{table*}[ht!]
    \centering
    \begin{tabular}{ccccc}
    \hline
    & \multicolumn{2}{c}{SFR} & \multicolumn{2}{c}{sSFR} \\ \hline
    $r_p$ & $m_1$ & $b_1$  & $m_2$ & $b_2$ \\
     \hline  
$5 R_e$ & $0.94 \pm 0.08$ & $-1.29 \pm 0.05$ & $0.69 \pm 0.04$ & $-1.37 \pm 0.04$ \\ \\
$7.5 R_e$ &  $0.56 \pm 0.05$ & $-1.23 \pm 0.05$ & $0.56 \pm 0.04$ & $-1.35 \pm 0.05$ \\ \\
$10 R_e$ & $0.60 \pm 0.08$ & $-1.26 \pm 0.08$ & $0.49 \pm 0.06$ & $-1.37 \pm 0.07$ \\  \hline
    \end{tabular}
    \vspace{0.5cm}
    \caption{{\bf The best-fit parameters of Eq.  \ref{eq:linear}.} The best-fit parameters for the relation between  H$\alpha$ emission surface brightness at different radii and SFR/sSFR of the central galaxy, as described in Eq. \ref{eq:linear}.}
    \label{tab:param}
\end{table*}

Also in Figure \ref{fig:fluxStats}, we include literature measurements for two galaxies for comparison.
The H$\alpha$ emission surface brightness of a nearby low-mass galaxy\cite{Zhang2024b} at $z=0.1723$ (M$_*\sim 10^{8.78\pm0.42}$ M$_\odot$ and SFR $\sim 0.1$ M$_\odot$ yr$^{-1}$) is significantly detected at  $5 R_e$ ($R_e\sim$ 1.34 kpc)\cite{Contini2016}, $7.5 R_e$, and $10 R_e$ in the {\it Hubble} Deep Field South (HDFS)\cite{MUSE-HDFS} using MUSE data. The consistency between the MaNGA and MUSE for galaxies with similar properties is reassuring.
The H$\beta$  surface brightness\cite{Nielsen2024} of a nearby starburst galaxy (IRAS 08339+6517\cite{Margon1988, Lopez2006}, M$_*\sim 10^{10.0\pm0.1}$ M$_\odot$ and SFR $\sim 12.1 \pm 0.1$ M$_\odot$ yr$^{-1}$) at $z=0.01911$ 
is significantly detected at  $5 R_e$ ($R_e\sim$ 1 kpc)\cite{Fisher2022}, $7.5 R_e$ and $10 R_e$. To place this object on the plot, we adopt H$\beta$/H$\alpha = 0.3$, a rough value consistent with theoretical expectations\cite{baker, hummer, osterbrock2006}. Again, this object also falls on the general trend. 

At $z\sim 0.05$, galaxies with SFR $\sim$ 1 M$_\odot$ yr$^{-1}$ exhibit a typical H$\alpha$ surface brightness of $\sim 10^{-17.4}$ \,erg\,cm$^{-2}$\,s$^{-1}$\,arcsec$^{-2}$ at $5R_e$, $\sim$ 10.0 kpc at $z=0.05$ estimated from the mean value of the entire sample. This corresponds to a LyC photons production rate of $1.9\times 10^{51}$ s$^{-1}$, based on a scaling relation between H$\alpha$ luminosity and LyC photon production\cite{Bouwens2016}, which is derived from simulations of starburst galaxies\cite{Leitherer1995} under the assumption of zero escape fraction into the intergalactic medium. Several mechanisms could produce these ionizing photons, including escaping radiation from central galaxies, the ultraviolet background (UVB) \cite{Haardt2012,Faucher2020}, shock ionization \cite{osterbrock2006,Allen2008,Draine2011}, or other feedback-driven processes. The UVB alone, however, is insufficient: with $\Gamma_{\rm HI}\approx4-10\times10^{-14}$ s$^{-1}$ \cite{FG2020}, it yields an H$\alpha$ surface brightness of only $\approx10^{-19}$ erg s$^{-1}$ cm$^{-2}$ arcsec$^{-2}$ at the disk-halo interface \cite{Fumagalli2017}, close to our $2\sigma$ detection limit ($\approx10^{-19.2}$ erg s$^{-1}$ cm$^{-2}$ arcsec$^{-2}$, estimated from the mean value of those $2\sigma$ detection limits). Since the detected H$\alpha$ emission (Figure \ref{fig:fluxStats}) far exceeds this limit, the UVB is not the dominant ionization source. 

We also conclude that shock ionization is also unlikely to dominate for several reasons. First, the stacked O$_{32}$ profile from our galaxy sample remains nearly constant at $\sim0.4$ from $0.2\,R_{\rm e}$ to $5\,R_{\rm e}$, indicating a remarkably uniform ionization parameter over this radial range. 
If fast radiative shocks dominate the inner CGM ionization, they would require nearly identical shock velocities and parameters across the entire $0.2-5\,R_{\rm e}$ range to maintain such a constant O$_{32}$. Second, a substantial fraction of galaxies show a slightly decreasing O$_{32}$ profile, which is inconsistent with the pure shock model predictions \cite{Allen2008,Dopita2017}. Third, our BPT diagrams \cite{bpt} for two galaxies for which we have data covering the same radial range lie in the star-forming region (details see Methods). Finally, shock models \cite{Allen2008,Dopita2017} that produce O$_{32}\sim0.4$ predict BPT line ratios (e.g., elevated [N {\small II}]/H$\alpha$ and [O {\small III}]/H$\beta$) that are inconsistent with our BPT diagnostics. 
Unfortunately there are concerns with several of these arguments. First, O$_{32}$ is sensitive to metallicity and we are stacking across a wide range of galaxy masses. Second, the O$_{32}$ profiles show some variance and become more difficult to measure at larger radii. Third, we have the radially sampled BPT diagrams only for two galaxies and again at large radii the uncertainties are large and do allow for ratio values that place the region slightly outside the star formation region of the diagram.
As such contributions from shocks cannot be definitively ruled out. Nevertheless, in light of these arguments and the connections we identify next between the H$\alpha$ emission at large radius and the star formation properties of these galaxies, we conclude that the most likely source of the ionizing radiation in the inner CGM region is star formation in the central galaxy.


We independently estimate the ionizing photon output from central star formation of the galaxies. Assuming solar metallicity, continuous star formation, and the Kroupa initial mass function\cite{Kroupa2001}, the production rate of ionizing photons can be estimated from the SFR using established scaling relations. For a SFR $\sim$ 1 M$_\odot$ yr$^{-1}$ and assuming $1-2$\% photons escape into the $5R_e$ region,  the resulting LyC photon rate is $1.4-2.7\times 10^{51}$ s$^{-1}$, which is consistent with that estimated from the observed H$\alpha$ emission flux. Note that the mean stellar metallicity of our galaxy sample is lower than the adopted solar metallicity, suggesting that the production rate of ionizing photons calculated above is actually under-estimated or that the assumed LyC escape fraction can be relaxed. Even in lower-SFR galaxies, the ionizing photon production estimated from SFR remains higher than that inferred from the observed H$\alpha$ surface brightness. Although uncertainties exist in estimating LyC photon production, due to assumptions about metallicity, rough estimation of the escape fraction, star formation history, and initial mass function, the simple calculations clearly demonstrate that the central galaxy’s star formation activity can produce a sufficient number of ionizing photons to fully account for the observed emission line fluxes.

One caveat is that the different stellar mass and SFR estimations have an approximately $0.1-0.2$ dex scatter, especially for galaxies with low SFR. We adopt different measurements of the stellar mass and SFR to test the scaling relation as expressed in Eq. \ref{eq:linear}. We extract the stellar mass\cite{Kauffmann2003a,Kauffmann2003b,Gallazzi2005} and SFR from the MPA-JHU catalog\cite{Tremonti2004, Brinchmann2004}. We also retrieve the stellar mass and SFR from the GSWLC (GALEX-SDSS-WISE LEGACY CATALOG)\cite{Salim2016, Salim2018} using the spectral energy distribution (SED) fitting technique. All model parameters are consistent within 1$\sigma$ for the three different radii except for the $b_1$ estimate, where the deviation lies between 1.5 and 2.5$\sigma$ for the different stellar mass and SFR estimates for the different radii. It further demonstrates that the strong correlation between the H$\alpha$ emission surface brightness and galaxy's SFR is an intrinsic property of the galaxies, regardless of the scatter of stellar mass and SFR estimations.

\begin{figure*}[ht!]
\begin{center}
\includegraphics[width = 0.875 \textwidth]{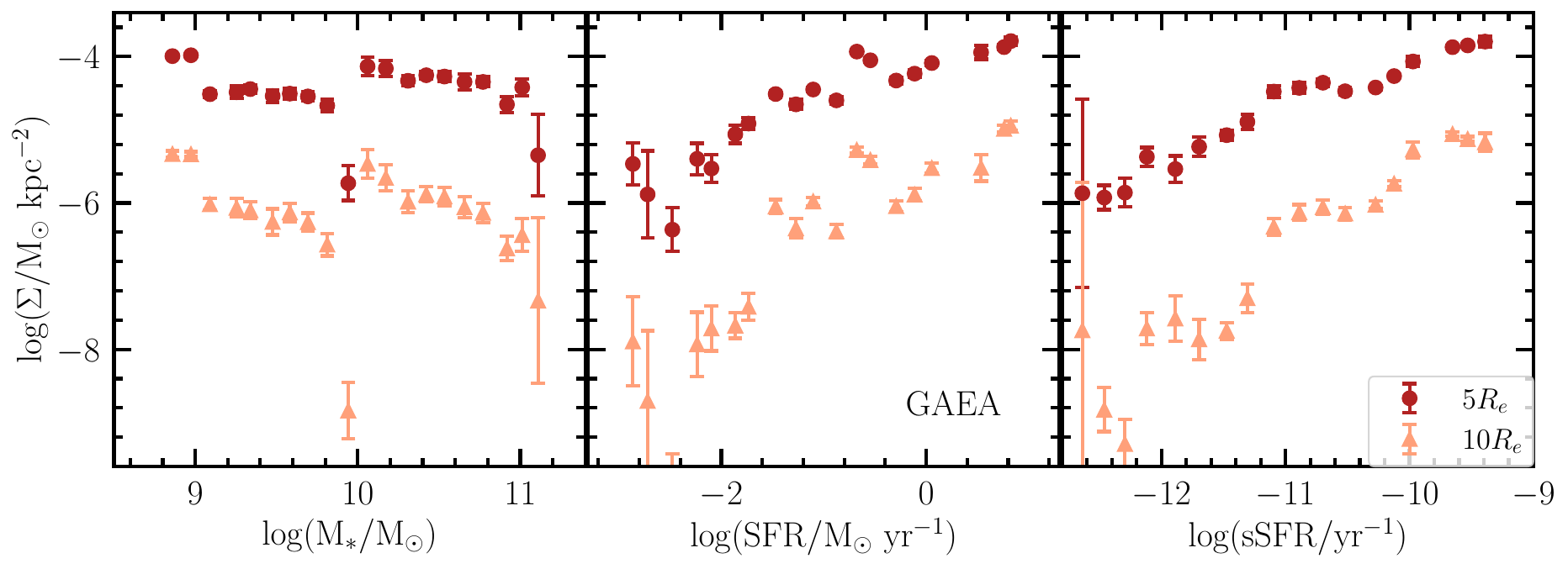}
\includegraphics[width = 0.875 \textwidth]{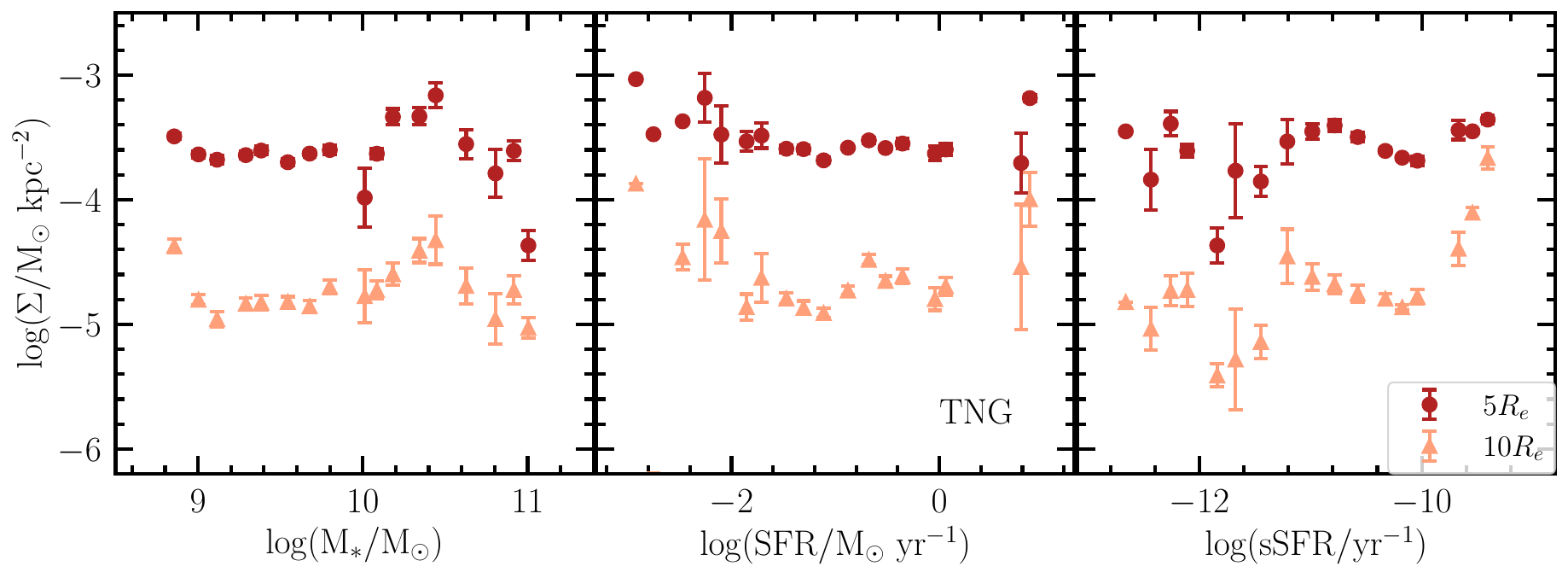}
\end{center}
\caption{{\bf The correlation between the cool gas and the galaxy's properties from simulations.} The cool gas surface density at $5 R_e$ (red  dot) and $10 R_e$ (coral  triangle) as a function of M$_*$, SFR and sSFR for GAEA (top) and TNG (bottom) simulations. The M$_*$, SFR of GAEA and TNG galaxies are already matched to our MaNGA galaxy sample within 0.1 dex.}  
\label{fig:fluxSimu}
\end{figure*}

\subsection{Implications from Simulations}


Cosmological simulations provide a useful tool to study the multi-phase gas of the galaxy and to understand the subgrid physics by the comparison to observations. 
In Figure \ref{fig:fluxSimu}, we present the cool gas (with temperature of $\sim 10^4$ K) surface density, $\Sigma$, at $5 R_e$ and $10 R_e$ as a function of M$_*$, SFR and sSFR for GAlaxy Evolution and Assembly (GAEA) semi-analytic model\cite{delucia2024}  and the IllustrisTNG (TNG) hydrodynamical simulation\cite{weinberger2017, springel2018, Nelson2018, Naiman2018, Marinacci2018, Pillepich2018a, pillepich2018b}, where GAEA and TNG mock/model galaxies are selected to have a stellar mass and SFR within $\pm$0.1 dex of each  galaxy in our MaNGA galaxy sample. This does not provide a direct comparison to the observations (cool gas surface density vs. emission line surface brightness, but under the simplest assumptions one might expect emission line surface brightness $\propto \Sigma^2$), since converting from gas surface density to emission line surface brightness involves understanding the ionization spectrum, the thermal equilibrium and recombination rates and GAEA does not provide temperature distribution. However, combining the fact that GAEA and TNG galaxies share the same properties as our MaNGA galaxy sample and there exists strong correlation between H$\alpha$ emission surface brightness and SFR (sSFR) as shown in Figure \ref{fig:fluxStats},  it is expected that GAEA and TNG galaxies should share similar H$\alpha$ emission surface brightness. The consistency in terms of the sense and magnitude of the correlations between our observations and the GAEA model is excellent, which might reflect the underlying physics driving the CGM line emission is the amount of cool gas. 
Additionally, Figure \ref{fig:fluxSimu} clearly demonstrates that the strength of star formation process (and sSFR) is the main regulators for the amount of cool gas at the circumgalactic scale.

In contrast, we find that the cool gas surface density at either $5R_e$ or $10 R_e$ does not correlate with any physical properties of the galaxies selected from TNG, as shown in the lower panel of Figure \ref{fig:fluxSimu}. These results are consistent with previous work, which found that the HI gas density of TNG galaxies at large radii ($30-70$ kpc) depends only weakly on SFR, while the central density varies dramatically with SFR \cite{Ma2022, shi2022}. In the TNG simulation, SFR is mainly suppressed by kinetic feedback from active galactic nuclei (AGN), which pushes gas outward and yields more extended HI density profiles and gas deficiency in the central region \cite{diemer2019, xie2025}, with minimal reduction of the HI gas reservoir. 
Consequently, the central galaxies in the TNG simulation have a larger fraction of cool gas at large radii than their counterparts in the GAEA model, and their surface density is independent of SFR. 
However, we do not detect any optical emission from a possible cool CGM at large radii for quenched massive galaxies ($M_* > 10^{10.5}$ M$_\odot$, sSFR $<10^{-12}$ yr$^{-1}$). Furthermore, using the HI-MaNGA sample\cite{Masters2019} we find that HI is detected in our sample for those galaxies with $M_* < 10^{10.5}$ M$_\odot$ and sSFR $>10^{-12}$ yr$^{-1}$, suggesting that massive quenched galaxies do not retain an HI reservoir, different from TNG simulations.  
One possibility, discussed in the recent literature\cite{peng2020,ZhangC2019,lu2022}, is that high angular momentum supports a stable neutral hydrogen gas at the outskirt region and prevents the gas from falling into the galaxy disk, thereby keeping galaxies quenched. Our discovery of the tight correlation between the H$\alpha$ emission surface brightness at all radii and SFR or sSFR for both high-mass and low-mass galaxies, however, disfavors this scenario. 




In summary, we present a mapping of the dominant baryonic component of galaxies. Although previous IFU observations have detected the CGM either in high-redshift\cite{Mackenzie2019,Kusakabe2022,Peroux2019} or lower-redshift extreme galaxies\cite{Yoshida2016}, this is the first systematic unbiased study to probe the inner cool CGM for a large sample of normal, low-redshift galaxies with a wide range of stellar mass, SFR and sSFR. Using the MaNGA IFS data, we find a steep turnover occurring at $(1-2) R_e$ in the detected H$\alpha$, [O {\small II}] and [O {\small III}]  emission surface brightness radial profiles, a shallower slope of the line emission radial profiles beyond this radius for low-SFR galaxies, and identify strong correlations in H$\alpha$, [O {\small II}] and [O {\small III}] surface brightness in the inner CGM with both the galaxy's SFR and sSFR at all radii, but not with its stellar mass. The ionizing photo escape fraction estimated for our MaNGA galaxy sample is $\sim (1-2)$\% out to a radius of the inner CGM or $\sim 5R_e$, and could reach as high as 35\% for the galaxy with highest SFR at a large radius of $r_p\sim$ 13 kpc. The star formation in the galaxies can provide enough escaping ionizing photons to fully account for the observed emission line fluxes. We compare our results against the prediction of two state-of-the-art theoretical models of galaxy formation and show that they provide quite different predictions for the relations between H$\alpha$ emission/gas density profiles and SFR. Our conclusions highlight the critical role these kind of observations can play in a better understanding of the fundamental physical processes responsible for galaxy evolution.

\newpage         

\section*{Materials and Methods}
\label{sec:method}

\medskip
\noindent{\bf MaNGA Survey and Data Sample} 

The MaNGA survey is an IFU spectroscopic program using the Sloan 2.5-m telescope\cite{Gunn2006}, targeting  $\sim 10,000$ galaxies in the nearby Universe ($0.01 \lesssim z \lesssim 0.15$) covering a wide range of stellar mass, SFR and colors\cite{Wake2017}. The primary goal of the MaNGA survey is to seek the physical origin of the mechanisms that drive  galaxy evolution by providing and analyzing  the stellar population and the gaseous component across the face of each galaxy. 
The fiber bundles vary in diameter from 12$^{\prime \prime}$ (19 fibers) to 32$^{\prime \prime}$ (127 fibers), typically covering $(1.5-2.5) \times$ the effective radius of the target, with a spatial resolution of 2.5$^{\prime \prime}$. The typical total integration time of the target is roughly 3 hours. The wavelength coverage is from 3600 \AA \ to 10350 \AA \ with spectral
resolving power $R\sim 2000$, corresponding to a velocity of 75 km s$^{-1}$ at 5100 \AA\cite{Law2016}. The spectroscopic datacubes are resampled on a $0.5\times 0.5^{\prime \prime}$ grid using the MaNGA’s Data Reduction Pipeline (DRP)\cite{Law2016}.


There are only 219 galaxies in the entire MaNGA sample for which the IFU bundle samples out to a radius of at least $5 R_e$ of the galaxy. 
Furthermore, we require that the IFU bundle consists of at least 91 fibers to provide sufficiently spatially resolved measurements of the CGM. 
We are left with 75 target fields, including one galaxy of a galaxy pair. The galaxy's stellar masses are provided in MaNGA DRP catalog\cite{Law2016} and SFRs are provided in MaNGA’s Data Analysis Pipeline (DAP) catalog\cite{Westfall2019, Belfiore2019}. The galaxy's stellar metallicity is adopted from the MaNGA {\it Pipe3D} catalog \cite{Sanchez2022}. 
Two galaxies with a bright bulge or nuclear source and one galaxy which is part of a galaxy pair are excluded from the galaxy sample, which will be elaborated below. So we are left with 72 galaxies. The stellar masses of the final galaxy sample span from $10^{8.6}$ M$_\odot$ to $10^{11}$ M$_\odot$, and the star formation rates of the final galaxy sample span from $10^{-3.5}$  M$_\odot$ yr$^{-1}$ to $10^{0.59}$ M$_\odot$ yr$^{-1}$, both of which span $\sim 3-4$  orders of magnitude. The mean (median) redshift, stellar mass, star formation rate, specific  star formation rate and stellar metallicity of this galaxy sample are 0.0528 (0.0384), $10^{9.91}$ ($10^{9.74}$) M$_\odot$, 0.016 (0.025) M$_\odot$ yr$^{-1}$, $10^{-11.70}$ ($10^{-12.11}$) yr$^{-1}$, 0.66 (0.81) $Z_\odot$. Our sample contains both star-forming and quiescent galaxies, and six of them are classified as AGNs according to the MaNGA-AGN catalog \cite{Comerford2020}. However, we do not find any difference for the emission line measurements or the other galaxy properties among those galaxies which are claimed to host AGNs and not host AGNs, except the galaxy with highest SFR. Because the targets are not selected based on any physical properties, except possibly surface brightness due to the comparably shallow observational sensitivity, and it spreads a wide range of stellar mass and star formation rate, our sample can be considered as an unbiased sample for the large-scale CGM study.



\begin{figure}[ht!]
\begin{center}
\includegraphics[width = 0.48 \textwidth]{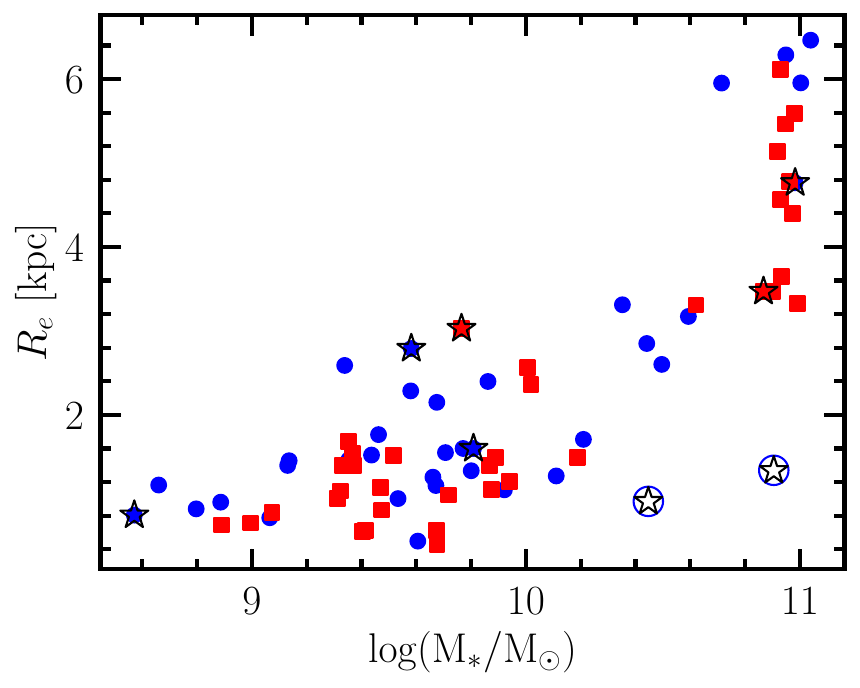}
\end{center}
\caption{{\bf The scatter plot of the stellar mass and effective radius of the entire sample}. The blue solid dots represent the ones with H$\alpha$ detection at $5 R_e$ and the red squares denote the ones without H$\alpha$ detection at $5 R_e$. The two blue open circles stand for the two galaxies which sit systematically below the galaxy M$_* - R_{e}$ scaling relation. The stars represent the AGNs.} 
\label{fig:sm-r}
\end{figure}

\medskip
\noindent{\bf Transition to the CGM}

A key challenge in the study of the CGM is the identification of the disk-halo boundary. At what radius are we detecting emission from the CGM rather than the galaxy disk? In our previous stacking work\cite{Zhang2018a,Zhang2018b}, we adopted a radial range defined in terms of the virial radii, $r_{\rm vir}$, of $0.05 < r/r_{\rm vir} < 0.25$. The outer radius was set to avoid contamination from neighboring galaxies, while the inner radius was set to avoid emission from the central galaxy. Neither of these boundaries is absolute or particularly well defined. In the best of cases, the boundary is defined by a clear change in the nature of the surface brightness emission profile. For example, a break at $\sim 50-60 $ kpc for Milky Way-like galaxies in the stacked surface brightness profile has been associated with contamination from nearby neighbors using simulations\cite{Zhang2018a}. This result helps define the outer limit of the range. In a recent study\cite{Nielsen2024}, there is a  tell-tale break in the profile at $\sim 4.0$ kpc (roughly twice the 90\% stellar radius) between a galaxy disk's exponential profile and the CGM power law profile for a starburst sub-$L^*$ galaxy (M$_* \sim 10^{10.0\pm0.1}$ M$_\odot$, SFR $\sim 12.1\pm0.1$  M$_\odot$ yr$^{-1}$) at  $z=0.01911$ that is associated with the disk-halo boundary. In this case, the virial radius of this galaxy is estimated to be 115 kpc\cite{Nielsen2024}, so this break occurs at $r/r_{\rm vir} \sim 0.04$, consistent with what was adopted as the inner boundary in our previous stacking analysis. 

For a more direct, empirical definition of the disk-halo boundary, we  translate our previous limit into units of  half-light radii. 
The previous work, based on scaling relations between galaxy's stellar mass and virial radius derived from the {\it UniverseMachine} simulation\cite{Behroozi2019}, suggests a mean value of $0.05\ r_{\rm vir}$ = 5.7 kpc for our MaNGA galaxy sample. To be somewhat conservative, we set the boundary at approximately twice this value, which corresponds to $\sim$ $5 R_e$ (11.4 kpc). To verify, we also evaluate $5 R_e$ in {\it UniverseMachine} simulation\cite{Behroozi2019} at $z=0.05$ and it is 12.1 kpc, which is consistent with that for our MaNGA galaxy sample. Thus we adopt a radius of $5  R_e$ to represent the inner edge of the CGM. The effective radius might be a misleading indicator if a galaxy has a  bright bulge or nuclear source. We identify two such galaxies among the 75 targets. These two galaxies sit systematically below the galaxy  stellar mass and  effective radius scaling relation\cite{Fernandez2013, Zhang2025} (Figure \ref{fig:sm-r}). In these cases, we do not trust the $5 R_e$ criterion and remove those two galaxies from our sample. Additionally,  there is one galaxy in our MaNGA sample that is part of a targeted galaxy pair and the galaxy is not centered in the FOV.  
Therefore, we also remove this galaxy  and we are left with 72 
galaxies.

A legitimate concern is that we have significantly underestimated the radius out to which disk H{\small II} regions produce emission lines. To provide some more intuition and support for our choice, $2R_{90}$ was adopted as the disk-halo boundary in a recent study \cite{Nielsen2024}, where $R_{90}$ is the radius containing 90\% of the optical flux. The concentration index $R_{90}/R_{\rm e}$ in the SDSS $g$-band can be obtained from the {\it Pipe3D} catalog. The mean (median) ratio is 2.78 (2.73), with a dispersion of 0.49 for the mean value. $2R_{90}$ is $\sim 5.6 R_e$, roughly consistent with our choice of $5R_{\rm e}$ as the disk-halo boundary. Furthermore, the independent measurement of outer disk star-forming clusters in a sample of nearby galaxies \cite{z07} shows that even among galaxies with extended, star-forming disks, e.g., M 83 and NGC 628, H{\small II} regions are found only within $2R_{90}$. There is no evidence for disk emission line sources outside of $\sim 5R_e$.

Finally, we test these assumptions and results using a toy model for the radial profile of the H$\alpha$ equivalent width. We measure the radial profile of the H$\alpha$ equivalent width (EW) for all 72 galaxies. We were able to measure EW profiles (requiring at least 3 positive measurements to define a profile) for 38. We will return to discuss the 34 for which we do not have profiles. For the majority of the 38 systems with profiles, the EW remains roughly constant within the galaxy disk and increases beyond it, driven primarily by a drop in the continuum emission, although there are some that drop at large radii.  To mimic this range of behavior, we model the profile as the combination of a star forming disk component that has constant EW out to $R_d$ and decreases suddenly but smoothly beyond this radius.
$${\rm EW}_{\rm disk}(R) = A_{\rm disk}/(1 + e^{-k(R-R_d)})$$ and a CGM component that is simply proportional to radius. Therefore, the asympotic behavior of the model is \begin{equation}
\mathrm{EW}(R) \approx
\begin{cases}
A_{\mathrm{disk}} + A_{\mathrm{CGM}}\, \left(\dfrac{R}{R_e}\right) & R \ll R_d \\[10pt]
A_{\mathrm{CGM}}\, \left(\dfrac{R}{R_e}\right)& R \gg R_d
\end{cases}
\end{equation}
and we set $k=10$ and fit for $A_{\rm disk}$, $R_d$, and $A_{\rm CGM}$. We fit the model for each galaxy, only using measured EW values $>10^{-5}$ because lower values are unreliable.

\begin{figure*}[ht!]
\begin{center}
\includegraphics[width = 0.95 \textwidth]{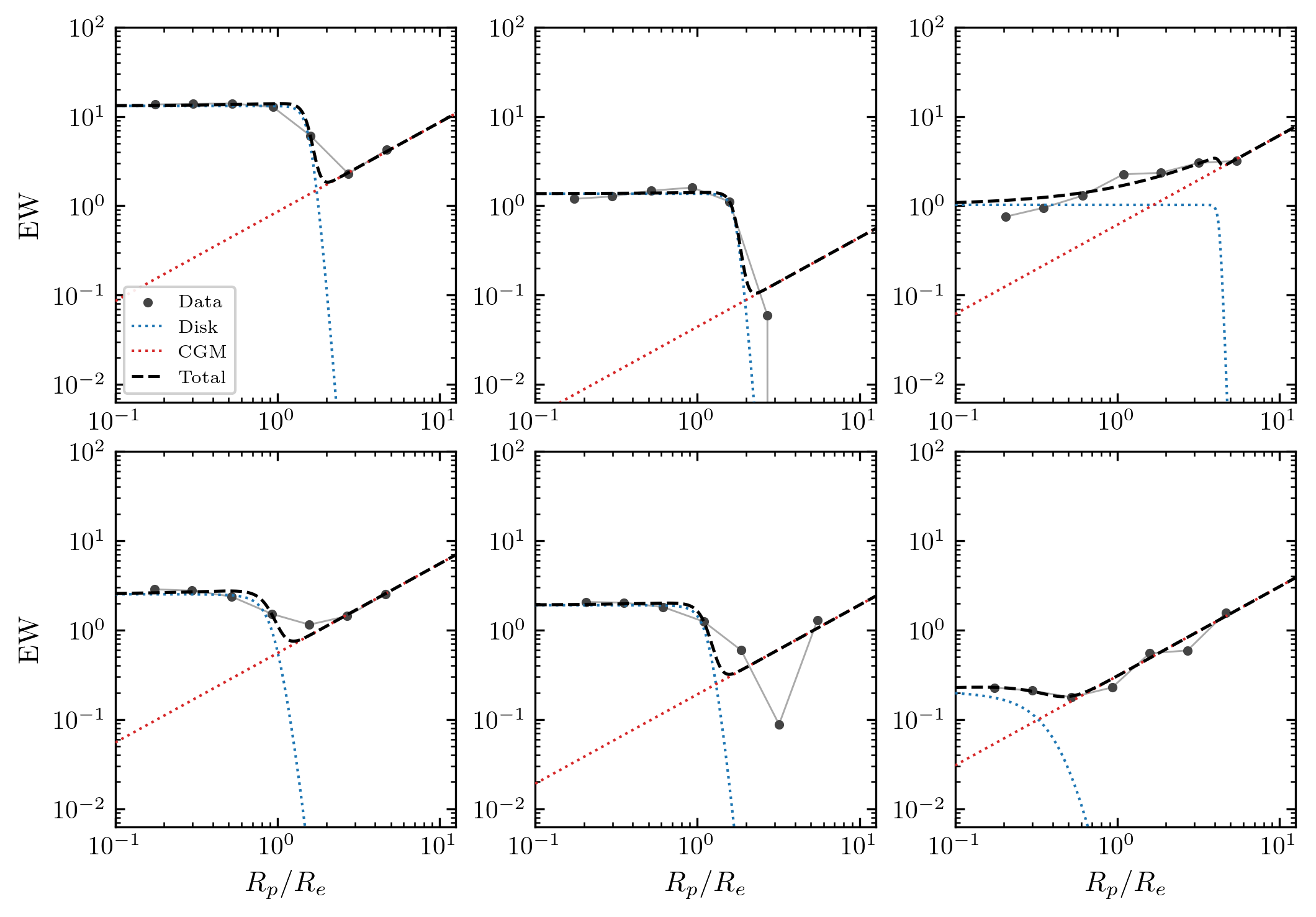}
\end{center}
\caption{{\bf The H$\alpha$ equivalent width radial profiles for six example galaxies}. The points and the connecting solid line are the measured values. The dotted lines show the disk and CGM modeled components separately and the dashed line shows the two components combining to provide the best fit of the data.} 
\label{fig:toy_model}
\end{figure*}

\begin{figure*}[ht!]
\begin{center}
\includegraphics[width = 0.95 \textwidth]{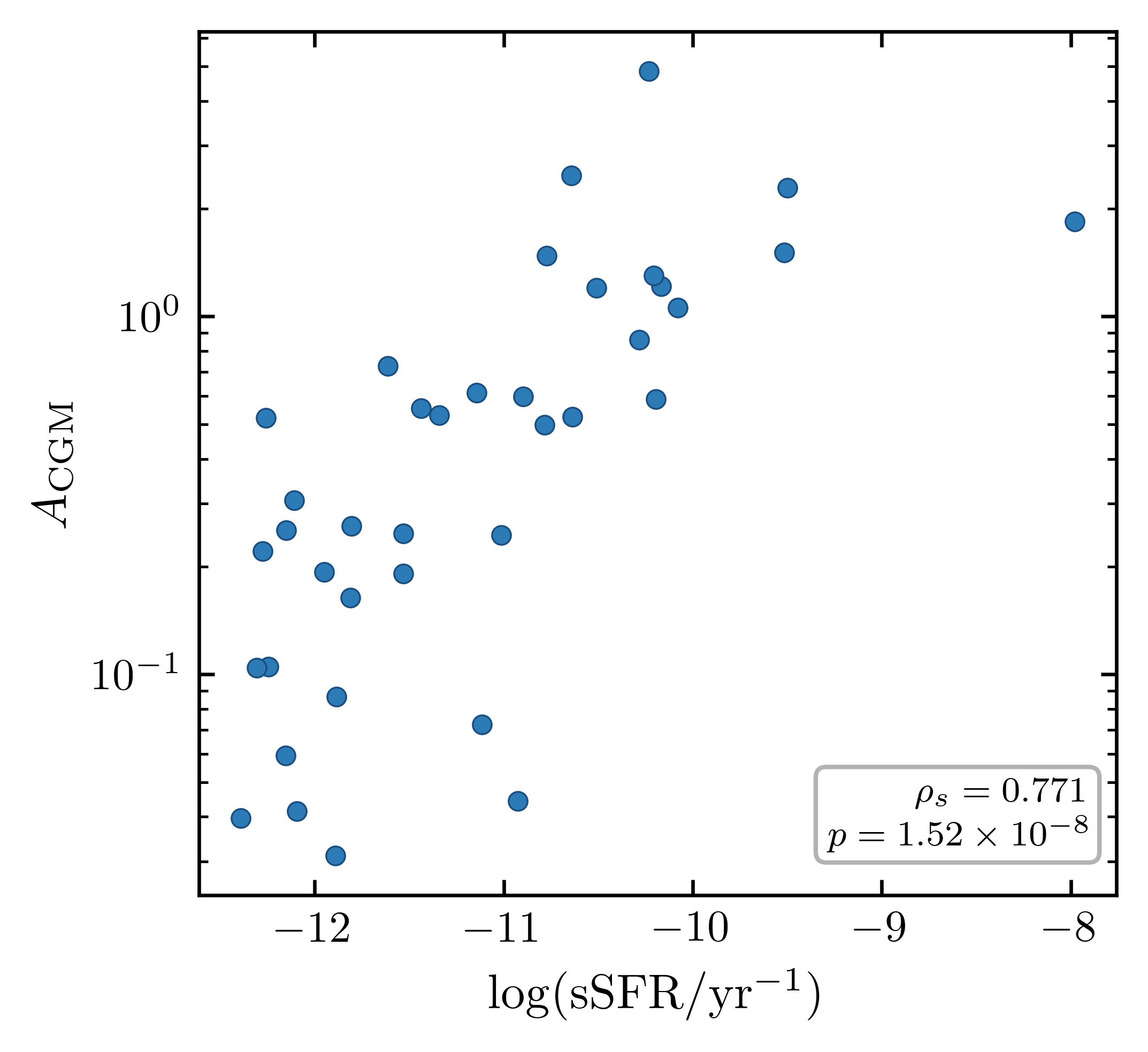}
\end{center}
\caption{{\bf The amplitude of the model CGM EW component, $A_{\rm CGM}$, vs. the specific star formation rate of the galaxy, sSFR.} The galaxies for which we were unable to measure the H$\alpha$ EW profile have a mean sSFR of $10^{-12.29}$ yr$^{-1}$.}
\label{fig:spearman}
\end{figure*}

We show examples of the fits, including a variety of behavior, in Figure \ref{fig:toy_model}. Our two key findings are that there is a strong correlation between $A_{\rm CGM}$ and sSFR (Spearmen rank correlation coefficient, $\rho_s$,  of 0.771 and a p value of $1.52\times 10^{-8}$; Figure \ref{fig:spearman}) and that $\langle R_d/R_e\rangle = 1.61 \pm 0.17$. The first of the two results supports our previous claims of a connection between the CGM emission and the star formation of the central galaxy and the second result supports our claim that the disk influence ends well 
before 5$R_e$ for most galaxies in the sample. Returning to the 34 galaxies for which we do not have EW profiles, the lack of a measured EW at any radius suggests that these systems have little if any CGM emission (at least to our sensitivity level). The mean sSFR for these galaxies is $10^{-12.29}$ yr$^{-1}$, placing them at the far left of the distribution in Figure \ref{fig:spearman} and further supporting the evident correlation. These galaxies are also interesting in the context of the role of shock ionization. It is circumstantial evidence against shocks as the dominant source of ionizing photons for the inner CGM that none of these low sSFR galaxies exhibits significant H$\alpha$ emission at large radii. 

\medskip
\noindent{\bf Data Analysis} 

To increase the signal-to-noise (S/N) sufficiently to detect the CGM, we bin the MaNGA data along both the spectral and spatial dimensions. In the spectral direction, we extend the approach we used with the SDSS spectral stacks\cite{Zhang2016}. We measure the residual emission flux above continuum within a filter window 
centered on the relevant central wavelength at the corresponding rest-frame for each spectrum in each MaNGA datacube. The velocity window is $\pm 150$ km s$^{-1}$ for galaxies with ${\rm M_*} \le 10^{9.5}$ M$_\odot$, $\pm 200$ km s$^{-1}$ for galaxies with $10^{9.5} < {\rm M_*/M_\odot} \le 10^{10.3}$,  $\pm 275$ km s$^{-1}$ for galaxies with $10^{10.3} < {\rm M_*/M_\odot} \le 10^{11}$ and $\pm 375$ km s$^{-1}$ for galaxies with ${\rm M_*} > 10^{11}$ M$_\odot$. And the tests demonstrate that the results are not sensitive to the slight change of the velocity windows. The residual flux of the spectrum in each spaxel is determined by fitting and subtracting a third order polynomial to a 300 \AA\ wide spectral section. We find that the estimation of the continuum from a third order polynomial is far from satisfactory for quiescent galaxies due to strong stellar absorptions, especially within the galaxy disk. Thus, we estimate the continuum from a linear combination of simple stellar population models (SSPs) as described further below. As done previously\cite{Zhang2016}, we require the continuum level within the 300 \AA\ spectral window for each spaxel spectrum to be less than 3 $\times$ $10^{-17}$ erg cm$^{-2}$ s$^{-1}$ \AA$^{-1}$ to limit the noise introduced by interloping strong emitters such as satellite galaxies or outer disk emission line regions, such as the HII regions\cite{thilker05,z07}.
In the spatial dimension, we 
bin $2\times 2$ spaxels 
to accumulate 
signal, which roughly matched the spaxel
size to the observing seeing, but not sacrifice too much spatial resolution. Note that we do not need to re-bin $2\times 2$ spaxels to construct the radial profile presented in Figure \ref{fig:radialProf} in order to retain the small-scale emission features.

We correct for the underlying stellar continuum within $3 R_e$ of each MaNGA datacube, including the stellar absorption lines that can affect the emission-line flux measurements, over the rest-frame wavelength range of $3700-8000$ \AA, using the penalized pixel-fitting (pPXF) package\cite{Cappellari2004, Cappellari2017, Cappellari2023} with a linear combination of stellar templates from the MILES library\cite{Sanchez2006, Falcon2011}. Beyond $3 R_e$ of each datacube, the signal to noise ratio in the continuum of the spectrum in each spaxel is insufficient to do this.
Instead, we still use a third order polynomial to estimate the continuum for each spectra. We measure our emission-line surface brightness within the annuli across the entire MaNGA FOV. For those galaxies with $b/a$ (minor axis to major axis ratio) less than 0.75, we correct for the annuli by adopting the ellipse with the same $b/a$. A  S/N of $\gtrsim 3$ is generally required for reliable CGM detection, and this threshold is adopted in the above discussion. To estimate the emission-line surface brightness or the line ratio uncertainties, we adopt the bootstrap method.  We repeat the bootstrap resampling 1,000 times and calculate the mean measurements of the resampled data for each iteration. From the distribution of measurements, we quote use the 16.5 and 83.5 percentiles as the uncertainty range. 


The linear regression analyses in this study were performed within a Bayesian framework, utilizing Markov Chain Monte Carlo (MCMC) sampling for parameter estimation.
In particular, we adopted the likelihood function for straight line fitting\cite{Sharma2017}, to model the H$\alpha$ surface brightness (log$f_{\rm H\alpha}$) dependence on either SFR or sSFR in the logarithmic scale. 
The likelihood function simultaneously accounts for measurement uncertainties in both the H$\alpha$ surface brightness and SFR (or sSFR).
Crucially, the model also incorporates an intrinsic scatter term ($\sigma_p$), accounting for the astrophysical dispersion beyond measurement uncertainties. The intrinsic scatter was prescribed a fixed value of $\sigma_p=0.2$ dex for all fittings.
The adopted priors are symmetric with respect to rotation (i.e., uniform for the inclination angle ($\theta$) rather than the line slope, slope = $\tan \theta$)\cite{Sharma2017} and uniform for the intercept. The MCMC sampling was completed with 50 walkers and 5000 steps per chain, with convergence assessed and confirmed using the Gelman-Rubin diagnostic\cite{Gelman1992}.
Best-fit parameters and their uncertainties reported in Table \ref{tab:param} correspond to the median and the central 68\% credible interval (the 16.5 and 83.5 percentiles, $\pm 1\sigma$) of the marginalized posterior distributions. The Python packages used in this work are all the newest version, including SciPy \cite{SciPy2020}, NumPy \cite{numpy2020}, AstroPy \cite{astropy}, Matplotlib \cite{Hunter2007}, Pandas \cite{pandas} and emcee \cite{emcee}.

\medskip
\noindent{\bf [O {\small II}]$\lambda \lambda$3727,3729 and [O {\small III}]$\lambda$5008 Emission}

\begin{figure*}[ht!]
\begin{center}
\includegraphics[width = 0.475 \textwidth]{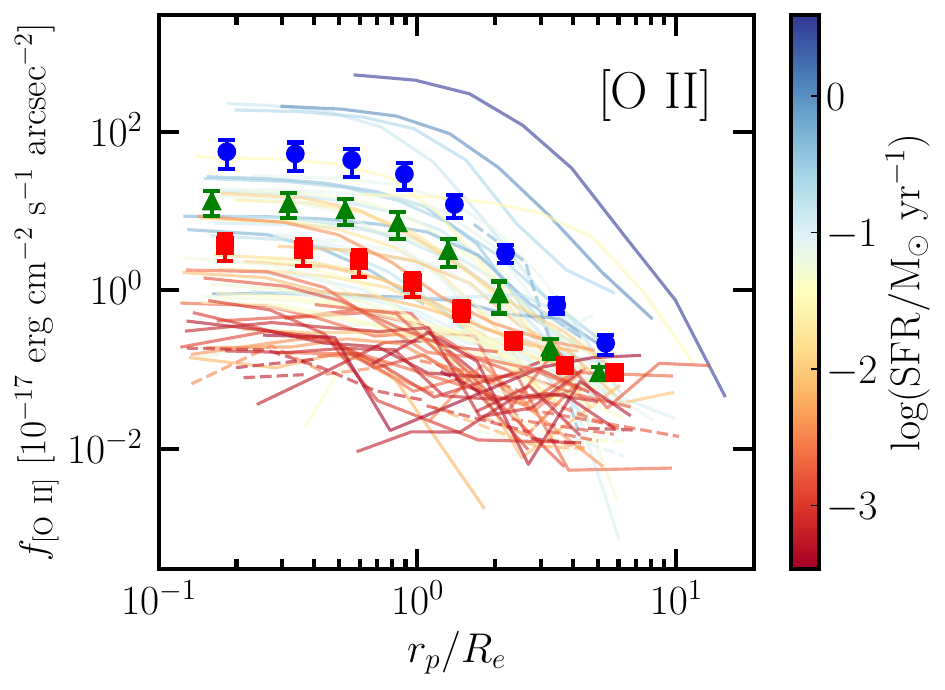}
\hspace{0.5 cm}
\includegraphics[width = 0.475 \textwidth]{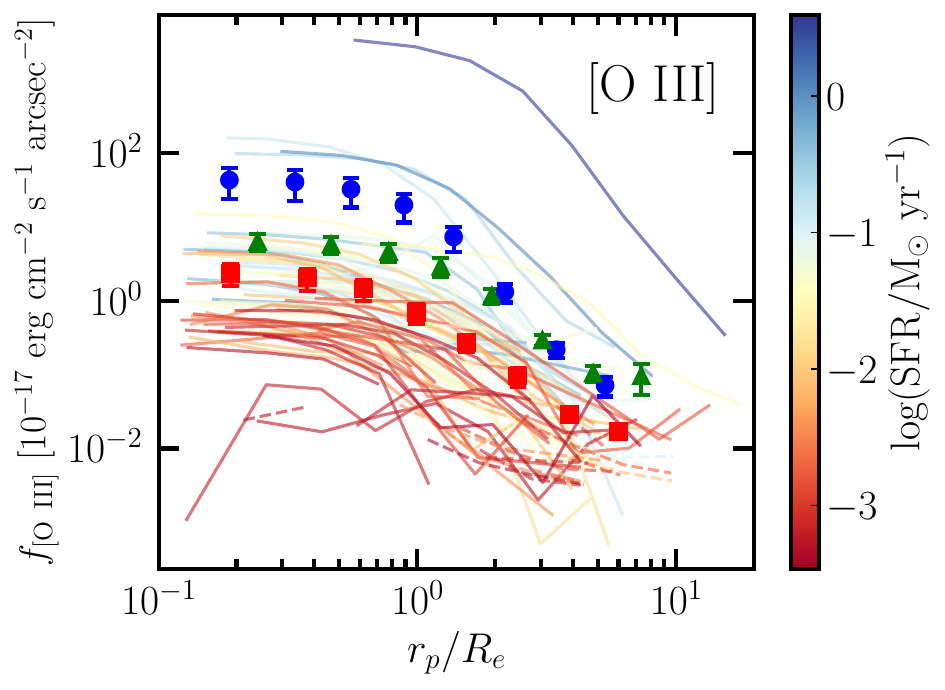}
\end{center}
\caption{{\bf The [O {\small II}] and [O {\small III}] emission SB radial profile.} The [O {\small II}] (left) and [O {\small III}] (right) emission surface brightness radial profile of all our MaNGA galaxies, color coded by  star formation rate. Dashed lines denote non-detections where we adopt the 2$\sigma$ upper limit. The blue, green, and red points represent the stacked radial profiles for galaxies within SFR bins of $-1< \log(\rm SFR/M_\odot \ yr^{-1})< 0 $, $-1.75 <\log(\rm SFR/M_\odot \ yr^{-1})<-1$ and $\log(\rm SFR/M_\odot \ yr^{-1})<-1.75$, respectively, using only galaxies with detections in all radial bins.} 
\label{fig:OIIOIIIradial}
\end{figure*}

The [O {\small II}] and [O {\small III}] emission line surface brightness  radial profiles for all our MaNGA galaxies are shown in Figure \ref{fig:OIIOIIIradial}. The general features of these profiles are consistent with those of the H$\alpha$ emission surface brightness radial profile.

\begin{figure*}[ht!]
\begin{center}
\includegraphics[width = 0.725 \textwidth]{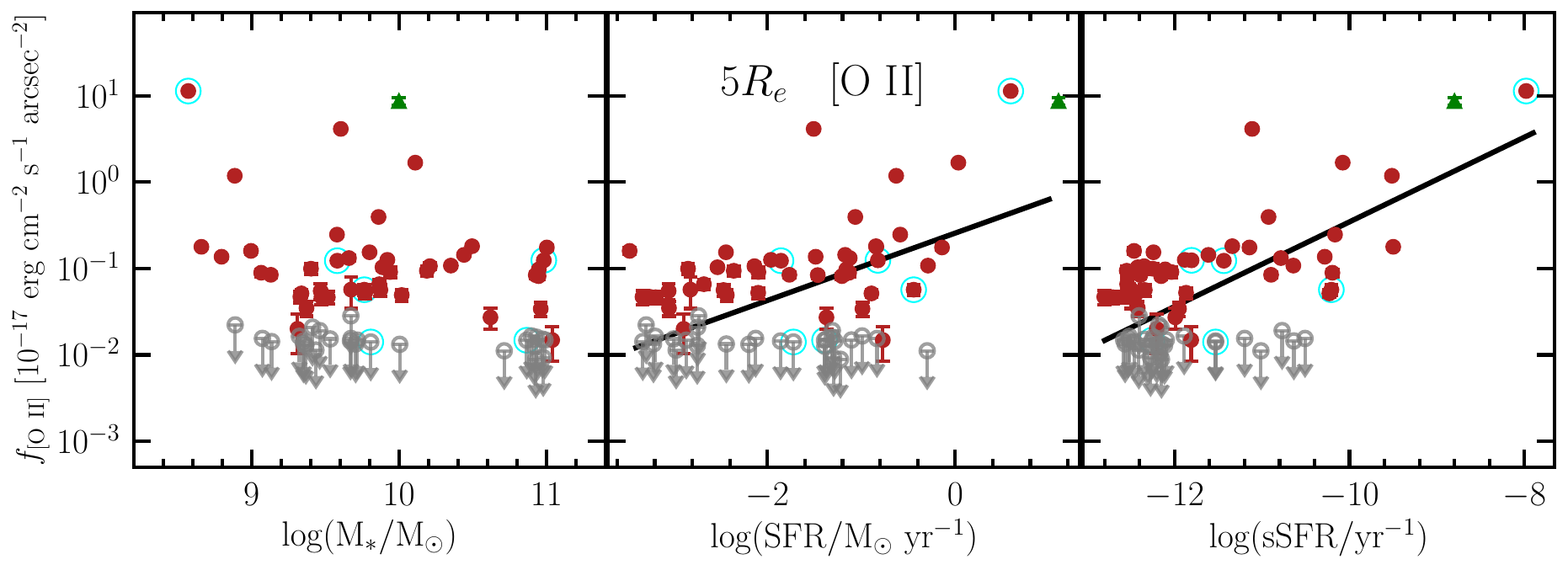}
\includegraphics[width = 0.725 \textwidth]{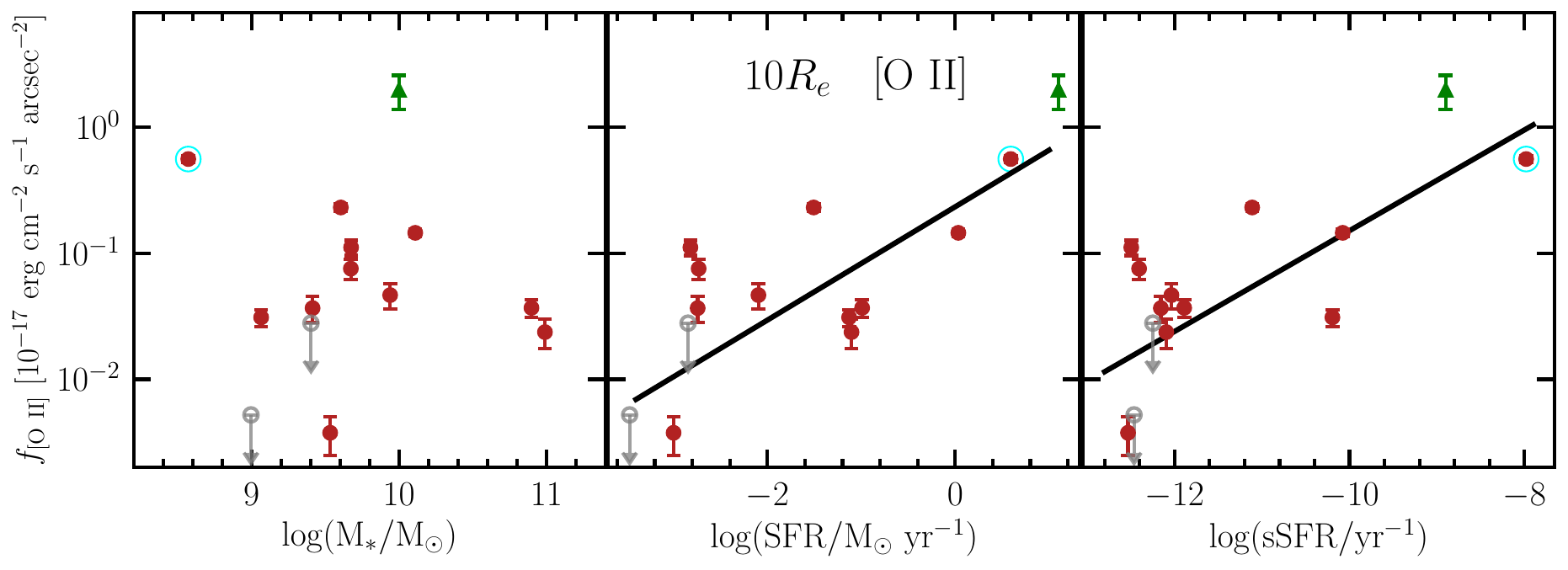}
\includegraphics[width = 0.725 \textwidth]{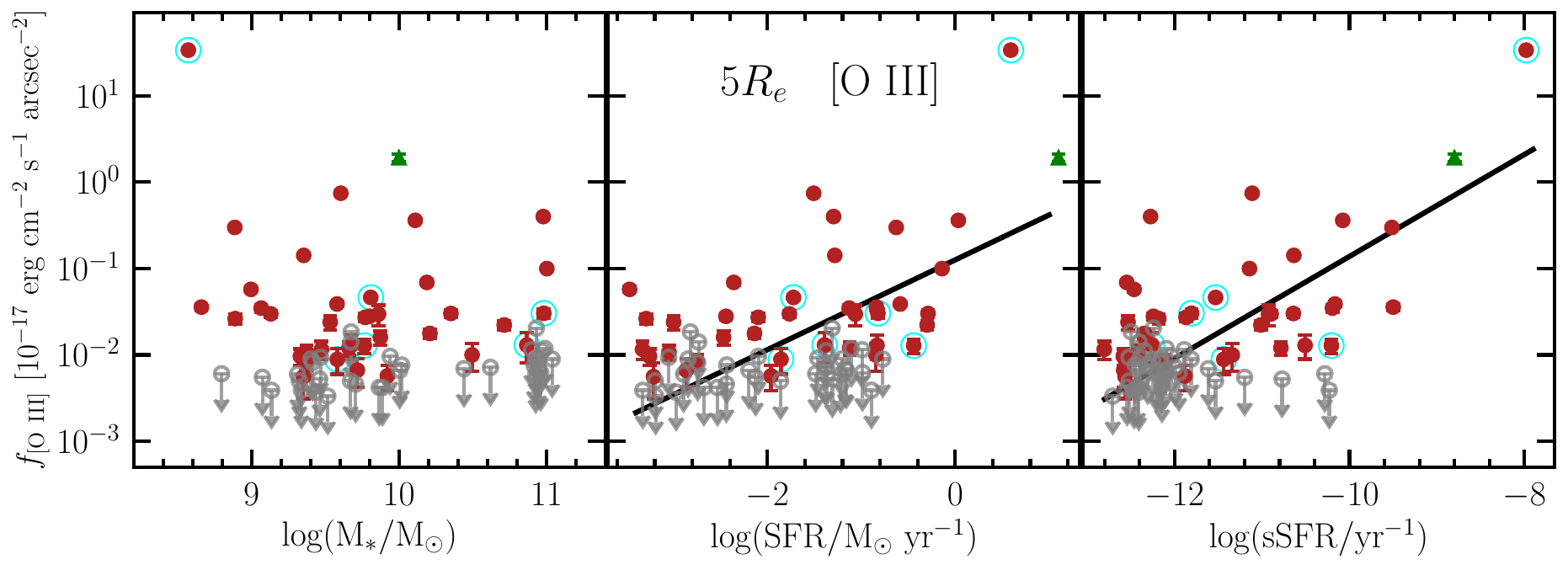}
\includegraphics[width = 0.725 \textwidth]{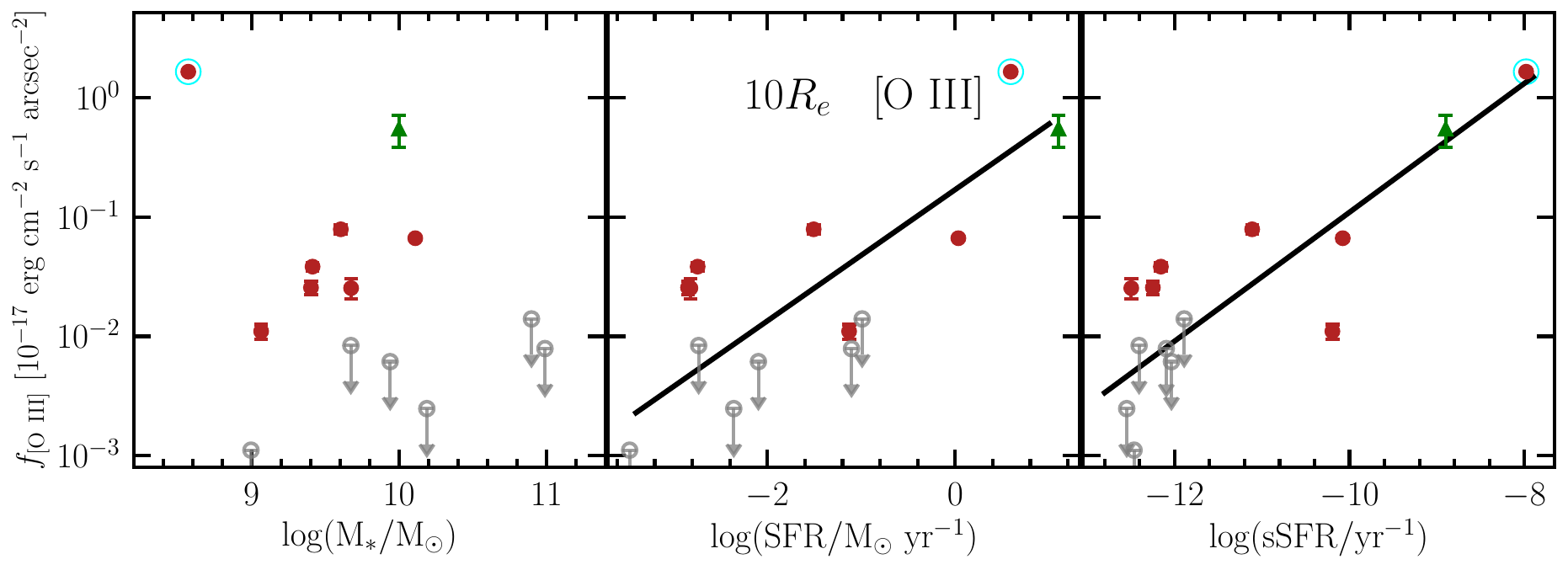}
\end{center}
\caption{{\bf The [O {\small II}]and [O {\small III}] emission SB.} The [O {\small II}] (top two) and [O {\small III}] (bottom two) emission surface brightness at $5 R_e$  and $10 R_e$ as a function of M$_*$, SFR and sSFR. The symbols are the same as those in Figure \ref{fig:fluxStats}.}  
\label{fig:OIIOIIIfluxStats}
\end{figure*}

In Figure \ref{fig:OIIOIIIfluxStats}, we present the [O {\small II}] and [O {\small III}] emission surface brightness measurements at $5R_e$ and $10 R_e$ as a function of M$_*$, SFR and sSFR. For those galaxies without [O {\small II}] or [O {\small III}] emission detection or a detection significance less than $2\sigma$, we adopt and plot the $2\sigma$ upper limit. Again, the [O {\small II}] and [O {\small III}]  emission  at large radius ($(5-10) R_e$) correlates  with the galaxy's SFR and sSFR, rather than with stellar mass. We find that the correlation slope between H$\alpha$ emission and SFR/sSFR at $5R_e$ is significantly steeper than the slopes derived from either [O {\small II}] or [O {\small III}] emission measurements at the same radius, which might suggest that the harder ionizing sources are lacking at this radius combining the fact that oxygen is primarily ionized by hard ionizing sources. We present the best-fit parameters in Table \ref{tab:param2}.

\begin{table*}[ht!]
    \centering
    \begin{tabular}{ccccc}
    \hline
    & \multicolumn{2}{c}{SFR} & \multicolumn{2}{c}{sSFR} \\ \hline
    $r_p$ & $m_1$ & $b_1$  & $m_2$ & $b_2$ \\
     \hline  
$5 R_e$ ([O {\small II}]) & $0.39 \pm 0.03$ & $-0.98 \pm 0.04$ & $0.49 \pm 0.03$ & $-0.95 \pm 0.04$ \\ \\
$10 R_e$ ([O {\small II}]) &  $0.45 \pm 0.06$ & $-1.08 \pm 0.08$ & $0.40 \pm 0.05$ & $-1.22 \pm 0.07$ \\ \\
$5R_e$ ([O {\small III}]) & $0.52 \pm 0.04$ & $-1.42 \pm 0.04$ & $0.59 \pm 0.04$ & $-1.45 \pm 0.04$ \\ \\
$10 R_e$ ([O {\small III}]) & $0.55 \pm 0.07$ & $-1.32 \pm 0.08$ & $0.54 \pm 0.06$ & $-1.50 \pm 0.07$ \\  \hline
    \end{tabular}
    \vspace{0.5cm}
    \caption{{\bf The best-fit parameters of Eq.  \ref{eq:linear} for [O {\small II}]and [O {\small III}] emission.} The best-fit parameters for the relation between  [O {\small II}]/[O {\small III}] emission surface brightness at different radii and SFR/sSFR of the central galaxy, as described in Eq. \ref{eq:linear}.}
    \label{tab:param2}
\end{table*}

\medskip
\noindent{\bf GAEA and TNG Simulations }

We use the latest version of GAlaxy Evolution and Assembly (GAEA) model\cite{delucia2024}, with more details in  \url{https://sites.google.com/inaf.it/gaea/home}, running on merger trees extracted from the Millennium Simulation\cite{Springel2005}, which has a box size of 683 cMpc. In this work we use about $1/50$ of the volume. This version of the model implements an updated and well-calibrated treatment for supermassive black hole (SMBH) accretion and AGN feedback\cite{fontanot2020}, plus a state-of-the-art treatment of enviromental effect (like tidal stripping and dynamical friction)\cite{xie2020}, which predicts good agreement with observational measurements on quenched fraction up to $z\sim 3$\cite{xie2024}.  In the semi-analytic approach, each galaxy includes several discrete components (bulge, disc, halo), where baryons are described by a multi-phase gas (hot gas, cold gas and stars). Hot gas is heated up to the virial temperature, which exceeds the temperature limit considered in this study for all resolved massive halos at $z=0$, and therefore is not accounted for in the measured cool gas surface density. Cold gas on the disk is assumed to have a temperature of $\le 10^4$ K, and follow exponential distribution of $\Sigma_g (r)= \frac{M_g}{2\pi r^2_g} e^{-\frac{r}{r_g}}$. Here, $M_g$ and $r_g$ are the cold gas mass and the scale length of the cold gas disk. 

IllustrisTNG (TNG) is a spatially resolved magnetohydrodynamical (MHD) simulation\cite{weinberger2017, springel2018, Nelson2018, Naiman2018, Marinacci2018, Pillepich2018a, pillepich2018b}  which is available at \url{https://www.tng-project.org/data/}. In this study, we use TNG100 from the TNG project. The side length of the simulation box is 106.5 Mpc. The baryon mass resolution is $1.4 \times 10^6\  {\rm M}_{\odot}$, the gravitational softening length of the dark matter and stars is 0.7 kpc at $z = 0$, and the minimum gas cell radius is 185 pc. We compute the effective temperature of each gas cell using the electron abundance and internal energy. The star-forming gas cells are excluded. Gas cells in TNG have a minimum temperature of $10^4$ K, therefore, we select gas cells with $10^4 \le T_{\rm eff} <10^{4.25}$ K as the cool gas. 



To compare to our MaNGA results, we measure the surface density of gas 
at $5 R_e$ and $10 R_e$ (within  annuli of $(4.5-5.5) R_e$ and $(9.5-10.5) R_e$, respectively) of the central galaxies  in both  the GAEA model and the TNG simulation. For the GAEA galaxies, the comparison is not straightforward because those models only include cold (able to form stars for the molecular
component) and hot (virial temperature) gas. To proceed, on one hand, we assume that the gas our observations trace is more closely related to the cold gas and that this cold gas has an exponential density profile beyond the galaxy disk. On the other hand, we include the cool gas (with temperature of $\sim$ 10,000 K) estimated according to the scaling relation between the mass cooling rate and the stacked H$\alpha$ emission flux\cite{Zhang2021}.
For the TNG simulation, we first identify the three axes of the distribution of stellar particles, and then determine the stellar disk using the long and intermediate axes, and finally project the cool gas cells of temperature  $T_{\rm eff} < 10^{4.25}$ K onto the stellar disk. We then calculate the surface density of the cool gas by summing up all the gas cells in each annulus. To match our MaNGA galaxy sample, we select analogues from both GAEA and TNG that have a stellar mass and SFR within $\pm$0.1 dex of each  galaxy in our MaNGA galaxy sample. Furthermore, we select only central galaxies because our MaNGA galaxies are not interacting with other galaxies or otherwise disturbed by the ambient environment. 
This leaves us with 3,481 galaxies from the GAEA simulation and 2,588 galaxies from the TNG simulation. We then calculate the median cool gas surface density for each bin in M$_*$, SFR and sSFR, respectively.


\medskip
\noindent{\bf The inner CGM of example galaxies}

To provide examples of the power of IFS data, 
we select 2 individual galaxies for which we detect  most or all of the principal optical  emission lines ([O {\small II}]$\lambda \lambda$3727,3729, H$\beta$, [O {\small III}]$\lambda$5008, H$\alpha$ and [N {\small II}]$\lambda$6585, all vacuum wavelengths). 


The first target we highlight is a starburst galaxy at $z = 0.04723$ with M$_* = 10^{8.57}$ M$_\odot$ and SFR = 3.90 M$_\odot$ yr$^{-1}$. The position angle of this target galaxy is 74$^\circ$ and the minor to major axis ratio is 0.92. The half light radius is 0.86$^{\prime \prime}$, corresponding to $\sim$ 0.81 kpc at the distance of the target galaxy. 
The virial radius of this galaxy is estimated to be $\sim$ 74 kpc according to the scaling relation between the stellar mass and the virial radius derived from the {\it UniverseMachine} simulation\cite{Behroozi2019}. For this galaxy, although it is star-bursting, the half light radius ($R_e$) is still a good measurement and consistent with identifying the location of the CGM in terms of fractional virial radii ($5R_e \sim 0.06\ r_{\rm vir}$).

\begin{figure*}[ht!]
\begin{center}
\includegraphics[width = 0.475 \textwidth]{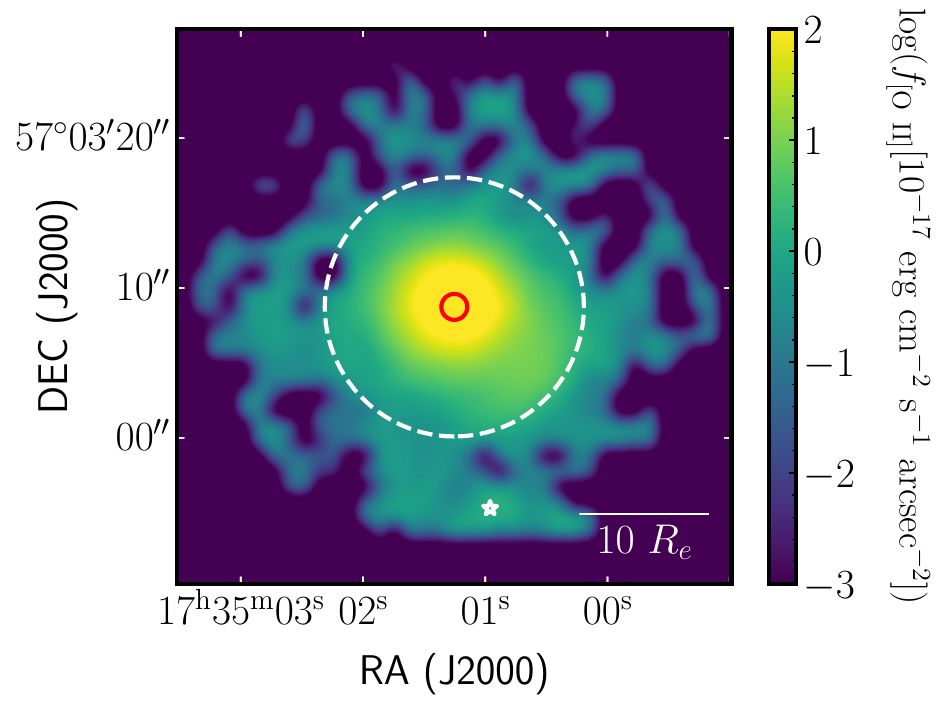}
\includegraphics[width = 0.475 \textwidth]{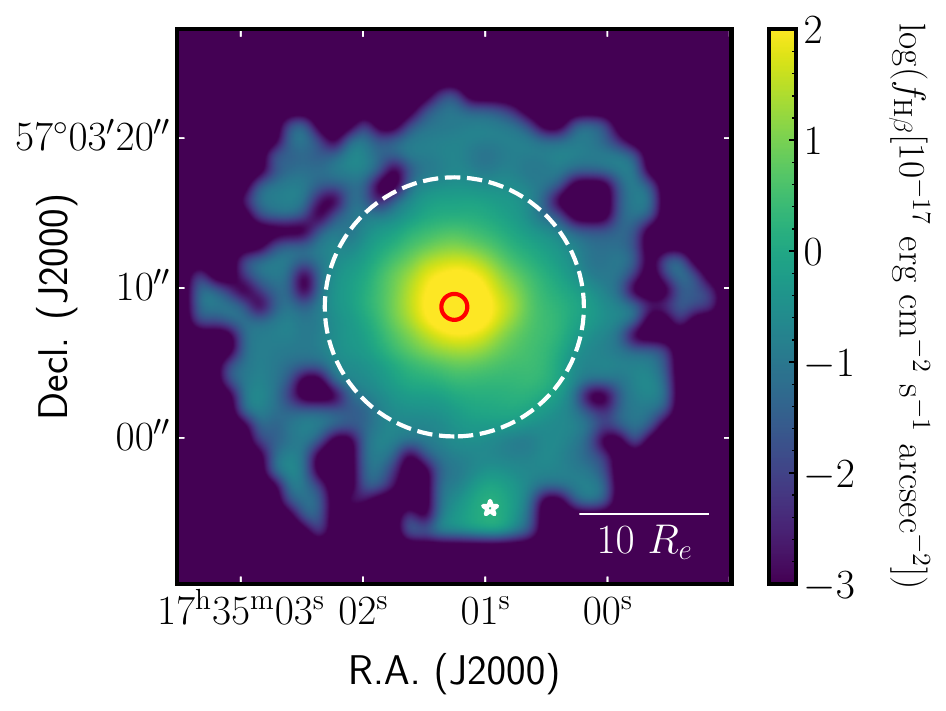}
\hspace{0.5cm}
\includegraphics[width = 0.475 \textwidth]{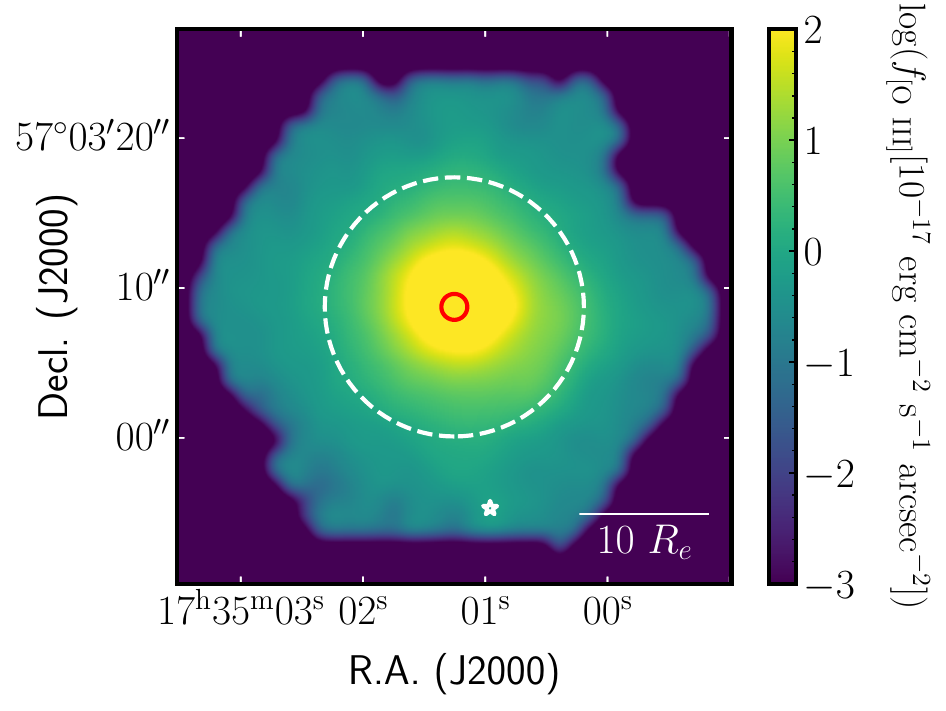}
\includegraphics[width = 0.475 \textwidth]{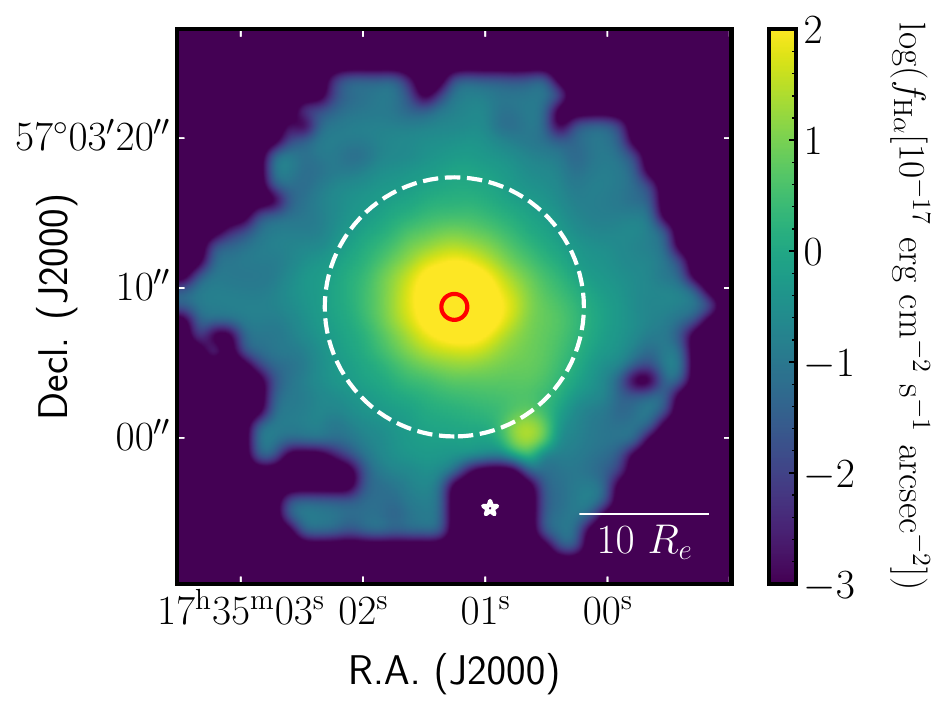}
\hspace{0.5cm}
\end{center}
\caption{{\bf Narrow-band images of the first example galaxy across the entire FOV.} Narrow-band images of the first target galaxy at $z = 0.04723$ with M$_* = 10^{8.57}$ M$_\odot$ and SFR = 3.90 M$_\odot$ yr$^{-1}$. Narrow-band images ($\pm 150$ km s$^{-1}$ or 6 \AA\  wide for H$\alpha$)  of [O {\small II}]$\lambda$$\lambda$3727, 3729  (top left),  H$\beta$ (top right),  [O {\small III}]$\lambda$5008 (bottom left) and  H$\alpha$  (bottom right) emission across the entire MaNGA FOV,  binned $2\times 2$ spaxels to further enhance the signal to noise (S/N) ratio. The small red circle represents the $R_e$ of the galaxy and the big white circle indicates $10  R_e$ of the galaxy. The white star marks the location of a bright foreground star. The hexagonal pattern of the MaNGA IFU FOV is clearest in the upper right panel.}  
\label{fig:fluxmap8626}
\end{figure*}

With significant detections of all the optical emission lines (Figure \ref{fig:fluxmap8626}), we can use the emission line ratios to estimate the metallicity and ionization of the CGM. The line ratio of R23 ($\equiv \log((\rm [O \  II]\lambda3727 + [O \ III] \lambda\lambda5008,4960)/H\beta$)) is 
widely adopted as a metallicity diagnostic\cite{Maiolino2019}, as it involves emission lines of both the main ionization stages of oxygen, O$^+$ and O$^{+2}$, and so is less affected by the ionization parameter than alternatives\cite{Kewley2002, Nagao2006}. We caution that the metallicity estimation using R23 as an indicator for the CGM at larger radius might be biased because the R23 method is calibrated using HII regions rather than the CGM and the ionization structure, temperature, and electron density will be quite different. Therefore, the metallicity estimates here are only used for qualitative purposes. 
Using the stacked spectra within annuli, we find that the metallicity within 8 kpc ($10R_e$ or 0.12 $r_{\rm vir}$) is quite homogeneous, with a value of 12 + $\log \rm (O/H) \sim$  8.0, consistent with the {\it Pipe3D} estimation at 2$R_e$ \cite{Sanchez2022}. We also measure a  metallicity gradient, with the gas metallicity decreasing from  $7.91^{+0.03}_{-0.03}$ at 10 kpc (within annulus with radius between 9 kpc and 11 kpc) to $7.78^{+0.08}_{-0.06} $ at 15 kpc (within annulus with radius between 14 kpc and 16 kpc).

\begin{figure}[ht]
\begin{center}
\includegraphics[width = 0.48 \textwidth]{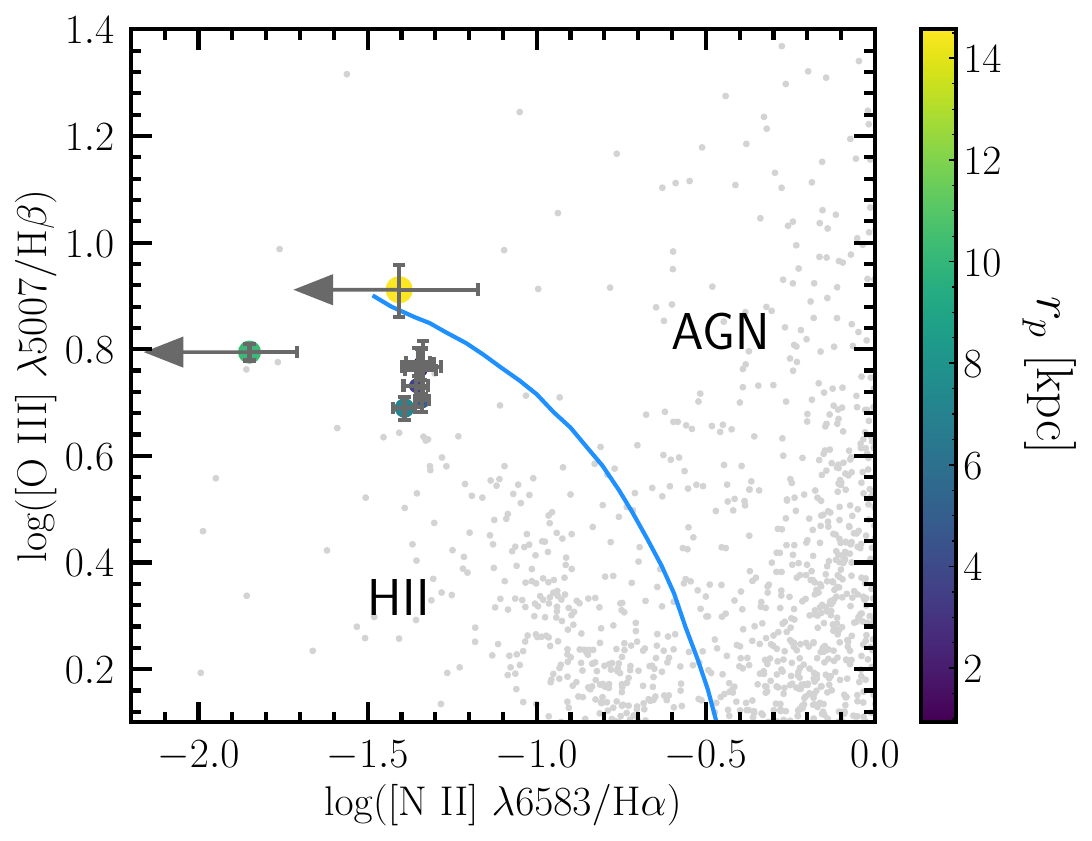}
\includegraphics[width = 0.42 \textwidth]{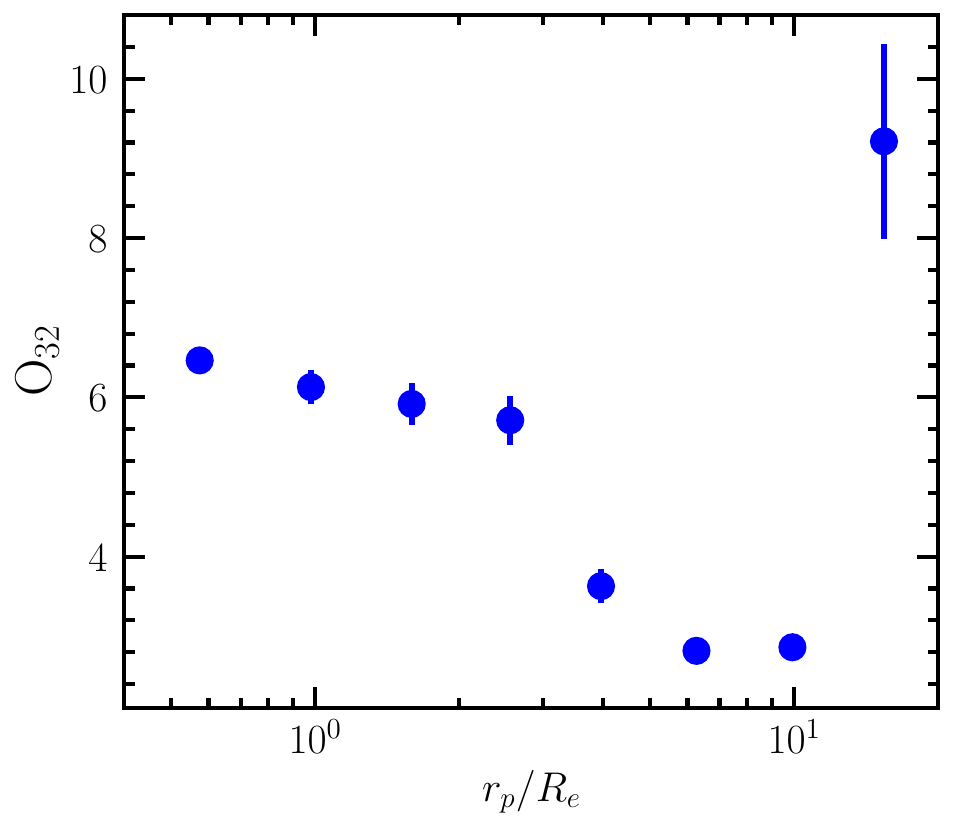}
\end{center}
\caption{{\bf The BPT emission line ratios and O$_{32}$ radial profile of the first example galaxy.} The BPT emission line ratios of the first representative galaxy for different radial bins (colored and sized by radius). Here the radial bins for the BPT emission line ratios are in equal linear step to resolve the large-scale features. The blue curve shows the boundary between the H{\small II} and AGN regions of the diagram\cite{Kauffmann2003} and the light grey points represent line ratios of individual SDSS galaxies\cite{sdss7}.}  
\label{fig:bpt8626}
\end{figure}

Furthermore, to discriminate between ``soft" ionizing sources such as massive stars (HII region) and ``hard" sources such as AGN and shocks, we use the line ratio diagram commonly  referred to as the BPT diagram\cite{bpt} that  compares [N {\small II}]$\lambda$6585/H$\alpha$ and [O {\small III}]$\lambda$5008/H$\beta$. We have no detection of [N {\small II}]$\lambda$6585 emission beyond 8 kpc, we adopt the upper limit.  The line ratios presented in BPT diagram as shown in Figure \ref{fig:bpt8626} for all regions within 16 kpc ($\sim 0.24 \ r_{\rm vir}$) are consistent with soft ionizing sources. The line ratios measured within an 8 kpc aperture  are fully consistent with the {\it Pipe3D} values ($\log \rm ([N\ II] ~\lambda 6583/H\alpha) = -1.31$, $\log \rm ([O\ III] ~\lambda 5007/H\beta) = 0.78$) derived from a $2.5^{\prime\prime}$ aperture. However, this result is not consistent with the classification reported in the MaNGA-AGN catalog, indicating a potential discrepancy between the emission-line-based classification and the mid-infrared color selection used by the catalog for this galaxy. The O$_{32}$ radial profile of this target galaxy is also presented in Figure \ref{fig:bpt8626}, which is the same as that presented in Figure \ref{fig:radialProf}. 

\begin{figure*}[ht!]
\begin{center}
\includegraphics[width = 0.475 \textwidth]{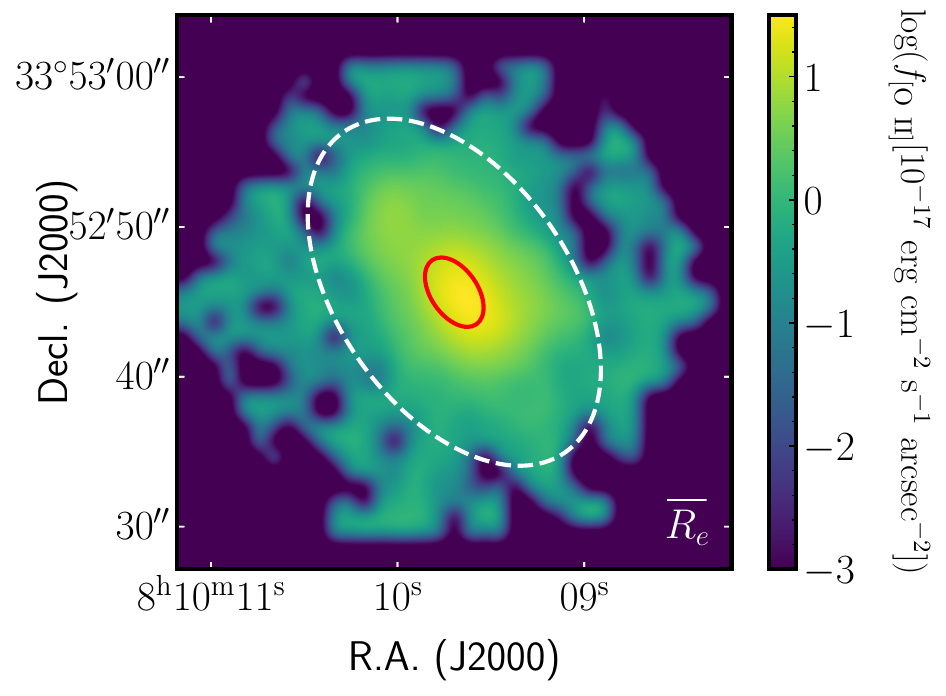}
\includegraphics[width = 0.475 \textwidth]{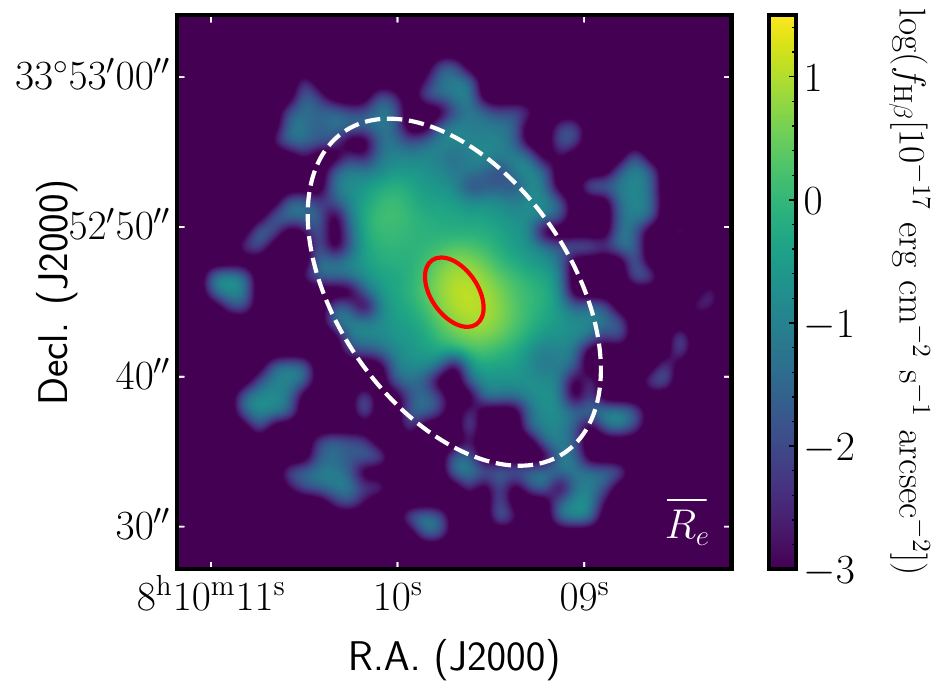}
\hspace{0.5cm}
\includegraphics[width = 0.475 \textwidth]{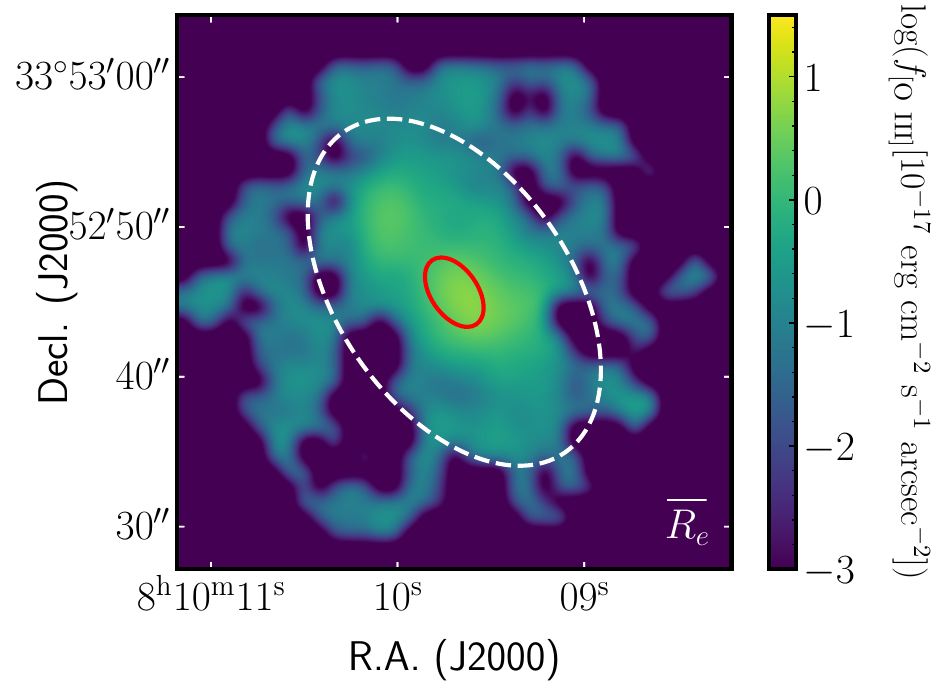}
\includegraphics[width = 0.475 \textwidth]{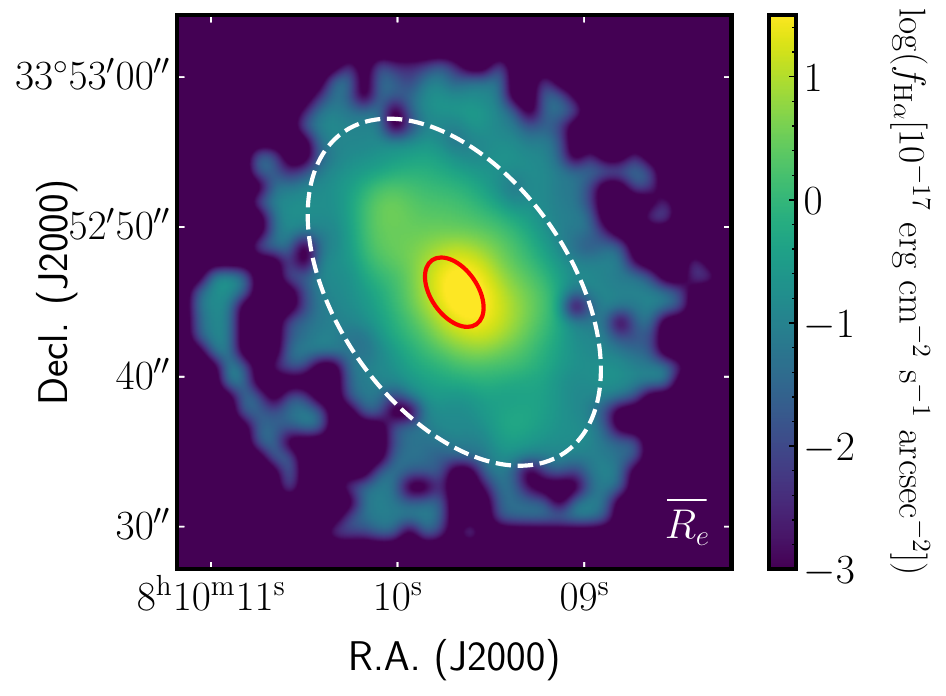}
\hspace{0.5cm}
\end{center}
\caption{{\bf Narrow-band images of the second example galaxy.} Narrow-band images  ($\pm 200$ km s$^{-1}$ or 9 \AA\  wide for H$\alpha$)  of [O {\small II}]$\lambda$$\lambda$3727, 3729  (top left),  H$\beta$ (top right),  [O {\small III}]$\lambda$5008 (bottom left) and  H$\alpha$  (bottom right) across the entire MaNGA FOV and binned $2\times 2$ spaxels for this target galaxy with M$_* = 10^{9.58}$ M$_\odot$ and SFR = 0.26 M$_\odot$ yr$^{-1}$ at $z = 0.04389$. The small red ellipse represents the $R_e$ of the galaxy and the big white ellipse indicates $5 R_e$ of the galaxy.}  
\label{fig:fluxmap10220}
\end{figure*}


The second target we highlight is a star-forming sub-$L^*$ galaxy 
at $z = 0.04389$. It has a stellar mass of $10^{9.58}$ M$_\odot$ and a SFR of 0.26 M$_\odot$ yr$^{-1}$. The position angle of this target galaxy is 55$^\circ$ and the minor to major axis ratio is 0.6. The half light radius is 2.62$^{\prime \prime}$, corresponding to $\sim$ 2.28 kpc at the distance of the target galaxy.
This target galaxy is a typical star-forming galaxy, which lies on the main sequence of galaxies in the  M$_* - $ SFR relation\cite{Brinchmann2004, Peng2010} and on
the stellar mass $-$ half light radius relation (M$_* - R_e$)\cite{Fernandez2013, Zhang2025}. We estimate that the virial radius of this galaxy is 107 kpc using the scaling relation between the stellar mass and the virial radius derived from the {\it UniverseMachine} simulation\cite{Behroozi2019}.
We detect [O {\small II}]$\lambda \lambda$3727,3729, H$\beta$, [O {\small III}]$\lambda$5008, H$\alpha$ out to at least $5 R_e$ (Figure \ref{fig:fluxmap10220}).
 Although H$\alpha$ is the strongest emission line for this galaxy, the H$\alpha$  surface brightness at 10 kpc ($9 < r < 11$ kpc, $\sim 5 R_e$) along the major axis of the galaxy is almost 3 times smaller than that of the first galaxy at the same radii. Even so, it is still significantly higher ($\gtrsim 5$) than the stacking results representing the average comparably massive galaxy \cite{Zhang2016, Zhang2018a}.  

\begin{figure*}[ht!]
\begin{center}
\includegraphics[width = 0.48 \textwidth]{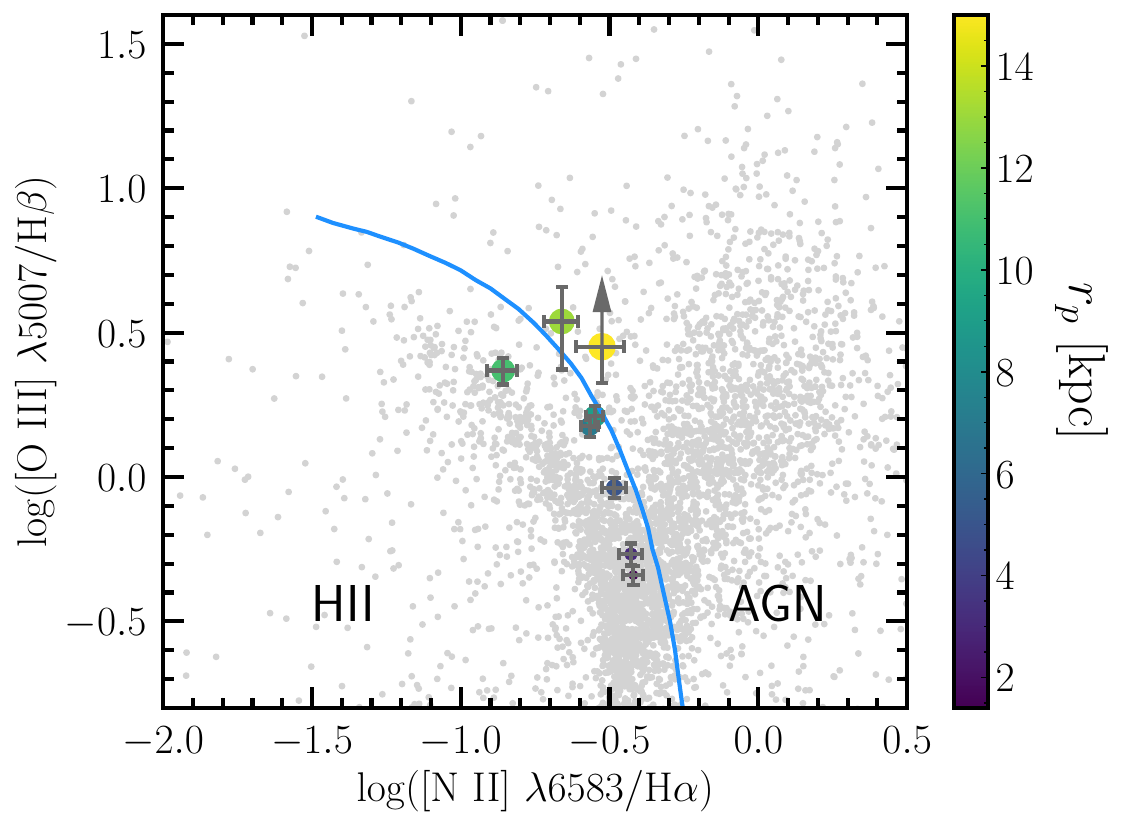}
\includegraphics[width = 0.42 \textwidth]{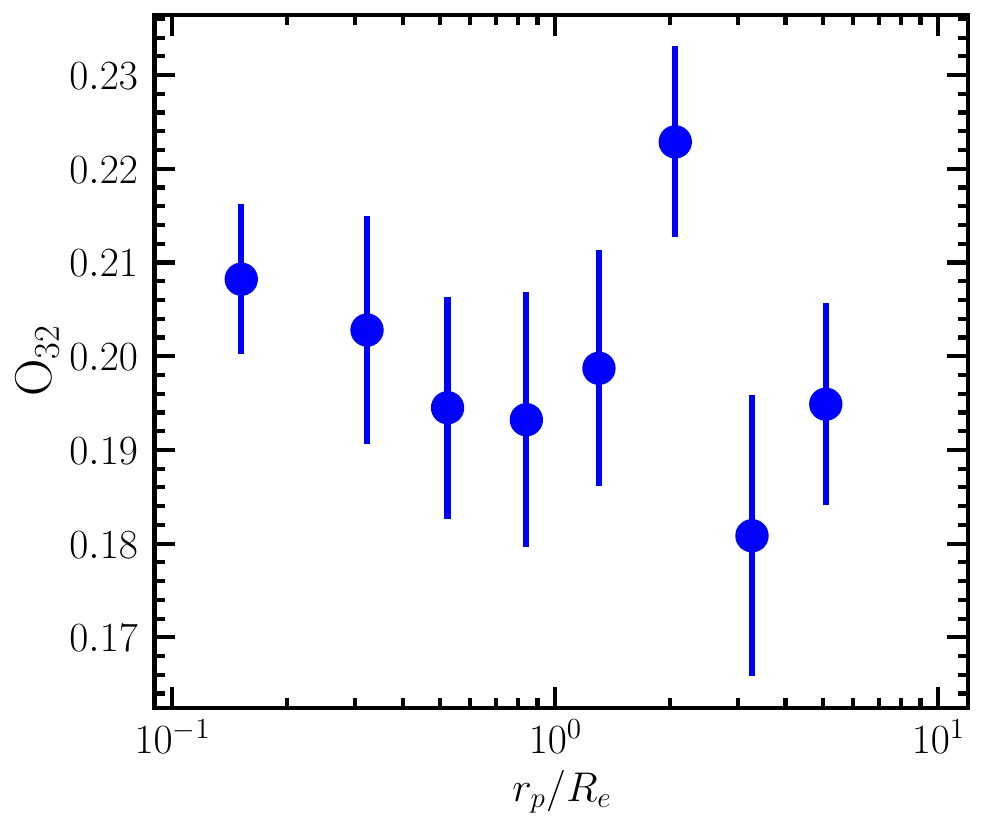}
\end{center}
\caption{{\bf The BPT emission line ratios and O$_{32}$ radial profile of the second example galaxy.} The BPT emission line ratios of the second representative galaxy for different radial bins (colored and sized by radius). Here the radial bins for the BPT emission line ratios are in equal linear step to resolve the large-scale features. The symbols are the same as in Figure \ref{fig:bpt8626}.}
\label{fig:bpt10220}
\end{figure*}

Again, we can use the emission lines for various diagnostics.
The metallicity at 
10 kpc ($9 < r < 11$ kpc, $\sim 5 R_e$) corresponds to [O/H] $= -0.7$ dex,  
roughly consistent with the estimates of the CGM metallicity from absorption lines studies\cite{prochaska2017, Pointon2019}. The metallicity is continuing to decrease due to the decreasing H$\beta$ flux beyond $\sim 5 R_e$.  
There is no detection of H$\beta$ emission beyond $\sim$ 12 kpc along the major axis of the galaxy, we adopt the upper limits of the H$\beta$ emission surface brightness to construct the BPT line ratios, or line ratio limits, for different radial bins. 
The line ratios at all radii are very close to the boundary between HII regions and AGNs. Specifically, they are consistent with soft ionizing sources for $r < 12$ kpc, and are consistent with hard ionizing sources such as AGNs or heat shocks outside this radius, as shown in Figure \ref{fig:bpt10220}. The O$_{32}$ values at all radii of this target galaxy as shown in Figure \ref{fig:bpt10220} are quite consistent, $\sim$ 0.2. 
\bibliographystyle{ScienceAdvances}

\bibliography{paper}

@ARTICLE{thilker05,
       author = {{Thilker}, David A. and {Bianchi}, Luciana and {Boissier}, Samuel and {Gil de Paz}, Armando and {Madore}, Barry F. and {Martin}, D. Christopher and {Meurer}, Gerhardt R. and {Neff}, Susan G. and {Rich}, R. Michael and {Schiminovich}, David and {Seibert}, Mark and {Wyder}, Ted K. and {Barlow}, Tom A. and {Byun}, Yong-Ik and {Donas}, Jose and {Forster}, Karl and {Friedman}, Peter G. and {Heckman}, Timothy M. and {Jelinsky}, Patrick N. and {Lee}, Young-Wook and {Malina}, Roger F. and {Milliard}, Bruno and {Morrissey}, Patrick and {Siegmund}, Oswald H.~W. and {Small}, Todd and {Szalay}, Alex S. and {Welsh}, Barry Y.},
        title = "{Recent Star Formation in the Extreme Outer Disk of M83}",
      journal = {\apjl},
         year = 2005,
        month = jan,
       volume = {619},
       number = {1},
        pages = {L79-L82},
          doi = {10.1086/425251},
archivePrefix = {arXiv},
       eprint = {astro-ph/0411306},
 primaryClass = {astro-ph},
       adsurl = {https://ui.adsabs.harvard.edu/abs/2005ApJ...619L..79T}
}

@ARTICLE{Kauffmann2003,
       author = {{Kauffmann}, Guinevere and {Heckman}, Timothy M. and {Tremonti}, Christy and {Brinchmann}, Jarle and {Charlot}, St{\'e}phane and {White}, Simon D.~M. and {Ridgway}, Susan E. and {Brinkmann}, Jon and {Fukugita}, Masataka and {Hall}, Patrick B. and {Ivezi{\'c}}, {\v{Z}}eljko and {Richards}, Gordon T. and {Schneider}, Donald P.},
        title = "{The host galaxies of active galactic nuclei}",
      journal = {\mnras},
         year = 2003,
        month = dec,
       volume = {346},
       number = {4},
        pages = {1055-1077},
          doi = {10.1111/j.1365-2966.2003.07154.x},
archivePrefix = {arXiv},
       eprint = {astro-ph/0304239},
 primaryClass = {astro-ph},
       adsurl = {https://ui.adsabs.harvard.edu/abs/2003MNRAS.346.1055K}
}

@ARTICLE{Kauffmann2003b,
       author = {{Kauffmann}, Guinevere and {Heckman}, Timothy M. and {White}, Simon D.~M. and {Charlot}, St{\'e}phane and {Tremonti}, Christy and {Peng}, Eric W. and {Seibert}, Mark and {Brinkmann}, Jon and {Nichol}, Robert C. and {SubbaRao}, Mark and {York}, Don},
        title = "{The dependence of star formation history and internal structure on stellar mass for {}10$^{5}$ low-redshift galaxies}",
      journal = {\mnras},
         year = 2003,
        month = may,
       volume = {341},
       number = {1},
        pages = {54-69},
          doi = {10.1046/j.1365-8711.2003.06292.x},
archivePrefix = {arXiv},
       eprint = {astro-ph/0205070},
 primaryClass = {astro-ph},
       adsurl = {https://ui.adsabs.harvard.edu/abs/2003MNRAS.341...54K}
}

@ARTICLE{Zhang2025,
       author = {{Zhang}, Huanian and {Ye}, Guangping and {Wu}, Rongyu and {Zaritsky}, Dennis},
        title = "{A Local Dwarf Galaxy Search Using Machine Learning}",
      journal = {\apjs},
         year = 2025,
        month = may,
       volume = {278},
       number = {1},
          eid = {18},
        pages = {18},
          doi = {10.3847/1538-4365/adbf0a},
archivePrefix = {arXiv},
       eprint = {2503.00109},
 primaryClass = {astro-ph.GA},
       adsurl = {https://ui.adsabs.harvard.edu/abs/2025ApJS..278...18Z}
}

@ARTICLE{Brinchmann2004,
   author = {{Brinchmann}, J. and {Charlot}, S. and {White}, S.~D.~M. and 
	{Tremonti}, C. and {Kauffmann}, G. and {Heckman}, T. and {Brinkmann}, J.
	},
    title = "{The physical properties of star-forming galaxies in the low-redshift Universe}",
  journal = {MNRAS},
   eprint = {astro-ph/0311060},
     year = 2004,
    month = jul,
   volume = 351,
    pages = {1151-1179},
      doi = {10.1111/j.1365-2966.2004.07881.x},
   adsurl = {http://adsabs.harvard.edu/abs/2004MNRAS.351.1151B}
}

@ARTICLE{Gallazzi2005,
   author = {{Gallazzi}, A. and {Charlot}, S. and {Brinchmann}, J. and {White}, S.~D.~M. and 
	{Tremonti}, C.~A.},
    title = "{The ages and metallicities of galaxies in the local universe}",
  journal = {MNRAS},
   eprint = {astro-ph/0506539},
     year = 2005,
    month = sep,
   volume = 362,
    pages = {41-58},
      doi = {10.1111/j.1365-2966.2005.09321.x},
   adsurl = {http://adsabs.harvard.edu/abs/2005MNRAS.362...41G}
}

@ARTICLE{Salim2016,
       author = {{Salim}, Samir and {Lee}, Janice C. and {Janowiecki}, Steven and {da Cunha}, Elisabete and {Dickinson}, Mark and {Boquien}, M{\'e}d{\'e}ric and {Burgarella}, Denis and {Salzer}, John J. and {Charlot}, St{\'e}phane},
        title = "{GALEX-SDSS-WISE Legacy Catalog (GSWLC): Star Formation Rates, Stellar Masses, and Dust Attenuations of 700,000 Low-redshift Galaxies}",
      journal = {\apjs},
         year = 2016,
        month = nov,
       volume = {227},
       number = {1},
          eid = {2},
        pages = {2},
          doi = {10.3847/0067-0049/227/1/2},
archivePrefix = {arXiv},
       eprint = {1610.00712},
 primaryClass = {astro-ph.GA},
       adsurl = {https://ui.adsabs.harvard.edu/abs/2016ApJS..227....2S}
}

@ARTICLE{Salim2018,
       author = {{Salim}, Samir and {Boquien}, M{\'e}d{\'e}ric and {Lee}, Janice C.},
        title = "{Dust Attenuation Curves in the Local Universe: Demographics and New Laws for Star-forming Galaxies and High-redshift Analogs}",
      journal = {\apj},
         year = 2018,
        month = may,
       volume = {859},
       number = {1},
          eid = {11},
        pages = {11},
          doi = {10.3847/1538-4357/aabf3c},
archivePrefix = {arXiv},
       eprint = {1804.05850},
 primaryClass = {astro-ph.GA},
       adsurl = {https://ui.adsabs.harvard.edu/abs/2018ApJ...859...11S}
}

@ARTICLE{Kauffmann2003a,
   author = {{Kauffmann}, G. and {Heckman}, T.~M. and {White}, S.~D.~M. and 
	{Charlot}, S. and {Tremonti}, C. and {Brinchmann}, J. and {Bruzual}, G. and 
	{Peng}, E.~W. and {Seibert}, M. and {Bernardi}, M. and {Blanton}, M. and 
	{Brinkmann}, J. and {Castander}, F. and {Cs{\'a}bai}, I. and 
	{Fukugita}, M. and {Ivezic}, Z. and {Munn}, J.~A. and {Nichol}, R.~C. and 
	{Padmanabhan}, N. and {Thakar}, A.~R. and {Weinberg}, D.~H. and 
	{York}, D.},
    title = "{Stellar masses and star formation histories for 10$^{5}$ galaxies from the Sloan Digital Sky Survey}",
  journal = {MNRAS},
   eprint = {astro-ph/0204055},
     year = 2003,
    month = may,
   volume = 341,
    pages = {33-53},
      doi = {10.1046/j.1365-8711.2003.06291.x},
   adsurl = {http://adsabs.harvard.edu/abs/2003MNRAS.341...33K}
}

@ARTICLE{Peng2010,
       author = {{Peng}, Ying-jie and {Lilly}, Simon J. and {Kova{\v{c}}}, Katarina and {Bolzonella}, Micol and {Pozzetti}, Lucia and {Renzini}, Alvio and {Zamorani}, Gianni and {Ilbert}, Olivier and {Knobel}, Christian and {Iovino}, Angela and {Maier}, Christian and {Cucciati}, Olga and {Tasca}, Lidia and {Carollo}, C. Marcella and {Silverman}, John and {Kampczyk}, Pawel and {de Ravel}, Loic and {Sanders}, David and {Scoville}, Nicholas and {Contini}, Thierry and {Mainieri}, Vincenzo and {Scodeggio}, Marco and {Kneib}, Jean-Paul and {Le F{\`e}vre}, Olivier and {Bardelli}, Sandro and {Bongiorno}, Angela and {Caputi}, Karina and {Coppa}, Graziano and {de la Torre}, Sylvain and {Franzetti}, Paolo and {Garilli}, Bianca and {Lamareille}, Fabrice and {Le Borgne}, Jean-Francois and {Le Brun}, Vincent and {Mignoli}, Marco and {Perez Montero}, Enrique and {Pello}, Roser and {Ricciardelli}, Elena and {Tanaka}, Masayuki and {Tresse}, Laurence and {Vergani}, Daniela and {Welikala}, Niraj and {Zucca}, Elena and {Oesch}, Pascal and {Abbas}, Ummi and {Barnes}, Luke and {Bordoloi}, Rongmon and {Bottini}, Dario and {Cappi}, Alberto and {Cassata}, Paolo and {Cimatti}, Andrea and {Fumana}, Marco and {Hasinger}, Gunther and {Koekemoer}, Anton and {Leauthaud}, Alexei and {Maccagni}, Dario and {Marinoni}, Christian and {McCracken}, Henry and {Memeo}, Pierdomenico and {Meneux}, Baptiste and {Nair}, Preethi and {Porciani}, Cristiano and {Presotto}, Valentina and {Scaramella}, Roberto},
        title = "{Mass and Environment as Drivers of Galaxy Evolution in SDSS and zCOSMOS and the Origin of the Schechter Function}",
      journal = {\apj},
         year = 2010,
        month = sep,
       volume = {721},
       number = {1},
        pages = {193-221},
          doi = {10.1088/0004-637X/721/1/193},
archivePrefix = {arXiv},
       eprint = {1003.4747},
 primaryClass = {astro-ph.CO},
       adsurl = {https://ui.adsabs.harvard.edu/abs/2010ApJ...721..193P}
}

@ARTICLE{Fernandez2013,
       author = {{Fern{\'a}ndez Lorenzo}, M. and {Sulentic}, J. and {Verdes-Montenegro}, L. and {Argudo-Fern{\'a}ndez}, M.},
        title = "{The stellar mass-size relation for the most isolated galaxies in the local Universe}",
      journal = {\mnras},
         year = 2013,
        month = sep,
       volume = {434},
       number = {1},
        pages = {325-335},
          doi = {10.1093/mnras/stt1020},
archivePrefix = {arXiv},
       eprint = {1306.1687},
 primaryClass = {astro-ph.CO},
       adsurl = {https://ui.adsabs.harvard.edu/abs/2013MNRAS.434..325F}
}

@ARTICLE{Tumlinson2013,
       author = {{Tumlinson}, Jason and {Thom}, Christopher and {Werk}, Jessica K. and {Prochaska}, J. Xavier and {Tripp}, Todd M. and {Katz}, Neal and {Dav{\'e}}, Romeel and {Oppenheimer}, Benjamin D. and {Meiring}, Joseph D. and {Ford}, Amanda Brady and {O'Meara}, John M. and {Peeples}, Molly S. and {Sembach}, Kenneth R. and {Weinberg}, David H.},
        title = "{The COS-Halos Survey: Rationale, Design, and a Census of Circumgalactic Neutral Hydrogen}",
      journal = {\apj},
         year = 2013,
        month = nov,
       volume = {777},
       number = {1},
          eid = {59},
        pages = {59},
          doi = {10.1088/0004-637X/777/1/59},
archivePrefix = {arXiv},
       eprint = {1309.6317},
 primaryClass = {astro-ph.CO},
       adsurl = {https://ui.adsabs.harvard.edu/abs/2013ApJ...777...59T}
}

@ARTICLE{Zhang2022,
       author = {{Zhang}, Huanian and {Zaritsky}, Dennis},
        title = "{The Anisotropic Circumgalactic Medium of Massive Early-type Galaxies}",
      journal = {\apj},
         year = 2022,
        month = dec,
       volume = {941},
       number = {1},
          eid = {18},
        pages = {18},
          doi = {10.3847/1538-4357/ac9c64},
archivePrefix = {arXiv},
       eprint = {2210.10043},
 primaryClass = {astro-ph.GA},
       adsurl = {https://ui.adsabs.harvard.edu/abs/2022ApJ...941...18Z}
}

@ARTICLE{Fisher2022,
       author = {{Fisher}, D.~B. and {Bolatto}, A.~D. and {Glazebrook}, K. and {Obreschkow}, D. and {Abraham}, R.~G. and {Kacprzak}, G.~G. and {Nielsen}, N.~M.},
        title = "{Extreme Variation in Star Formation Efficiency across a Compact, Starburst Disk Galaxy}",
      journal = {\apj},
         year = 2022,
        month = apr,
       volume = {928},
       number = {2},
          eid = {169},
        pages = {169},
          doi = {10.3847/1538-4357/ac51c8},
archivePrefix = {arXiv},
       eprint = {2202.00024},
 primaryClass = {astro-ph.GA},
       adsurl = {https://ui.adsabs.harvard.edu/abs/2022ApJ...928..169F}
}

@ARTICLE{Zhang2024a,
       author = {{Zhang}, Huanian and {Li}, Miao and {Zaritsky}, Dennis},
        title = "{The Anisotropic Circumgalactic Medium of Sub-L* Galaxies}",
      journal = {\apj},
         year = 2024,
        month = oct,
       volume = {974},
       number = {2},
          eid = {148},
        pages = {148},
          doi = {10.3847/1538-4357/ad738c},
archivePrefix = {arXiv},
       eprint = {2408.07102},
 primaryClass = {astro-ph.GA},
       adsurl = {https://ui.adsabs.harvard.edu/abs/2024ApJ...974..148Z}
}

@ARTICLE{Burchett2021,
       author = {{Burchett}, Joseph N. and {Rubin}, Kate H.~R. and {Prochaska}, J. Xavier and {Coil}, Alison L. and {Vaught}, Ryan Rickards and {Hennawi}, Joseph F.},
        title = "{Circumgalactic Mg II Emission from an Isotropic Starburst Galaxy Outflow Mapped by KCWI}",
      journal = {\apj},
         year = 2021,
        month = mar,
       volume = {909},
       number = {2},
          eid = {151},
        pages = {151},
          doi = {10.3847/1538-4357/abd4e0},
archivePrefix = {arXiv},
       eprint = {2005.03017},
 primaryClass = {astro-ph.GA},
       adsurl = {https://ui.adsabs.harvard.edu/abs/2021ApJ...909..151B}
}

@ARTICLE{Zabl2021,
       author = {{Zabl}, Johannes and {Bouch{\'e}}, Nicolas F. and {Wisotzki}, Lutz and {Schaye}, Joop and {Leclercq}, Floriane and {Garel}, Thibault and {Wendt}, Martin and {Schroetter}, Ilane and {Muzahid}, Sowgat and {Cantalupo}, Sebastiano and {Contini}, Thierry and {Bacon}, Roland and {Brinchmann}, Jarle and {Richard}, Johan},
        title = "{MusE GAs FLOw and Wind (MEGAFLOW) VIII. Discovery of a MgII emission halo probed by a quasar sightline}",
      journal = {\mnras},
         year = 2021,
        month = nov,
       volume = {507},
       number = {3},
        pages = {4294-4315},
          doi = {10.1093/mnras/stab2165},
archivePrefix = {arXiv},
       eprint = {2105.14090},
 primaryClass = {astro-ph.GA},
       adsurl = {https://ui.adsabs.harvard.edu/abs/2021MNRAS.507.4294Z}
}

@ARTICLE{Comerford2020,
       author = {{Comerford}, Julia M. and {Negus}, James and {M{\"u}ller-S{\'a}nchez}, Francisco and {Eracleous}, Michael and {Wylezalek}, Dominika and {Storchi-Bergmann}, Thaisa and {Greene}, Jenny E. and {Barrows}, R. Scott and {Nevin}, Rebecca and {Roy}, Namrata and {Stemo}, Aaron},
        title = "{A Catalog of 406 AGNs in MaNGA: A Connection between Radio-mode AGNs and Star Formation Quenching}",
      journal = {\apj},
         year = 2020,
        month = oct,
       volume = {901},
       number = {2},
          eid = {159},
        pages = {159},
          doi = {10.3847/1538-4357/abb2ae},
archivePrefix = {arXiv},
       eprint = {2008.11210},
 primaryClass = {astro-ph.GA},
       adsurl = {https://ui.adsabs.harvard.edu/abs/2020ApJ...901..159C}
}

@ARTICLE{Sanchez2022,
       author = {{S{\'a}nchez}, S.~F. and {Barrera-Ballesteros}, J.~K. and {Lacerda}, E. and {Mej{\'\i}a-Narvaez}, A. and {Camps-Fari{\~n}a}, A. and {Bruzual}, Gustavo and {Espinosa-Ponce}, C. and {Rodr{\'\i}guez-Puebla}, A. and {Calette}, A.~R. and {Ibarra-Medel}, H. and {Avila-Reese}, V. and {Hernandez-Toledo}, H. and {Bershady}, M.~A. and {Cano-Diaz}, M. and {Munguia-Cordova}, A.~M.},
        title = "{SDSS-IV MaNGA: pyPipe3D Analysis Release for 10,000 Galaxies}",
      journal = {\apjs},
         year = 2022,
        month = oct,
       volume = {262},
       number = {2},
          eid = {36},
        pages = {36},
          doi = {10.3847/1538-4365/ac7b8f},
archivePrefix = {arXiv},
       eprint = {2206.07062},
 primaryClass = {astro-ph.GA},
       adsurl = {https://ui.adsabs.harvard.edu/abs/2022ApJS..262...36S}
}

@ARTICLE{emcee,
       author = {{Foreman-Mackey}, Daniel and {Hogg}, David W. and {Lang}, Dustin and {Goodman}, Jonathan},
        title = "{emcee: The MCMC Hammer}",
      journal = {\pasp},
         year = 2013,
        month = mar,
       volume = {125},
       number = {925},
        pages = {306},
          doi = {10.1086/670067},
archivePrefix = {arXiv},
       eprint = {1202.3665},
 primaryClass = {astro-ph.IM},
       adsurl = {https://ui.adsabs.harvard.edu/abs/2013PASP..125..306F}
}

@ARTICLE{astropy,
       author = {{Astropy Collaboration} and {Price-Whelan}, Adrian M. and {Lim}, Pey Lian and {Earl}, Nicholas and {Starkman}, Nathaniel and {Bradley}, Larry and {Shupe}, David L. and {Patil}, Aarya A. and {Corrales}, Lia and {Brasseur}, C.~E. and {N{\"o}the}, Maximilian and {Donath}, Axel and {Tollerud}, Erik and {Morris}, Brett M. and {Ginsburg}, Adam and {Vaher}, Eero and {Weaver}, Benjamin A. and {Tocknell}, James and {Jamieson}, William and {van Kerkwijk}, Marten H. and {Robitaille}, Thomas P. and {Merry}, Bruce and {Bachetti}, Matteo and {G{\"u}nther}, H. Moritz and {Aldcroft}, Thomas L. and {Alvarado-Montes}, Jaime A. and {Archibald}, Anne M. and {B{\'o}di}, Attila and {Bapat}, Shreyas and {Barentsen}, Geert and {Baz{\'a}n}, Juanjo and {Biswas}, Manish and {Boquien}, M{\'e}d{\'e}ric and {Burke}, D.~J. and {Cara}, Daria and {Cara}, Mihai and {Conroy}, Kyle E. and {Conseil}, Simon and {Craig}, Matthew W. and {Cross}, Robert M. and {Cruz}, Kelle L. and {D'Eugenio}, Francesco and {Dencheva}, Nadia and {Devillepoix}, Hadrien A.~R. and {Dietrich}, J{\"o}rg P. and {Eigenbrot}, Arthur Davis and {Erben}, Thomas and {Ferreira}, Leonardo and {Foreman-Mackey}, Daniel and {Fox}, Ryan and {Freij}, Nabil and {Garg}, Suyog and {Geda}, Robel and {Glattly}, Lauren and {Gondhalekar}, Yash and {Gordon}, Karl D. and {Grant}, David and {Greenfield}, Perry and {Groener}, Austen M. and {Guest}, Steve and {Gurovich}, Sebastian and {Handberg}, Rasmus and {Hart}, Akeem and {Hatfield-Dodds}, Zac and {Homeier}, Derek and {Hosseinzadeh}, Griffin and {Jenness}, Tim and {Jones}, Craig K. and {Joseph}, Prajwel and {Kalmbach}, J. Bryce and {Karamehmetoglu}, Emir and {Ka{\l}uszy{\'n}ski}, Miko{\l}aj and {Kelley}, Michael S.~P. and {Kern}, Nicholas and {Kerzendorf}, Wolfgang E. and {Koch}, Eric W. and {Kulumani}, Shankar and {Lee}, Antony and {Ly}, Chun and {Ma}, Zhiyuan and {MacBride}, Conor and {Maljaars}, Jakob M. and {Muna}, Demitri and {Murphy}, N.~A. and {Norman}, Henrik and {O'Steen}, Richard and {Oman}, Kyle A. and {Pacifici}, Camilla and {Pascual}, Sergio and {Pascual-Granado}, J. and {Patil}, Rohit R. and {Perren}, Gabriel I. and {Pickering}, Timothy E. and {Rastogi}, Tanuj and {Roulston}, Benjamin R. and {Ryan}, Daniel F. and {Rykoff}, Eli S. and {Sabater}, Jose and {Sakurikar}, Parikshit and {Salgado}, Jes{\'u}s and {Sanghi}, Aniket and {Saunders}, Nicholas and {Savchenko}, Volodymyr and {Schwardt}, Ludwig and {Seifert-Eckert}, Michael and {Shih}, Albert Y. and {Jain}, Anany Shrey and {Shukla}, Gyanendra and {Sick}, Jonathan and {Simpson}, Chris and {Singanamalla}, Sudheesh and {Singer}, Leo P. and {Singhal}, Jaladh and {Sinha}, Manodeep and {Sip{\H{o}}cz}, Brigitta M. and {Spitler}, Lee R. and {Stansby}, David and {Streicher}, Ole and {{\v{S}}umak}, Jani and {Swinbank}, John D. and {Taranu}, Dan S. and {Tewary}, Nikita and {Tremblay}, Grant R. and {de Val-Borro}, Miguel and {Van Kooten}, Samuel J. and {Vasovi{\'c}}, Zlatan and {Verma}, Shresth and {de Miranda Cardoso}, Jos{\'e} Vin{\'\i}cius and {Williams}, Peter K.~G. and {Wilson}, Tom J. and {Winkel}, Benjamin and {Wood-Vasey}, W.~M. and {Xue}, Rui and {Yoachim}, Peter and {Zhang}, Chen and {Zonca}, Andrea and {Astropy Project Contributors}},
        title = "{The Astropy Project: Sustaining and Growing a Community-oriented Open-source Project and the Latest Major Release (v5.0) of the Core Package}",
      journal = {\apj},
         year = 2022,
        month = aug,
       volume = {935},
       number = {2},
          eid = {167},
        pages = {167},
          doi = {10.3847/1538-4357/ac7c74},
archivePrefix = {arXiv},
       eprint = {2206.14220},
 primaryClass = {astro-ph.IM},
       adsurl = {https://ui.adsabs.harvard.edu/abs/2022ApJ...935..167A}
}

@Article{numpy2020,
 title         = {Array programming with {NumPy}},
 author        = {Charles R. Harris and K. Jarrod Millman and St{\'{e}}fan J.
                 van der Walt and Ralf Gommers and Pauli Virtanen and David
                 Cournapeau and Eric Wieser and Julian Taylor and Sebastian
                 Berg and Nathaniel J. Smith and Robert Kern and Matti Picus
                 and Stephan Hoyer and Marten H. van Kerkwijk and Matthew
                 Brett and Allan Haldane and Jaime Fern{\'{a}}ndez del
                 R{\'{i}}o and Mark Wiebe and Pearu Peterson and Pierre
                 G{\'{e}}rard-Marchant and Kevin Sheppard and Tyler Reddy and
                 Warren Weckesser and Hameer Abbasi and Christoph Gohlke and
                 Travis E. Oliphant},
 year          = {2020},
 month         = sep,
 journal       = {Nature},
 volume        = {585},
 number        = {7825},
 pages         = {357--362},
 doi           = {10.1038/s41586-020-2649-2},
 publisher     = {Springer Science and Business Media {LLC}},
 url           = {https://doi.org/10.1038/s41586-020-2649-2}
}

@ARTICLE{SciPy2020,
  author  = {Virtanen, Pauli and Gommers, Ralf and Oliphant, Travis E. and
            Haberland, Matt and Reddy, Tyler and Cournapeau, David and
            Burovski, Evgeni and Peterson, Pearu and Weckesser, Warren and
            Bright, Jonathan and {van der Walt}, St{\'e}fan J. and
            Brett, Matthew and Wilson, Joshua and Millman, K. Jarrod and
            Mayorov, Nikolay and Nelson, Andrew R. J. and Jones, Eric and
            Kern, Robert and Larson, Eric and Carey, C J and
            Polat, {\.I}lhan and Feng, Yu and Moore, Eric W. and
            {VanderPlas}, Jake and Laxalde, Denis and Perktold, Josef and
            Cimrman, Robert and Henriksen, Ian and Quintero, E. A. and
            Harris, Charles R. and Archibald, Anne M. and
            Ribeiro, Ant{\^o}nio H. and Pedregosa, Fabian and
            {van Mulbregt}, Paul and {SciPy 1.0 Contributors}},
  title   = {{{SciPy} 1.0: Fundamental Algorithms for Scientific
            Computing in Python}},
  journal = {Nature Methods},
  year    = {2020},
  volume  = {17},
  pages   = {261--272},
  adsurl  = {https://rdcu.be/b08Wh},
  doi     = {10.1038/s41592-019-0686-2},
}

@ARTICLE{Leitet2013,
       author = {{Leitet}, E. and {Bergvall}, N. and {Hayes}, M. and {Linn{\'e}}, S. and {Zackrisson}, E.},
        title = "{Escape of Lyman continuum radiation from local galaxies. Detection of leakage from the young starburst Tol 1247-232}",
      journal = {\aap},
         year = 2013,
        month = may,
       volume = {553},
          eid = {A106},
        pages = {A106},
          doi = {10.1051/0004-6361/201118370},
archivePrefix = {arXiv},
       eprint = {1302.6971},
 primaryClass = {astro-ph.CO},
       adsurl = {https://ui.adsabs.harvard.edu/abs/2013A&A...553A.106L}
}

@ARTICLE{Hunter2007,
  author={Hunter, John D.},
  journal={Computing in Science \& Engineering}, 
  title={Matplotlib: A 2D Graphics Environment}, 
  year={2007},
  volume={9},
  number={3},
  pages={90-95},
  doi={10.1109/MCSE.2007.55}}

@ARTICLE{pandas,
       author = {{McKinney}, W.},
      journal = {Proceedings of the 9th Python in Science Conference},
         year = 2010,
        pages = {51}
}

@ARTICLE{Haardt2012,
       author = {{Haardt}, Francesco and {Madau}, Piero},
        title = "{Radiative Transfer in a Clumpy Universe. IV. New Synthesis Models of the Cosmic UV/X-Ray Background}",
      journal = {\apj},
         year = 2012,
        month = feb,
       volume = {746},
       number = {2},
          eid = {125},
        pages = {125},
          doi = {10.1088/0004-637X/746/2/125},
       adsurl = {https://ui.adsabs.harvard.edu/abs/2012ApJ...746..125H}
}

@ARTICLE{Faucher2020,
       author = {{Faucher-Gigu{\`e}re}, Claude-Andr{\'e}},
        title = "{A cosmic UV/X-ray background model update}",
      journal = {\mnras},
         year = 2020,
        month = apr,
       volume = {493},
       number = {2},
        pages = {1614-1632},
          doi = {10.1093/mnras/staa302},
archivePrefix = {arXiv},
       eprint = {1903.08657},
 primaryClass = {astro-ph.CO},
       adsurl = {https://ui.adsabs.harvard.edu/abs/2020MNRAS.493.1614F}
}

@ARTICLE{Allen2008,
       author = {{Allen}, Mark G. and {Groves}, Brent A. and {Dopita}, Michael A. and {Sutherland}, Ralph S. and {Kewley}, Lisa J.},
        title = "{The MAPPINGS III Library of Fast Radiative Shock Models}",
      journal = {\apjs},
         year = 2008,
        month = sep,
       volume = {178},
       number = {1},
        pages = {20-55},
          doi = {10.1086/589652},
archivePrefix = {arXiv},
       eprint = {0805.0204},
 primaryClass = {astro-ph},
       adsurl = {https://ui.adsabs.harvard.edu/abs/2008ApJS..178...20A}
}

@ARTICLE{Menard2010,
       author = {{M{\'e}nard}, Brice and {Scranton}, Ryan and {Fukugita}, Masataka and {Richards}, Gordon},
        title = "{Measuring the galaxy-mass and galaxy-dust correlations through magnification and reddening}",
      journal = {\mnras},
         year = 2010,
        month = jun,
       volume = {405},
       number = {2},
        pages = {1025-1039},
          doi = {10.1111/j.1365-2966.2010.16486.x},
archivePrefix = {arXiv},
       eprint = {0902.4240},
 primaryClass = {astro-ph.CO},
       adsurl = {https://ui.adsabs.harvard.edu/abs/2010MNRAS.405.1025M}
}

@ARTICLE{WISE_Wright_2010,
       author = {{Wright}, Edward L. and {Eisenhardt}, Peter R.~M. and {Mainzer}, Amy K. and {Ressler}, Michael E. and {Cutri}, Roc M. and {Jarrett}, Thomas and {Kirkpatrick}, J. Davy and {Padgett}, Deborah and {McMillan}, Robert S. and {Skrutskie}, Michael and {Stanford}, S.~A. and {Cohen}, Martin and {Walker}, Russell G. and {Mather}, John C. and {Leisawitz}, David and {Gautier}, III, Thomas N. and {McLean}, Ian and {Benford}, Dominic and {Lonsdale}, Carol J. and {Blain}, Andrew and {Mendez}, Bryan and {Irace}, William R. and {Duval}, Valerie and {Liu}, Fengchuan and {Royer}, Don and {Heinrichsen}, Ingolf and {Howard}, Joan and {Shannon}, Mark and {Kendall}, Martha and {Walsh}, Amy L. and {Larsen}, Mark and {Cardon}, Joel G. and {Schick}, Scott and {Schwalm}, Mark and {Abid}, Mohamed and {Fabinsky}, Beth and {Naes}, Larry and {Tsai}, Chao-Wei},
        title = "{The Wide-field Infrared Survey Explorer (WISE): Mission Description and Initial On-orbit Performance}",
      journal = {\aj},
         year = 2010,
        month = dec,
       volume = {140},
       number = {6},
        pages = {1868-1881},
          doi = {10.1088/0004-6256/140/6/1868},
archivePrefix = {arXiv},
       eprint = {1008.0031},
 primaryClass = {astro-ph.IM},
       adsurl = {https://ui.adsabs.harvard.edu/abs/2010AJ....140.1868W}
}

@ARTICLE{WISE_Mainzer_2011,
       author = {{Mainzer}, A. and {Bauer}, J. and {Grav}, T. and {Masiero}, J. and {Cutri}, R.~M. and {Dailey}, J. and {Eisenhardt}, P. and {McMillan}, R.~S. and {Wright}, E. and {Walker}, R. and {Jedicke}, R. and {Spahr}, T. and {Tholen}, D. and {Alles}, R. and {Beck}, R. and {Brandenburg}, H. and {Conrow}, T. and {Evans}, T. and {Fowler}, J. and {Jarrett}, T. and {Marsh}, K. and {Masci}, F. and {McCallon}, H. and {Wheelock}, S. and {Wittman}, M. and {Wyatt}, P. and {DeBaun}, E. and {Elliott}, G. and {Elsbury}, D. and {Gautier}, IV, T. and {Gomillion}, S. and {Leisawitz}, D. and {Maleszewski}, C. and {Micheli}, M. and {Wilkins}, A.},
        title = "{Preliminary Results from NEOWISE: An Enhancement to the Wide-field Infrared Survey Explorer for Solar System Science}",
      journal = {\apj},
         year = 2011,
        month = apr,
       volume = {731},
       number = {1},
          eid = {53},
        pages = {53},
          doi = {10.1088/0004-637X/731/1/53},
archivePrefix = {arXiv},
       eprint = {1102.1996},
 primaryClass = {astro-ph.EP},
       adsurl = {https://ui.adsabs.harvard.edu/abs/2011ApJ...731...53M}
}

@ARTICLE{Peek2015,
       author = {{Peek}, J.~E.~G. and {M{\'e}nard}, Brice and {Corrales}, Lia},
        title = "{Dust in the Circumgalactic Medium of Low-redshift Galaxies}",
      journal = {\apj},
         year = 2015,
        month = nov,
       volume = {813},
       number = {1},
          eid = {7},
        pages = {7},
          doi = {10.1088/0004-637X/813/1/7},
archivePrefix = {arXiv},
       eprint = {1411.3333},
 primaryClass = {astro-ph.GA},
       adsurl = {https://ui.adsabs.harvard.edu/abs/2015ApJ...813....7P}
}

@BOOK{Draine2011,
       author = {{Draine}, Bruce T.},
        title = "{Physics of the Interstellar and Intergalactic Medium}",
         year = 2011,
       adsurl = {https://ui.adsabs.harvard.edu/abs/2011piim.book.....D}
}

@ARTICLE{Pessa2024,
       author = {{Pessa}, Ismael and {Wisotzki}, Lutz and {Urrutia}, Tanya and {Pharo}, John and {Augustin}, Ramona and {Bouch{\'e}}, Nicolas F. and {Feltre}, Anna and {Guo}, Yucheng and {Kozlova}, Daria and {Krajnovic}, Davor and {Kusakabe}, Haruka and {Leclercq}, Floriane and {Salas}, H{\'e}ctor and {Schaye}, Joop and {Verhamme}, Anne},
        title = "{A galactic outflow traced by its extended Mg II emission out to a {\ensuremath{\sim}}30 kpc radius in the Hubble Ultra Deep Field with MUSE}",
      journal = {\aap},
         year = 2024,
        month = nov,
       volume = {691},
          eid = {A5},
        pages = {A5},
          doi = {10.1051/0004-6361/202450547},
archivePrefix = {arXiv},
       eprint = {2408.16067},
 primaryClass = {astro-ph.GA},
       adsurl = {https://ui.adsabs.harvard.edu/abs/2024A&A...691A...5P}
}

@ARTICLE{Nelson2018,
       author = {{Nelson}, Dylan and {Kauffmann}, Guinevere and {Pillepich}, Annalisa and {Genel}, Shy and {Springel}, Volker and {Pakmor}, R{\"u}diger and {Hernquist}, Lars and {Weinberger}, Rainer and {Torrey}, Paul and {Vogelsberger}, Mark and {Marinacci}, Federico},
        title = "{The abundance, distribution, and physical nature of highly ionized oxygen O VI, O VII, and O VIII in IllustrisTNG}",
      journal = {\mnras},
         year = 2018,
        month = jun,
       volume = {477},
       number = {1},
        pages = {450-479},
          doi = {10.1093/mnras/sty656},
archivePrefix = {arXiv},
       eprint = {1712.00016},
 primaryClass = {astro-ph.GA},
       adsurl = {https://ui.adsabs.harvard.edu/abs/2018MNRAS.477..450N}
}

@ARTICLE{Margon1988,
       author = {{Margon}, Bruce and {Anderson}, Scott F. and {Mateo}, Mario and {Fich}, Michel and {Massey}, Philip},
        title = "{An Exceptionally Bright, Compact Starburst Nucleus}",
      journal = {\apj},
         year = 1988,
        month = nov,
       volume = {334},
        pages = {597},
          doi = {10.1086/166863},
       adsurl = {https://ui.adsabs.harvard.edu/abs/1988ApJ...334..597M}
}

@ARTICLE{Lopez2006,
       author = {{L{\'o}pez-S{\'a}nchez}, {\'A}. R. and {Esteban}, C. and {Garc{\'\i}a-Rojas}, J.},
        title = "{Star formation and stellar populations in the Wolf-Rayet(?) luminous compact blue galaxy IRAS 08339+6517}",
      journal = {\aap},
         year = 2006,
        month = apr,
       volume = {449},
       number = {3},
        pages = {997-1017},
          doi = {10.1051/0004-6361:20053119},
archivePrefix = {arXiv},
       eprint = {astro-ph/0512035},
 primaryClass = {astro-ph},
       adsurl = {https://ui.adsabs.harvard.edu/abs/2006A&A...449..997L}
}

@ARTICLE{CGM2023,
       author = {{Faucher-Gigu{\`e}re}, Claude-Andr{\'e} and {Oh}, S. Peng},
        title = "{Key Physical Processes in the Circumgalactic Medium}",
      journal = {\araa},
         year = 2023,
        month = aug,
       volume = {61},
        pages = {131-195},
          doi = {10.1146/annurev-astro-052920-125203},
archivePrefix = {arXiv},
       eprint = {2301.10253},
 primaryClass = {astro-ph.GA},
       adsurl = {https://ui.adsabs.harvard.edu/abs/2023ARA&A..61..131F}
}

@ARTICLE{Peroux2020,
       author = {{P{\'e}roux}, C{\'e}line and {Howk}, J. Christopher},
        title = "{The Cosmic Baryon and Metal Cycles}",
      journal = {\araa},
         year = 2020,
        month = aug,
       volume = {58},
        pages = {363-406},
          doi = {10.1146/annurev-astro-021820-120014},
archivePrefix = {arXiv},
       eprint = {2011.01935},
 primaryClass = {astro-ph.GA},
       adsurl = {https://ui.adsabs.harvard.edu/abs/2020ARA&A..58..363P}
}

@ARTICLE{Zhang2024b,
       author = {{Zhang}, Huanian and {Zaritsky}, Dennis},
        title = "{A MUSE source-blind survey for emission from the circumgalactic medium}",
      journal = {Science Advances},
         year = 2024,
        month = nov,
       volume = {10},
       number = {47},
          eid = {eadp8629},
        pages = {eadp8629},
          doi = {10.1126/sciadv.adp8629},
archivePrefix = {arXiv},
       eprint = {2410.05392},
 primaryClass = {astro-ph.GA},
       adsurl = {https://ui.adsabs.harvard.edu/abs/2024SciA...10P8629Z}
}

@ARTICLE{manga,
       author = {{Bundy}, Kevin and {Bershady}, Matthew A. and {Law}, David R. and {Yan}, Renbin and {Drory}, Niv and {MacDonald}, Nicholas and {Wake}, David A. and {Cherinka}, Brian and {S{\'a}nchez-Gallego}, Jos{\'e} R. and {Weijmans}, Anne-Marie and {Thomas}, Daniel and {Tremonti}, Christy and {Masters}, Karen and {Coccato}, Lodovico and {Diamond-Stanic}, Aleksandar M. and {Arag{\'o}n-Salamanca}, Alfonso and {Avila-Reese}, Vladimir and {Badenes}, Carles and {Falc{\'o}n-Barroso}, J{\'e}sus and {Belfiore}, Francesco and {Bizyaev}, Dmitry and {Blanc}, Guillermo A. and {Bland-Hawthorn}, Joss and {Blanton}, Michael R. and {Brownstein}, Joel R. and {Byler}, Nell and {Cappellari}, Michele and {Conroy}, Charlie and {Dutton}, Aaron A. and {Emsellem}, Eric and {Etherington}, James and {Frinchaboy}, Peter M. and {Fu}, Hai and {Gunn}, James E. and {Harding}, Paul and {Johnston}, Evelyn J. and {Kauffmann}, Guinevere and {Kinemuchi}, Karen and {Klaene}, Mark A. and {Knapen}, Johan H. and {Leauthaud}, Alexie and {Li}, Cheng and {Lin}, Lihwai and {Maiolino}, Roberto and {Malanushenko}, Viktor and {Malanushenko}, Elena and {Mao}, Shude and {Maraston}, Claudia and {McDermid}, Richard M. and {Merrifield}, Michael R. and {Nichol}, Robert C. and {Oravetz}, Daniel and {Pan}, Kaike and {Parejko}, John K. and {Sanchez}, Sebastian F. and {Schlegel}, David and {Simmons}, Audrey and {Steele}, Oliver and {Steinmetz}, Matthias and {Thanjavur}, Karun and {Thompson}, Benjamin A. and {Tinker}, Jeremy L. and {van den Bosch}, Remco C.~E. and {Westfall}, Kyle B. and {Wilkinson}, David and {Wright}, Shelley and {Xiao}, Ting and {Zhang}, Kai},
        title = "{Overview of the SDSS-IV MaNGA Survey: Mapping nearby Galaxies at Apache Point Observatory}",
      journal = {\apj},
         year = 2015,
        month = jan,
       volume = {798},
       number = {1},
          eid = {7},
        pages = {7},
          doi = {10.1088/0004-637X/798/1/7},
archivePrefix = {arXiv},
       eprint = {1412.1482},
 primaryClass = {astro-ph.GA},
       adsurl = {https://ui.adsabs.harvard.edu/abs/2015ApJ...798....7B}
}

@ARTICLE{Drory2015,
       author = {{Drory}, N. and {MacDonald}, N. and {Bershady}, M.~A. and {Bundy}, K. and {Gunn}, J. and {Law}, D.~R. and {Smith}, M. and {Stoll}, R. and {Tremonti}, C.~A. and {Wake}, D.~A. and {Yan}, R. and {Weijmans}, A.~M. and {Byler}, N. and {Cherinka}, B. and {Cope}, F. and {Eigenbrot}, A. and {Harding}, P. and {Holder}, D. and {Huehnerhoff}, J. and {Jaehnig}, K. and {Jansen}, T.~C. and {Klaene}, M. and {Paat}, A.~M. and {Percival}, J. and {Sayres}, C.},
        title = "{The MaNGA Integral Field Unit Fiber Feed System for the Sloan 2.5 m Telescope}",
      journal = {\aj},
         year = 2015,
        month = feb,
       volume = {149},
       number = {2},
          eid = {77},
        pages = {77},
          doi = {10.1088/0004-6256/149/2/77},
archivePrefix = {arXiv},
       eprint = {1412.1535},
 primaryClass = {astro-ph.IM},
       adsurl = {https://ui.adsabs.harvard.edu/abs/2015AJ....149...77D}
}

@ARTICLE{Westfall2019,
       author = {{Westfall}, Kyle B. and {Cappellari}, Michele and {Bershady}, Matthew A. and {Bundy}, Kevin and {Belfiore}, Francesco and {Ji}, Xihan and {Law}, David R. and {Schaefer}, Adam and {Shetty}, Shravan and {Tremonti}, Christy A. and {Yan}, Renbin and {Andrews}, Brett H. and {Brownstein}, Joel R. and {Cherinka}, Brian and {Coccato}, Lodovico and {Drory}, Niv and {Maraston}, Claudia and {Parikh}, Taniya and {S{\'a}nchez-Gallego}, Jos{\'e} R. and {Thomas}, Daniel and {Weijmans}, Anne-Marie and {Barrera-Ballesteros}, Jorge and {Du}, Cheng and {Goddard}, Daniel and {Li}, Niu and {Masters}, Karen and {Ibarra Medel}, H{\'e}ctor Javier and {S{\'a}nchez}, Sebasti{\'a}n F. and {Yang}, Meng and {Zheng}, Zheng and {Zhou}, Shuang},
        title = "{The Data Analysis Pipeline for the SDSS-IV MaNGA IFU Galaxy Survey: Overview}",
      journal = {\aj},
         year = 2019,
        month = dec,
       volume = {158},
       number = {6},
          eid = {231},
        pages = {231},
          doi = {10.3847/1538-3881/ab44a2},
archivePrefix = {arXiv},
       eprint = {1901.00856},
 primaryClass = {astro-ph.GA},
       adsurl = {https://ui.adsabs.harvard.edu/abs/2019AJ....158..231W}
}

@ARTICLE{Belfiore2019,
       author = {{Belfiore}, Francesco and {Westfall}, Kyle B. and {Schaefer}, Adam and {Cappellari}, Michele and {Ji}, Xihan and {Bershady}, Matthew A. and {Tremonti}, Christy and {Law}, David R. and {Yan}, Renbin and {Bundy}, Kevin and {Shetty}, Shravan and {Drory}, Niv and {Thomas}, Daniel and {Emsellem}, Eric and {S{\'a}nchez}, Sebasti{\'a}n F.},
        title = "{The Data Analysis Pipeline for the SDSS-IV MaNGA IFU Galaxy Survey: Emission-line Modeling}",
      journal = {\aj},
         year = 2019,
        month = oct,
       volume = {158},
       number = {4},
          eid = {160},
        pages = {160},
          doi = {10.3847/1538-3881/ab3e4e},
archivePrefix = {arXiv},
       eprint = {1901.00866},
 primaryClass = {astro-ph.GA},
       adsurl = {https://ui.adsabs.harvard.edu/abs/2019AJ....158..160B}
}

@ARTICLE{Lokhorst2022,
       author = {{Lokhorst}, Deborah and {Abraham}, Roberto and {Pasha}, Imad and {van Dokkum}, Pieter and {Chen}, Seery and {Miller}, Tim and {Danieli}, Shany and {Greco}, Johnny and {Zhang}, Jielai and {Merritt}, Allison and {Conroy}, Charlie},
        title = "{A Giant Shell of Ionized Gas Discovered near M82 with the Dragonfly Spectral Line Mapper Pathfinder}",
      journal = {\apj},
         year = 2022,
        month = mar,
       volume = {927},
       number = {2},
          eid = {136},
        pages = {136},
          doi = {10.3847/1538-4357/ac50b6},
archivePrefix = {arXiv},
       eprint = {2203.04933},
 primaryClass = {astro-ph.GA},
       adsurl = {https://ui.adsabs.harvard.edu/abs/2022ApJ...927..136L}
}

@ARTICLE{Rupke2019,
       author = {{Rupke}, David S.~N. and {Coil}, Alison and {Geach}, James E. and {Tremonti}, Christy and {Diamond-Stanic}, Aleksandar M. and {George}, Erin R. and {Hickox}, Ryan C. and {Kepley}, Amanda A. and {Leung}, Gene and {Moustakas}, John and {Rudnick}, Gregory and {Sell}, Paul H.},
        title = "{A 100-kiloparsec wind feeding the circumgalactic medium of a massive compact galaxy}",
      journal = {\nat},
         year = 2019,
        month = oct,
       volume = {574},
       number = {7780},
        pages = {643-646},
          doi = {10.1038/s41586-019-1686-1},
archivePrefix = {arXiv},
       eprint = {1910.13507},
 primaryClass = {astro-ph.GA},
       adsurl = {https://ui.adsabs.harvard.edu/abs/2019Natur.574..643R}
}

@ARTICLE{Kroupa2001,
       author = {{Kroupa}, Pavel},
        title = "{On the variation of the initial mass function}",
      journal = {\mnras},
         year = 2001,
        month = apr,
       volume = {322},
       number = {2},
        pages = {231-246},
          doi = {10.1046/j.1365-8711.2001.04022.x},
archivePrefix = {arXiv},
       eprint = {astro-ph/0009005},
 primaryClass = {astro-ph},
       adsurl = {https://ui.adsabs.harvard.edu/abs/2001MNRAS.322..231K}
}

@ARTICLE{Contini2016,
       author = {{Contini}, T. and {Epinat}, B. and {Bouch{\'e}}, N. and {Brinchmann}, J. and {Boogaard}, L.~A. and {Ventou}, E. and {Bacon}, R. and {Richard}, J. and {Weilbacher}, P.~M. and {Wisotzki}, L. and {Krajnovi{\'c}}, D. and {Vielfaure}, J. -B. and {Emsellem}, E. and {Finley}, H. and {Inami}, H. and {Schaye}, J. and {Swinbank}, M. and {Gu{\'e}rou}, A. and {Martinsson}, T. and {Michel-Dansac}, L. and {Schroetter}, I. and {Shirazi}, M. and {Soucail}, G.},
        title = "{Deep MUSE observations in the HDFS. Morpho-kinematics of distant star-forming galaxies down to {}10$^{8}$M$_{{\ensuremath{\odot}}}$}",
      journal = {\ana},
         year = 2016,
        month = jun,
       volume = {591},
          eid = {A49},
        pages = {A49},
          doi = {10.1051/0004-6361/201527866},
archivePrefix = {arXiv},
       eprint = {1512.00246},
 primaryClass = {astro-ph.GA},
       adsurl = {https://ui.adsabs.harvard.edu/abs/2016A&A...591A..49C}
}

@ARTICLE{Guo2023,
       author = {{Guo}, Yucheng and {Bacon}, Roland and {Bouch{\'e}}, Nicolas F. and {Wisotzki}, Lutz and {Schaye}, Joop and {Blaizot}, J{\'e}r{\'e}my and {Verhamme}, Anne and {Cantalupo}, Sebastiano and {Boogaard}, Leindert A. and {Brinchmann}, Jarle and {Cherrey}, Maxime and {Kusakabe}, Haruka and {Langan}, Ivanna and {Leclercq}, Floriane and {Matthee}, Jorryt and {Michel-Dansac}, L{\'e}o and {Schroetter}, Ilane and {Wendt}, Martin},
        title = "{Bipolar outflows out to 10 kpc for massive galaxies at redshift z {\ensuremath{\approx}} 1}",
      journal = {\nat},
         year = 2023,
        month = dec,
       volume = {624},
       number = {7990},
        pages = {53-56},
          doi = {10.1038/s41586-023-06718-w},
archivePrefix = {arXiv},
       eprint = {2312.05167},
 primaryClass = {astro-ph.GA},
       adsurl = {https://ui.adsabs.harvard.edu/abs/2023Natur.624...53G}
}

@ARTICLE{Jaskot2013,
       author = {{Jaskot}, A.~E. and {Oey}, M.~S.},
        title = "{The Origin and Optical Depth of Ionizing Radiation in the ``Green Pea'' Galaxies}",
      journal = {\apj},
         year = 2013,
        month = apr,
       volume = {766},
       number = {2},
          eid = {91},
        pages = {91},
          doi = {10.1088/0004-637X/766/2/91},
archivePrefix = {arXiv},
       eprint = {1301.0530},
 primaryClass = {astro-ph.CO},
       adsurl = {https://ui.adsabs.harvard.edu/abs/2013ApJ...766...91J}
}

@ARTICLE{z07,
       author = {{Zaritsky}, Dennis and {Christlein}, Daniel},
        title = "{On the Extended Knotted Disks of Galaxies}",
      journal = {\aj},
         year = 2007,
        month = jul,
       volume = {134},
       number = {1},
        pages = {135-141},
          doi = {10.1086/518238},
archivePrefix = {arXiv},
       eprint = {astro-ph/0703487},
 primaryClass = {astro-ph},
       adsurl = {https://ui.adsabs.harvard.edu/abs/2007AJ....134..135Z}
}

@ARTICLE{baker,
   author = {{Baker}, J.~G. and {Menzel}, D.~H.},
    title = "{Physical Processes in Gaseous Nebulae. III. The Balmer Decrement.}",
  journal = {\apj},
     year = 1938,
    month = jul,
   volume = 88,
    pages = {52},
      doi = {10.1086/143959},
   adsurl = {http://adsabs.harvard.edu/abs/1938ApJ....88...52B}
}

@ARTICLE{Nielsen2024,
       author = {{Nielsen}, Nikole M. and {Fisher}, Deanne B. and {Kacprzak}, Glenn G. and {Chisholm}, John and {Martin}, D. Christopher and {Reichardt Chu}, Bronwyn and {Sandstrom}, Karin M. and {Rickards Vaught}, Ryan J.},
        title = "{An emission map of the disk-circumgalactic medium transition in starburst IRAS 08339+6517}",
      journal = {Nature Astronomy},
         year = 2024,
        month = dec,
       volume = {8},
        pages = {1602-1609},
          doi = {10.1038/s41550-024-02365-x},
archivePrefix = {arXiv},
       eprint = {2311.00856},
 primaryClass = {astro-ph.GA},
       adsurl = {https://ui.adsabs.harvard.edu/abs/2024NatAs...8.1602N}
}

@ARTICLE{Wake2017,
       author = {{Wake}, David A. and {Bundy}, Kevin and {Diamond-Stanic}, Aleksandar M. and {Yan}, Renbin and {Blanton}, Michael R. and {Bershady}, Matthew A. and {S{\'a}nchez-Gallego}, Jos{\'e} R. and {Drory}, Niv and {Jones}, Amy and {Kauffmann}, Guinevere and {Law}, David R. and {Li}, Cheng and {MacDonald}, Nicholas and {Masters}, Karen and {Thomas}, Daniel and {Tinker}, Jeremy and {Weijmans}, Anne-Marie and {Brownstein}, Joel R.},
        title = "{The SDSS-IV MaNGA Sample: Design, Optimization, and Usage Considerations}",
      journal = {\aj},
         year = 2017,
        month = sep,
       volume = {154},
       number = {3},
          eid = {86},
        pages = {86},
          doi = {10.3847/1538-3881/aa7ecc},
archivePrefix = {arXiv},
       eprint = {1707.02989},
 primaryClass = {astro-ph.GA},
       adsurl = {https://ui.adsabs.harvard.edu/abs/2017AJ....154...86W}
}

@ARTICLE{Law2016,
       author = {{Law}, David R. and {Cherinka}, Brian and {Yan}, Renbin and {Andrews}, Brett H. and {Bershady}, Matthew A. and {Bizyaev}, Dmitry and {Blanc}, Guillermo A. and {Blanton}, Michael R. and {Bolton}, Adam S. and {Brownstein}, Joel R. and {Bundy}, Kevin and {Chen}, Yanmei and {Drory}, Niv and {D'Souza}, Richard and {Fu}, Hai and {Jones}, Amy and {Kauffmann}, Guinevere and {MacDonald}, Nicholas and {Masters}, Karen L. and {Newman}, Jeffrey A. and {Parejko}, John K. and {S{\'a}nchez-Gallego}, Jos{\'e} R. and {S{\'a}nchez}, Sebastian F. and {Schlegel}, David J. and {Thomas}, Daniel and {Wake}, David A. and {Weijmans}, Anne-Marie and {Westfall}, Kyle B. and {Zhang}, Kai},
        title = "{The Data Reduction Pipeline for the SDSS-IV MaNGA IFU Galaxy Survey}",
      journal = {\aj},
         year = 2016,
        month = oct,
       volume = {152},
       number = {4},
          eid = {83},
        pages = {83},
          doi = {10.3847/0004-6256/152/4/83},
archivePrefix = {arXiv},
       eprint = {1607.08619},
 primaryClass = {astro-ph.IM},
       adsurl = {https://ui.adsabs.harvard.edu/abs/2016AJ....152...83L}
}

@ARTICLE{Gunn2006,
       author = {{Gunn}, James E. and {Siegmund}, Walter A. and {Mannery}, Edward J. and {Owen}, Russell E. and {Hull}, Charles L. and {Leger}, R. French and {Carey}, Larry N. and {Knapp}, Gillian R. and {York}, Donald G. and {Boroski}, William N. and {Kent}, Stephen M. and {Lupton}, Robert H. and {Rockosi}, Constance M. and {Evans}, Michael L. and {Waddell}, Patrick and {Anderson}, John E. and {Annis}, James and {Barentine}, John C. and {Bartoszek}, Larry M. and {Bastian}, Steven and {Bracker}, Stephen B. and {Brewington}, Howard J. and {Briegel}, Charles I. and {Brinkmann}, Jon and {Brown}, Yorke J. and {Carr}, Michael A. and {Czarapata}, Paul C. and {Drennan}, Craig C. and {Dombeck}, Thomas and {Federwitz}, Glenn R. and {Gillespie}, Bruce A. and {Gonzales}, Carlos and {Hansen}, Sten U. and {Harvanek}, Michael and {Hayes}, Jeffrey and {Jordan}, Wendell and {Kinney}, Ellyne and {Klaene}, Mark and {Kleinman}, S.~J. and {Kron}, Richard G. and {Kresinski}, Jurek and {Lee}, Glenn and {Limmongkol}, Siriluk and {Lindenmeyer}, Carl W. and {Long}, Daniel C. and {Loomis}, Craig L. and {McGehee}, Peregrine M. and {Mantsch}, Paul M. and {Neilsen}, Jr., Eric H. and {Neswold}, Richard M. and {Newman}, Peter R. and {Nitta}, Atsuko and {Peoples}, Jr., John and {Pier}, Jeffrey R. and {Prieto}, Peter S. and {Prosapio}, Angela and {Rivetta}, Claudio and {Schneider}, Donald P. and {Snedden}, Stephanie and {Wang}, Shu-i.},
        title = "{The 2.5 m Telescope of the Sloan Digital Sky Survey}",
      journal = {\aj},
         year = 2006,
        month = apr,
       volume = {131},
       number = {4},
        pages = {2332-2359},
          doi = {10.1086/500975},
archivePrefix = {arXiv},
       eprint = {astro-ph/0602326},
 primaryClass = {astro-ph},
       adsurl = {https://ui.adsabs.harvard.edu/abs/2006AJ....131.2332G}
}

@ARTICLE{hummer,
   author = {{Hummer}, D.~G. and {Storey}, P.~J.},
    title = "{Recombination-line intensities for hydrogenic ions. I - Case B calculations for H I and He II}",
  journal = {\mnras},
     year = 1987,
    month = feb,
   volume = 224,
    pages = {801-820},
      doi = {10.1093/mnras/224.3.801},
   adsurl = {http://adsabs.harvard.edu/abs/1987MNRAS.224..801H}
}

@BOOK{osterbrock2006,
   author = {{Osterbrock}, D.~E. and {Ferland}, G.~J.},
    title = "{Astrophysics of gaseous nebulae and active galactic nuclei}",
booktitle = {Astrophysics of gaseous nebulae and active galactic nuclei, 2nd.~ed.~by D.E.~Osterbrock and G.J.~Ferland.~Sausalito, CA: University Science Books, 2006},
     year = 2006,
   adsurl = {http://adsabs.harvard.edu/abs/2006agna.book.....O}
}

@ARTICLE{Pointon2019,
       author = {{Pointon}, Stephanie K. and {Kacprzak}, Glenn G. and {Nielsen}, Nikole M. and {Muzahid}, Sowgat and {Murphy}, Michael T. and {Churchill}, Christopher W. and {Charlton}, Jane C.},
        title = "{Relationship between the Metallicity of the Circumgalactic Medium and Galaxy Orientation}",
      journal = {\apj},
         year = 2019,
        month = sep,
       volume = {883},
       number = {1},
          eid = {78},
        pages = {78},
          doi = {10.3847/1538-4357/ab3b0e},
archivePrefix = {arXiv},
       eprint = {1907.05557},
 primaryClass = {astro-ph.GA},
       adsurl = {https://ui.adsabs.harvard.edu/abs/2019ApJ...883...78P}
}

@ARTICLE{Nagao2006,
       author = {{Nagao}, T. and {Maiolino}, R. and {Marconi}, A.},
        title = "{Gas metallicity diagnostics in star-forming galaxies}",
      journal = {\ana},
         year = 2006,
        month = nov,
       volume = {459},
       number = {1},
        pages = {85-101},
          doi = {10.1051/0004-6361:20065216},
archivePrefix = {arXiv},
       eprint = {astro-ph/0603580},
 primaryClass = {astro-ph},
       adsurl = {https://ui.adsabs.harvard.edu/abs/2006A&A...459...85N}
}

@ARTICLE{Kewley2002,
       author = {{Kewley}, L.~J. and {Dopita}, M.~A.},
        title = "{Using Strong Lines to Estimate Abundances in Extragalactic H II Regions and Starburst Galaxies}",
      journal = {\apjs},
         year = 2002,
        month = sep,
       volume = {142},
       number = {1},
        pages = {35-52},
          doi = {10.1086/341326},
archivePrefix = {arXiv},
       eprint = {astro-ph/0206495},
 primaryClass = {astro-ph},
       adsurl = {https://ui.adsabs.harvard.edu/abs/2002ApJS..142...35K}
}

@ARTICLE{bpt,
   author = {{Baldwin}, J.~A. and {Phillips}, M.~M. and {Terlevich}, R.},
    title = "{Classification parameters for the emission-line spectra of extragalactic objects}",
  journal = {\pasp},
     year = 1981,
    month = feb,
   volume = 93,
    pages = {5-19},
      doi = {10.1086/130766},
   adsurl = {http://adsabs.harvard.edu/abs/1981PASP...93....5B}
}

@ARTICLE{Maiolino2019,
       author = {{Maiolino}, R. and {Mannucci}, F.},
        title = "{De re metallica: the cosmic chemical evolution of galaxies}",
      journal = {\aapr},
         year = 2019,
        month = feb,
       volume = {27},
       number = {1},
          eid = {3},
        pages = {3},
          doi = {10.1007/s00159-018-0112-2},
archivePrefix = {arXiv},
       eprint = {1811.09642},
 primaryClass = {astro-ph.GA},
       adsurl = {https://ui.adsabs.harvard.edu/abs/2019A&ARv..27....3M}
}

@ARTICLE{Tremonti2004,
       author = {{Tremonti}, Christy A. and {Heckman}, Timothy M. and {Kauffmann}, Guinevere and {Brinchmann}, Jarle and {Charlot}, St{\'e}phane and {White}, Simon D.~M. and {Seibert}, Mark and {Peng}, Eric W. and {Schlegel}, David J. and {Uomoto}, Alan and {Fukugita}, Masataka and {Brinkmann}, Jon},
        title = "{The Origin of the Mass-Metallicity Relation: Insights from 53,000 Star-forming Galaxies in the Sloan Digital Sky Survey}",
      journal = {\apj},
         year = 2004,
        month = oct,
       volume = {613},
       number = {2},
        pages = {898-913},
          doi = {10.1086/423264},
archivePrefix = {arXiv},
       eprint = {astro-ph/0405537},
 primaryClass = {astro-ph},
       adsurl = {https://ui.adsabs.harvard.edu/abs/2004ApJ...613..898T}
}

@ARTICLE{sdss7,
       author = {{Abazajian}, Kevork N. and {Adelman-McCarthy}, Jennifer K. and {Ag{\"u}eros}, Marcel A. and {Allam}, Sahar S. and {Allende Prieto}, Carlos and {An}, Deokkeun and {Anderson}, Kurt S.~J. and {Anderson}, Scott F. and {Annis}, James and {Bahcall}, Neta A. and {Bailer-Jones}, C.~A.~L. and {Barentine}, J.~C. and {Bassett}, Bruce A. and {Becker}, Andrew C. and {Beers}, Timothy C. and {Bell}, Eric F. and {Belokurov}, Vasily and {Berlind}, Andreas A. and {Berman}, Eileen F. and {Bernardi}, Mariangela and {Bickerton}, Steven J. and {Bizyaev}, Dmitry and {Blakeslee}, John P. and {Blanton}, Michael R. and {Bochanski}, John J. and {Boroski}, William N. and {Brewington}, Howard J. and {Brinchmann}, Jarle and {Brinkmann}, J. and {Brunner}, Robert J. and {Budav{\'a}ri}, Tam{\'a}s and {Carey}, Larry N. and {Carliles}, Samuel and {Carr}, Michael A. and {Castander}, Francisco J. and {Cinabro}, David and {Connolly}, A.~J. and {Csabai}, Istv{\'a}n and {Cunha}, Carlos E. and {Czarapata}, Paul C. and {Davenport}, James R.~A. and {de Haas}, Ernst and {Dilday}, Ben and {Doi}, Mamoru and {Eisenstein}, Daniel J. and {Evans}, Michael L. and {Evans}, N.~W. and {Fan}, Xiaohui and {Friedman}, Scott D. and {Frieman}, Joshua A. and {Fukugita}, Masataka and {G{\"a}nsicke}, Boris T. and {Gates}, Evalyn and {Gillespie}, Bruce and {Gilmore}, G. and {Gonzalez}, Belinda and {Gonzalez}, Carlos F. and {Grebel}, Eva K. and {Gunn}, James E. and {Gy{\"o}ry}, Zsuzsanna and {Hall}, Patrick B. and {Harding}, Paul and {Harris}, Frederick H. and {Harvanek}, Michael and {Hawley}, Suzanne L. and {Hayes}, Jeffrey J.~E. and {Heckman}, Timothy M. and {Hendry}, John S. and {Hennessy}, Gregory S. and {Hindsley}, Robert B. and {Hoblitt}, J. and {Hogan}, Craig J. and {Hogg}, David W. and {Holtzman}, Jon A. and {Hyde}, Joseph B. and {Ichikawa}, Shin-ichi and {Ichikawa}, Takashi and {Im}, Myungshin and {Ivezi{\'c}}, {\v{Z}}eljko and {Jester}, Sebastian and {Jiang}, Linhua and {Johnson}, Jennifer A. and {Jorgensen}, Anders M. and {Juri{\'c}}, Mario and {Kent}, Stephen M. and {Kessler}, R. and {Kleinman}, S.~J. and {Knapp}, G.~R. and {Konishi}, Kohki and {Kron}, Richard G. and {Krzesinski}, Jurek and {Kuropatkin}, Nikolay and {Lampeitl}, Hubert and {Lebedeva}, Svetlana and {Lee}, Myung Gyoon and {Lee}, Young Sun and {French Leger}, R. and {L{\'e}pine}, S{\'e}bastien and {Li}, Nolan and {Lima}, Marcos and {Lin}, Huan and {Long}, Daniel C. and {Loomis}, Craig P. and {Loveday}, Jon and {Lupton}, Robert H. and {Magnier}, Eugene and {Malanushenko}, Olena and {Malanushenko}, Viktor and {Mandelbaum}, Rachel and {Margon}, Bruce and {Marriner}, John P. and {Mart{\'\i}nez-Delgado}, David and {Matsubara}, Takahiko and {McGehee}, Peregrine M. and {McKay}, Timothy A. and {Meiksin}, Avery and {Morrison}, Heather L. and {Mullally}, Fergal and {Munn}, Jeffrey A. and {Murphy}, Tara and {Nash}, Thomas and {Nebot}, Ada and {Neilsen}, Jr., Eric H. and {Newberg}, Heidi Jo and {Newman}, Peter R. and {Nichol}, Robert C. and {Nicinski}, Tom and {Nieto-Santisteban}, Maria and {Nitta}, Atsuko and {Okamura}, Sadanori and {Oravetz}, Daniel J. and {Ostriker}, Jeremiah P. and {Owen}, Russell and {Padmanabhan}, Nikhil and {Pan}, Kaike and {Park}, Changbom and {Pauls}, George and {Peoples}, Jr., John and {Percival}, Will J. and {Pier}, Jeffrey R. and {Pope}, Adrian C. and {Pourbaix}, Dimitri and {Price}, Paul A. and {Purger}, Norbert and {Quinn}, Thomas and {Raddick}, M. Jordan and {Re Fiorentin}, Paola and {Richards}, Gordon T. and {Richmond}, Michael W. and {Riess}, Adam G. and {Rix}, Hans-Walter and {Rockosi}, Constance M. and {Sako}, Masao and {Schlegel}, David J. and {Schneider}, Donald P. and {Scholz}, Ralf-Dieter and {Schreiber}, Matthias R. and {Schwope}, Axel D. and {Seljak}, Uro{\v{s}} and {Sesar}, Branimir and {Sheldon}, Erin and {Shimasaku}, Kazu and {Sibley}, Valena C. and {Simmons}, A.~E. and {Sivarani}, Thirupathi and {Allyn Smith}, J. and {Smith}, Martin C. and {Smol{\v{c}}i{\'c}}, Vernesa and {Snedden}, Stephanie A. and {Stebbins}, Albert and {Steinmetz}, Matthias and {Stoughton}, Chris and {Strauss}, Michael A. and {SubbaRao}, Mark and {Suto}, Yasushi and {Szalay}, Alexander S. and {Szapudi}, Istv{\'a}n and {Szkody}, Paula and {Tanaka}, Masayuki and {Tegmark}, Max and {Teodoro}, Luis F.~A. and {Thakar}, Aniruddha R. and {Tremonti}, Christy A. and {Tucker}, Douglas L. and {Uomoto}, Alan and {Vanden Berk}, Daniel E. and {Vandenberg}, Jan and {Vidrih}, S. and {Vogeley}, Michael S. and {Voges}, Wolfgang and {Vogt}, Nicole P. and {Wadadekar}, Yogesh and {Watters}, Shannon and {Weinberg}, David H. and {West}, Andrew A. and {White}, Simon D.~M. and {Wilhite}, Brian C. and {Wonders}, Alainna C. and {Yanny}, Brian and {Yocum}, D.~R.},
        title = "{The Seventh Data Release of the Sloan Digital Sky Survey}",
      journal = {\apjs},
         year = 2009,
        month = jun,
       volume = {182},
       number = {2},
        pages = {543-558},
          doi = {10.1088/0067-0049/182/2/543},
archivePrefix = {arXiv},
       eprint = {0812.0649},
 primaryClass = {astro-ph},
       adsurl = {https://ui.adsabs.harvard.edu/abs/2009ApJS..182..543A}
}

@ARTICLE{Dopita2017,
       author = {{Dopita}, Michael A. and {Sutherland}, Ralph S.},
        title = "{Effects of Preionization in Radiative Shocks. II. Application to the Herbig-Haro Objects}",
      journal = {\apjs},
         year = 2017,
        month = apr,
       volume = {229},
       number = {2},
          eid = {35},
        pages = {35},
          doi = {10.3847/1538-4365/aa6542},
archivePrefix = {arXiv},
       eprint = {1703.02700},
 primaryClass = {astro-ph.GA},
       adsurl = {https://ui.adsabs.harvard.edu/abs/2017ApJS..229...35D}
}

@ARTICLE{riess,
       author = {{Riess}, Adam G. and {Casertano}, Stefano and {Yuan}, Wenlong and
         {Macri}, Lucas and {Bucciarelli}, Beatrice and {Lattanzi}, Mario G. and
         {MacKenty}, John W. and {Bowers}, J. Bradley and {Zheng}, WeiKang and
         {Filippenko}, Alexei V. and {Huang}, Caroline and
         {Anderson}, Richard I.},
        title = "{Milky Way Cepheid Standards for Measuring Cosmic Distances and Application to Gaia DR2: Implications for the Hubble Constant}",
      journal = {\apj},
         year = "2018",
        month = "Jul",
       volume = {861},
       number = {2},
          eid = {126},
        pages = {126},
          doi = {10.3847/1538-4357/aac82e},
archivePrefix = {arXiv},
       eprint = {1804.10655},
 primaryClass = {astro-ph.CO},
       adsurl = {https://ui.adsabs.harvard.edu/abs/2018ApJ...861..126R}
}

@ARTICLE{Planck2018,
       author = {{Planck Collaboration} and {Akrami}, Y. and {Arroja}, F. and {Ashdown},
        M. and {Aumont}, J. and {Baccigalupi}, C. and {Ballardini}, M.
        and {Banday}, A.~J. and {Barreiro}, R.~B. and {Bartolo}, N. },
        title = "{Planck 2018 results. I. Overview and the cosmological legacy of Planck}",
      journal = {arXiv e-prints},
         year = 2018,
        month = Jul,
          eid = {arXiv:1807.06205},
        pages = {arXiv:1807.06205},
archivePrefix = {arXiv},
       eprint = {1807.06205},
 primaryClass = {astro-ph.CO},
       adsurl = {https://ui.adsabs.harvard.edu/\#abs/2018arXiv180706205P}
}

@ARTICLE{Zhang2016,
   author = {{Zhang}, H. and {Zaritsky}, D. and {Zhu}, G. and {M{\'e}nard}, B. and 
	{Hogg}, D.~W.},
    title = "{Hydrogen Emission from the Ionized Gaseous Halos of Low-redshift Galaxies}",
  journal = {ApJ},
archivePrefix = "arXiv",
   eprint = {1611.00004},
     year = 2016,
    month = dec,
   volume = 833,
      eid = {276},
    pages = {276},
      doi = {10.3847/1538-4357/833/2/276},
   adsurl = {http://adsabs.harvard.edu/abs/2016ApJ...833..276Z}
}

@ARTICLE{Zhang2018a,
   author = {{Zhang}, H. and {Zaritsky}, D. and {Behroozi}, P.},
    title = "{Emission from the Ionized Gaseous Halos of Low-redshift Galaxies and Their Neighbors}",
  journal = {ApJ},
archivePrefix = "arXiv",
   eprint = {1805.08217},
     year = 2018,
    month = jul,
   volume = 861,
      eid = {34},
    pages = {34 (Paper II)},
      doi = {10.3847/1538-4357/aac6b7},
   adsurl = {http://adsabs.harvard.edu/abs/2018ApJ...861...34Z}
}

@ARTICLE{Zhang2018b,
       author = {{Zhang}, Huanian and {Zaritsky}, Dennis and {Werk}, Jessica and
        {Behroozi}, Peter},
        title = "{Emission Line Ratios for the Circumgalactic Medium and the {\textquotedblleft}Bimodal{\textquotedblright} Nature of Galaxies}",
      journal = {ApJL},
         year = 2018,
        month = Oct,
       volume = {866},
          eid = {L4},
        pages = {L4 (Paper III)},
          doi = {10.3847/2041-8213/aae37e},
archivePrefix = {arXiv},
       eprint = {1809.09113},
 primaryClass = {astro-ph.GA},
       adsurl = {https://ui.adsabs.harvard.edu/\#abs/2018ApJ...866L...4Z}
}

@ARTICLE{Zhang2019,
       author = {{Zhang}, Huanian and {Zaritsky}, Dennis and {Behroozi}, Peter and
         {Werk}, Jessica},
        title = "{On the Effect of Environment on Line Emission from the Circumgalactic Medium}",
      journal = {ApJ},
         year = 2019,
        month = jul,
       volume = {880},
       number = {1},
          eid = {28},
        pages = {28 (Paper IV)},
          doi = {10.3847/1538-4357/ab2761},
archivePrefix = {arXiv},
       eprint = {1906.01643},
 primaryClass = {astro-ph.GA},
       adsurl = {https://ui.adsabs.harvard.edu/abs/2019ApJ...880...28Z}
}

@ARTICLE{Zhang2020a,
       author = {{Zhang}, Huanian and {Yang}, Xiaohu and {Zaritsky}, Dennis and
         {Behroozi}, Peter and {Werk}, Jessica},
        title = "{H$\alpha$ Emission and the Dependence of the Circumgalactic Cool Gas Fraction on Halo Mass}",
      journal = {ApJ},
         year = 2020,
        month = jan,
        volume = {880},
       number = {1},
          eid = {33},
        pages = {33 (Paper V)},
        doi = {10.3847/1538-4357/ab55ed},
archivePrefix = {arXiv},
       eprint = {1911.02032},
 primaryClass = {astro-ph.GA},
       adsurl = {https://ui.adsabs.harvard.edu/abs/2019arXiv191102032Z}
}

@ARTICLE{Zhang2020b,
       author = {{Zhang}, Huanian and {Fang}, Taotao and {Zaritsky}, Dennis and
         {Behroozi}, Peter and {Werk}, Jessica and {Yang}, Xiaohu},
        title = "{Observing the Effects of Galaxy Interactions on the Circumgalactic Medium}",
      journal = {ApJL},
         year = 2020,
        month = apr,
       volume = {893},
       number = {1},
          eid = {L3},
        pages = {L3 (Paper VI)},
          doi = {10.3847/2041-8213/ab8068},
archivePrefix = {arXiv},
       eprint = {2003.08393},
 primaryClass = {astro-ph.GA},
       adsurl = {https://ui.adsabs.harvard.edu/abs/2020ApJ...893L...3Z}
}

@ARTICLE{Zhang2021,
       author = {{Zhang}, Huanian and {Zaritsky}, Dennis and {Olsen}, Karen Pardos and {Behroozi}, Peter and {Werk}, Jessica and {Kennicutt}, Robert and {Xie}, Lizhi and {Yang}, Xiaohu and {Fang}, Taotao and {De Lucia}, Gabriella and {Hirschmann}, Michaela and {Fontanot}, Fabio},
        title = "{An Empirical Determination of the Dependence of the Circumgalactic Mass Cooling Rate and Feedback Mass Loading Factor on Galactic Stellar Mass}",
      journal = {ApJ},
         year = 2021,
        month = aug,
       volume = {916},
       number = {2},
          eid = {101},
        pages = {101},
          doi = {10.3847/1538-4357/ac0433},
archivePrefix = {arXiv},
       eprint = {2104.12777},
 primaryClass = {astro-ph.GA},
       adsurl = {https://ui.adsabs.harvard.edu/abs/2021ApJ...916..101Z}
}

@ARTICLE{MUSE-HDFS,
       author = {{Bacon}, R. and {Brinchmann}, J. and {Richard}, J. and {Contini}, T. and {Drake}, A. and {Franx}, M. and {Tacchella}, S. and {Vernet}, J. and {Wisotzki}, L. and {Blaizot}, J. and {Bouch{\'e}}, N. and {Bouwens}, R. and {Cantalupo}, S. and {Carollo}, C.~M. and {Carton}, D. and {Caruana}, J. and {Cl{\'e}ment}, B. and {Dreizler}, S. and {Epinat}, B. and {Guiderdoni}, B. and {Herenz}, C. and {Husser}, T. -O. and {Kamann}, S. and {Kerutt}, J. and {Kollatschny}, W. and {Krajnovic}, D. and {Lilly}, S. and {Martinsson}, T. and {Michel-Dansac}, L. and {Patricio}, V. and {Schaye}, J. and {Shirazi}, M. and {Soto}, K. and {Soucail}, G. and {Steinmetz}, M. and {Urrutia}, T. and {Weilbacher}, P. and {de Zeeuw}, T.},
        title = "{The MUSE 3D view of the Hubble Deep Field South}",
      journal = {Astron. \& Astroph.},
         year = 2015,
        month = mar,
       volume = {575},
          eid = {A75},
        pages = {A75},
          doi = {10.1051/0004-6361/201425419},
archivePrefix = {arXiv},
       eprint = {1411.7667},
 primaryClass = {astro-ph.GA},
       adsurl = {https://ui.adsabs.harvard.edu/abs/2015A&A...575A..75B}
}

@ARTICLE{Herenz2023,
       author = {{Herenz}, E.~C. and {Inoue}, J. and {Salas}, H. and {Koenigs}, B. and {Moya-Sierralta}, C. and {Cannon}, J.~M. and {Hayes}, M. and {Papaderos}, P. and {{\"O}stlin}, G. and {Bik}, A. and {Le Reste}, A. and {Kusakabe}, H. and {Monreal-Ibero}, A. and {Puschnig}, J.},
        title = "{A {\ensuremath{\sim}}15 kpc outflow cone piercing through the halo of the blue compact metal-poor galaxy SBS 0335-052E}",
      journal = {\ana},
         year = 2023,
        month = feb,
       volume = {670},
          eid = {A121},
        pages = {A121},
          doi = {10.1051/0004-6361/202244930},
archivePrefix = {arXiv},
       eprint = {2212.01239},
 primaryClass = {astro-ph.GA},
       adsurl = {https://ui.adsabs.harvard.edu/abs/2023A&A...670A.121H}
}

@ARTICLE{Yoshida2016,
       author = {{Yoshida}, Michitoshi and {Yagi}, Masafumi and {Ohyama}, Youichi and {Komiyama}, Yutaka and {Kashikawa}, Nobunari and {Tanaka}, Hisashi and {Okamura}, Sadanori},
        title = "{Giant H{\ensuremath{\alpha}} Nebula Surrounding the Starburst Merger NGC 6240}",
      journal = {\apj},
         year = 2016,
        month = mar,
       volume = {820},
       number = {1},
          eid = {48},
        pages = {48},
          doi = {10.3847/0004-637X/820/1/48},
archivePrefix = {arXiv},
       eprint = {1512.09208},
 primaryClass = {astro-ph.GA},
       adsurl = {https://ui.adsabs.harvard.edu/abs/2016ApJ...820...48Y}
}

@ARTICLE{Mackenzie2019,
       author = {{Mackenzie}, Ruari and {Fumagalli}, Michele and {Theuns}, Tom and {Hatton}, David J. and {Garel}, Thibault and {Cantalupo}, Sebastiano and {Christensen}, Lise and {Fynbo}, Johan P.~U. and {Kanekar}, Nissim and {M{\o}ller}, Palle and {O'Meara}, John and {Prochaska}, J. Xavier and {Rafelski}, Marc and {Shanks}, Tom and {Trayford}, James},
        title = "{Linking gas and galaxies at high redshift: MUSE surveys the environments of six damped Ly{\ensuremath{\alpha}} systems at z {\ensuremath{\approx}} 3}",
      journal = {\mnras},
         year = 2019,
        month = aug,
       volume = {487},
       number = {4},
        pages = {5070-5096},
          doi = {10.1093/mnras/stz1501},
archivePrefix = {arXiv},
       eprint = {1904.07254},
 primaryClass = {astro-ph.GA},
       adsurl = {https://ui.adsabs.harvard.edu/abs/2019MNRAS.487.5070M}
}

@ARTICLE{Kusakabe2022,
       author = {{Kusakabe}, Haruka and {Verhamme}, Anne and {Blaizot}, J{\'e}r{\'e}my and {Garel}, Thibault and {Wisotzki}, Lutz and {Leclercq}, Floriane and {Bacon}, Roland and {Schaye}, Joop and {Gallego}, Sofia G. and {Kerutt}, Josephine and {Matthee}, Jorryt and {Maseda}, Michael and {Nanayakkara}, Themiya and {Pell{\'o}}, Roser and {Richard}, Johan and {Tresse}, Laurence and {Urrutia}, Tanya and {Vitte}, Elo{\"\i}se},
        title = "{The MUSE eXtremely Deep Field: Individual detections of Ly{\ensuremath{\alpha}} haloes around rest-frame UV-selected galaxies at z = 2.9-4.4}",
      journal = {\ana},
         year = 2022,
        month = apr,
       volume = {660},
          eid = {A44},
        pages = {A44},
          doi = {10.1051/0004-6361/202142302},
archivePrefix = {arXiv},
       eprint = {2201.07257},
 primaryClass = {astro-ph.GA},
       adsurl = {https://ui.adsabs.harvard.edu/abs/2022A&A...660A..44K}
}

@ARTICLE{Peroux2019,
       author = {{P{\'e}roux}, C{\'e}line and {Zwaan}, Martin A. and {Klitsch}, Anne and {Augustin}, Ramona and {Hamanowicz}, Aleksandra and {Rahmani}, Hadi and {Pettini}, Max and {Kulkarni}, Varsha and {Straka}, Lorrie A. and {Biggs}, Andy D. and {York}, Donald G. and {Milliard}, Bruno},
        title = "{Multiphase circumgalactic medium probed with MUSE and ALMA}",
      journal = {\mnras},
         year = 2019,
        month = may,
       volume = {485},
       number = {2},
        pages = {1595-1613},
          doi = {10.1093/mnras/stz202},
archivePrefix = {arXiv},
       eprint = {1901.05217},
 primaryClass = {astro-ph.GA},
       adsurl = {https://ui.adsabs.harvard.edu/abs/2019MNRAS.485.1595P}
}

@ARTICLE{Kewley2019,
       author = {{Kewley}, Lisa J. and {Nicholls}, David C. and {Sutherland}, Ralph S.},
        title = "{Understanding Galaxy Evolution Through Emission Lines}",
      journal = {\araa},
         year = 2019,
        month = aug,
       volume = {57},
        pages = {511-570},
          doi = {10.1146/annurev-astro-081817-051832},
archivePrefix = {arXiv},
       eprint = {1910.09730},
 primaryClass = {astro-ph.GA},
       adsurl = {https://ui.adsabs.harvard.edu/abs/2019ARA&A..57..511K}
}

@ARTICLE{Dutta2023,
       author = {{Dutta}, Rajeshwari and {Fossati}, Matteo and {Fumagalli}, Michele and {Revalski}, Mitchell and {Lofthouse}, Emma K. and {Nelson}, Dylan and {Papini}, Giulia and {Rafelski}, Marc and {Cantalupo}, Sebastiano and {Arrigoni Battaia}, Fabrizio and {Dayal}, Pratika and {Longobardi}, Alessia and {P{\'e}roux}, Celine and {Prichard}, Laura J. and {Prochaska}, J. Xavier},
        title = "{Metal line emission from galaxy haloes at z {\ensuremath{\approx}} 1}",
      journal = {\mnras},
         year = 2023,
        month = jun,
       volume = {522},
       number = {1},
        pages = {535-558},
          doi = {10.1093/mnras/stad1002},
archivePrefix = {arXiv},
       eprint = {2302.09087},
 primaryClass = {astro-ph.GA},
       adsurl = {https://ui.adsabs.harvard.edu/abs/2023MNRAS.522..535D}
}

@ARTICLE{Dutta2024,
       author = {{Dutta}, Rajeshwari and {Fumagalli}, Michele and {Fossati}, Matteo and {Rafelski}, Marc and {Revalski}, Mitchell and {Arrigoni Battaia}, Fabrizio and {D'Odorico}, Valentina and {P{\'e}roux}, Celine and {Prichard}, Laura J. and {Swinbank}, A. Mark},
        title = "{Metal line emission around z $<$ 1 galaxies}",
      journal = {\aap},
         year = 2024,
        month = nov,
       volume = {691},
          eid = {A236},
        pages = {A236},
          doi = {10.1051/0004-6361/202450733},
archivePrefix = {arXiv},
       eprint = {2409.02182},
 primaryClass = {astro-ph.GA},
       adsurl = {https://ui.adsabs.harvard.edu/abs/2024A&A...691A.236D}
}

@ARTICLE{Behroozi2019,
       author = {{Behroozi}, Peter and {Wechsler}, Risa H. and {Hearin}, Andrew P. and {Conroy}, Charlie},
        title = "{UNIVERSEMACHINE: The correlation between galaxy growth and dark matter halo assembly from z = 0-10}",
      journal = {MNRAS},
         year = 2019,
        month = sep,
       volume = {488},
       number = {3},
        pages = {3143-3194},
          doi = {10.1093/mnras/stz1182},
archivePrefix = {arXiv},
       eprint = {1806.07893},
 primaryClass = {astro-ph.GA},
       adsurl = {https://ui.adsabs.harvard.edu/abs/2019MNRAS.488.3143B}
}

@ARTICLE{SDSSDR16,
       author = {{Ahumada}, Romina and {Allende Prieto}, Carlos and
         {Almeida}, Andr{\'e}s and {Anders}, Friedrich and {Anderson}},
        title = "{The 16th Data Release of the Sloan Digital Sky Surveys: First Release from the APOGEE-2 Southern Survey and Full Release of eBOSS Spectra}",
      journal = {ApJS},
         year = 2020,
        month = jul,
       volume = {249},
       number = {1},
          eid = {3},
        pages = {3},
          doi = {10.3847/1538-4365/ab929e},
archivePrefix = {arXiv},
       eprint = {1912.02905},
 primaryClass = {astro-ph.GA},
       adsurl = {https://ui.adsabs.harvard.edu/abs/2020ApJS..249....3A}
}

@ARTICLE{Werk2014,
       author = {{Werk}, Jessica K. and {Prochaska}, J. Xavier and {Tumlinson}, Jason and {Peeples}, Molly S. and {Tripp}, Todd M. and {Fox}, Andrew J. and {Lehner}, Nicolas and {Thom}, Christopher and {O'Meara}, John M. and {Ford}, Amanda Brady and {Bordoloi}, Rongmon and {Katz}, Neal and {Tejos}, Nicolas and {Oppenheimer}, Benjamin D. and {Dav{\'e}}, Romeel and {Weinberg}, David H.},
        title = "{The COS-Halos Survey: Physical Conditions and Baryonic Mass in the Low-redshift Circumgalactic Medium}",
      journal = {\apj},
         year = 2014,
        month = sep,
       volume = {792},
       number = {1},
          eid = {8},
        pages = {8},
          doi = {10.1088/0004-637X/792/1/8},
archivePrefix = {arXiv},
       eprint = {1403.0947},
 primaryClass = {astro-ph.CO},
       adsurl = {https://ui.adsabs.harvard.edu/abs/2014ApJ...792....8W}
}

@ARTICLE{prochaska2017,
       author = {{Prochaska}, J. Xavier and {Werk}, Jessica K. and {Worseck}, G{\'a}bor and {Tripp}, Todd M. and {Tumlinson}, Jason and {Burchett}, Joseph N. and {Fox}, Andrew J. and {Fumagalli}, Michele and {Lehner}, Nicolas and {Peeples}, Molly S. and {Tejos}, Nicolas},
        title = "{The COS-Halos Survey: Metallicities in the Low-redshift Circumgalactic Medium}",
      journal = {\apj},
         year = 2017,
        month = mar,
       volume = {837},
       number = {2},
          eid = {169},
        pages = {169},
          doi = {10.3847/1538-4357/aa6007},
archivePrefix = {arXiv},
       eprint = {1702.02618},
 primaryClass = {astro-ph.GA},
       adsurl = {https://ui.adsabs.harvard.edu/abs/2017ApJ...837..169P}
}

@ARTICLE{Behroozi2010,
   author = {{Behroozi}, P.~S. and {Conroy}, C. and {Wechsler}, R.~H.},
    title = "{A Comprehensive Analysis of Uncertainties Affecting the Stellar Mass-Halo Mass Relation for 0 $<$ z $<$ 4}",
  journal = {\apj},
archivePrefix = "arXiv",
   eprint = {1001.0015},
 primaryClass = "astro-ph.CO",
     year = 2010,
    month = jul,
   volume = 717,
    pages = {379-403},
      doi = {10.1088/0004-637X/717/1/379},
   adsurl = {http://adsabs.harvard.edu/abs/2010ApJ...717..379B}
}

@ARTICLE{McGaugh2010,
   author = {{McGaugh}, S.~S. and {Schombert}, J.~M. and {de Blok}, W.~J.~G. and 
	{Zagursky}, M.~J.},
    title = "{The Baryon Content of Cosmic Structures}",
  journal = {\apjl},
archivePrefix = "arXiv",
   eprint = {0911.2700},
 primaryClass = "astro-ph.CO",
     year = 2010,
    month = jan,
   volume = 708,
    pages = {L14-L17},
      doi = {10.1088/2041-8205/708/1/L14},
   adsurl = {http://adsabs.harvard.edu/abs/2010ApJ...708L..14M}
}

@ARTICLE{CGM2017,
   author = {{Tumlinson}, J. and {Peeples}, M.~S. and {Werk}, J.~K.},
    title = "{The Circumgalactic Medium}",
  journal = {\araa},
archivePrefix = "arXiv",
   eprint = {1709.09180},
     year = 2017,
    month = aug,
   volume = 55,
    pages = {389-432},
      doi = {10.1146/annurev-astro-091916-055240},
   adsurl = {http://adsabs.harvard.edu/abs/2017ARA%26A..55..389T}
}

@INPROCEEDINGS{MUSE2010,
       author = {{Bacon}, R. and {Accardo}, M. and {Adjali}, L. and {Anwand}, H. and {Bauer}, S. and {Biswas}, I. and {Blaizot}, J. and {Boudon}, D. and {Brau-Nogue}, S. and {Brinchmann}, J. and {Caillier}, P. and {Capoani}, L. and {Carollo}, C.~M. and {Contini}, T. and {Couderc}, P. and {Daguis{\'e}}, E. and {Deiries}, S. and {Delabre}, B. and {Dreizler}, S. and {Dubois}, J. and {Dupieux}, M. and {Dupuy}, C. and {Emsellem}, E. and {Fechner}, T. and {Fleischmann}, A. and {Fran{\c{c}}ois}, M. and {Gallou}, G. and {Gharsa}, T. and {Glindemann}, A. and {Gojak}, D. and {Guiderdoni}, B. and {Hansali}, G. and {Hahn}, T. and {Jarno}, A. and {Kelz}, A. and {Koehler}, C. and {Kosmalski}, J. and {Laurent}, F. and {Le Floch}, M. and {Lilly}, S.~J. and {Lizon}, J. -L. and {Loupias}, M. and {Manescau}, A. and {Monstein}, C. and {Nicklas}, H. and {Olaya}, J. -C. and {Pares}, L. and {Pasquini}, L. and {P{\'e}contal-Rousset}, A. and {Pell{\'o}}, R. and {Petit}, C. and {Popow}, E. and {Reiss}, R. and {Remillieux}, A. and {Renault}, E. and {Roth}, M. and {Rupprecht}, G. and {Serre}, D. and {Schaye}, J. and {Soucail}, G. and {Steinmetz}, M. and {Streicher}, O. and {Stuik}, R. and {Valentin}, H. and {Vernet}, J. and {Weilbacher}, P. and {Wisotzki}, L. and {Yerle}, N.},
        title = "{The MUSE second-generation VLT instrument}",
    booktitle = {Ground-based and Airborne Instrumentation for Astronomy III},
         year = 2010,
       editor = {{McLean}, Ian S. and {Ramsay}, Suzanne K. and {Takami}, Hideki},
       series = {Society of Photo-Optical Instrumentation Engineers (SPIE) Conference Series},
       volume = {7735},
        month = jul,
          eid = {773508},
        pages = {773508},
          doi = {10.1117/12.856027},
       adsurl = {https://ui.adsabs.harvard.edu/abs/2010SPIE.7735E..08B}
}


\section*{Acknowledgments}
HZ acknowledges financial support from the start-up funding of the Huazhong University of Science and Technology, the National Science Foundation of China grant (No. 12303007) and the China Manned Space Program (CMS-CSST-2025-A06). DZ and HZ acknowledge financial support from NSF AST-2006785 and the University of Arizona. LZX acknowledges support from the National Natural Science Foundation of China (grant number 12041302), the Ministry of Science and Technology of China (grant No. 2020SKA0110100). LCH was supported by the National Science Foundation of China (12233001), the National Key R\&D Program of China (2022YFF0503401), and the China Manned Space Program (CMS-CSST-2025-A09). The authors gratefully acknowledge the SDSS-IV team for providing a valuable resource to the community.
Funding for SDSS-IV has been provided by the Alfred P. Sloan Foundation, the Participating I institutions, the National Science Foundation, and the U.S. Department of Energy Office of Science. The SDSS-IV web site is \url{http://www.sdss4.org/}.

SDSS-IV is managed by the Astrophysical Research Consortium for the Participating Institutions of the SDSS-III Collaboration including the University of Arizona, the Brazilian Participation Group, Brookhaven National Laboratory, Carnegie Mellon University, University of Florida, the French Participation Group, the German Participation Group, Harvard University, the Instituto de Astrofisica de Canarias, the Michigan State/Notre Dame/JINA Participation Group, Johns Hopkins University, Lawrence Berkeley National Laboratory, Max Planck Institute for Astrophysics, Max Planck Institute for Extraterrestrial Physics, New Mexico State University, New York University, Ohio State University, Pennsylvania State University, University of Portsmouth, Princeton University, the Spanish Participation Group, University of Tokyo, University of Utah, Vanderbilt University, University of Virginia, University of Washington, and Yale University.

\section*{Funding} 
This work was supported by the start-up funding of the Huazhong University of Science and Technology, the National Science Foundation of China grant (No. 12303007), the China Manned Space Program (CMS-CSST-2025-A06), the NSF AST-2006785, the National Natural Science Foundation of China (grant number 12041302), the Ministry of Science and Technology of China (grant No. 2020SKA0110100), the National Science Foundation of China (12233001), the National Key R\&D Program of China (2022YFF0503401), and the China Manned Space Program (CMS-CSST-2025-A09).

\section*{Author Contributions}
 HZ led the data analysis of the MaNGA sample, LX led the analysis using simulations, HZ, DZ, LX, ZQ and LH led the interpretation of the results, LH provided the initial suggestion for the program. HZ, DZ, LX and ZQ led the manuscirpt writing, and LH, MB, FF, MH, CL, GL, EW contributed to the manuscript writing and constructive comments and suggestions. HYZ helped download datacubes and catalogs.

\section*{Competing  Interests}
The authors declare that they have no competing interests.

\section*{Data Code, and Materials Availability}
 
All data and code needed to evaluate and reproduce the results in the paper are present in the paper and/or the Supplementary Materials. This study did not generate new materials. All spectra data used in this paper are publicly available in the SDSS database \url{https://data.sdss.org/sas/dr17/manga/spectro/redux/v3_1_1/}.

\end{document}